\documentclass[%
 pra,reprint,nofootinbib,
 amsmath,amssymb,
 aps,
floatfix,superscriptaddress
]{revtex4-2}
\usepackage{graphicx}
\usepackage{dcolumn}
\usepackage{bm}
\usepackage{siunitx}
\usepackage{physics}
\usepackage{xcolor}
\usepackage[caption=false]{subfig}
\usepackage[export]{adjustbox}
\usepackage{hyperref}
\hypersetup{
        colorlinks=false
}

\usepackage{nicematrix}
\NiceMatrixOptions{
code-for-first-row = \color{black} ,
code-for-last-row = \color{black} ,
code-for-first-col = \color{black} ,
code-for-last-col = \color{black}
}
\usepackage{float}

\AtBeginDocument{\RenewCommandCopy\qty\SI} 
\makeatletter
\let\save@mathaccent\mathaccent
\newcommand*\if@single[3]{%
  \setbox0\hbox{${\mathaccent"0362{#1}}^H$}%
  \setbox2\hbox{${\mathaccent"0362{\kern0pt#1}}^H$}%
  \ifdim\ht0=\ht2 #3\else #2\fi
  }
\newcommand*\rel@kern[1]{\kern#1\dimexpr\macc@kerna}
\newcommand*\wideaccent[2]{\@ifnextchar^{{\wide@accent{#1}{#2}{0}}}{\wide@accent{#1}{#2}{1}}}
\newcommand*\wide@accent[3]{\if@single{#2}{\wide@accent@{#1}{#2}{#3}{1}}{\wide@accent@{#1}{#2}{#3}{2}}}
\newcommand*\wide@accent@[4]{%
  \begingroup
  \def\mathaccent##1##2{%
    \let\mathaccent\save@mathaccent
    \if#42 \let\macc@nucleus\first@char \fi
    \setbox\z@\hbox{$\macc@style{\macc@nucleus}_{}$}%
    \setbox\tw@\hbox{$\macc@style{\macc@nucleus}{}_{}$}%
    \dimen@\wd\tw@
    \advance\dimen@-\wd\z@
    \divide\dimen@ 3
    \@tempdima\wd\tw@
    \advance\@tempdima-\scriptspace
    \divide\@tempdima 10
    \advance\dimen@-\@tempdima
    \ifdim\dimen@>\z@ \dimen@0pt\fi
    \rel@kern{0.6}\kern-\dimen@
    \if#41
      #1{\rel@kern{-0.6}\kern\dimen@\macc@nucleus\rel@kern{0.4}\kern\dimen@}%
      \advance\dimen@0.4\dimexpr\macc@kerna
      \let\final@kern#3%
      \ifdim\dimen@<\z@ \let\final@kern1\fi
      \if\final@kern1 \kern-\dimen@\fi
    \else
      #1{\rel@kern{-0.6}\kern\dimen@#2}%
    \fi
  }%
  \macc@depth\@ne
  \let\math@bgroup\@empty \let\math@egroup\macc@set@skewchar
  \mathsurround\z@ \frozen@everymath{\mathgroup\macc@group\relax}%
  \macc@set@skewchar\relax
  \let\mathaccentV\macc@nested@a
  \if#41
    \macc@nested@a\relax111{#2}%
  \else
    \def\gobble@till@marker##1\endmarker{}%
    \futurelet\first@char\gobble@till@marker#2\endmarker
    \ifcat\noexpand\first@char A\else
      \def\first@char{}%
    \fi
    \macc@nested@a\relax111{\first@char}%
  \fi
  \endgroup
}
\makeatother

\newcommand{\bop}{\hat{b}}
\newcommand{\bdag}{\hat{b}^\dagger}

\newcommand{\aop}{\hat{a}}
\newcommand{\adag}{\hat{a}^\dagger}
\newcommand{\adagT}{\hat{a}^{\dagger 2}}

\newcommand{\adagn}{\hat{a}^{\dagger n}}
\newcommand{\sigP}{\hat{\sigma}_+}
\newcommand{\sigM}{\hat{\sigma}_-}
\newcommand{\sigZ}{\hat{\sigma}_z}
\newcommand{\sigX}{\hat{\sigma}_x}

\newcommand{\lamBar}{\overline{\lambda}}
\newcommand{\xmn}{\hat{X}_{-n}}
\newcommand{\xpn}{\hat{X}_{+n}}
\newcommand{\xpmn}{\hat{X}_{\pm n}}
\newcommand{\ymn}{\hat{Y}_{-n}}
\newcommand{\ypn}{\hat{Y}_{+n}}
\newcommand{\ypmn}{\hat{Y}_{\pm n}}

\newcommand{\adagk}{\hat{a}^{\dagger}_k}
\newcommand{\aopj}{\hat{a}_j}
\newcommand{\aopk}{\hat{a}_k}

\begin{document}

\preprint{APS/123-QED}

\title{Dispersive regime of multiphoton qubit-oscillator interactions}

\author{Mohammad Ayyash}\thanks{mmayyash@uwaterloo.ca}

\affiliation{%
 Red Blue Quantum Inc., 72 Ellis Crescent North, Waterloo, Ontario N2J 3N8, Canada}
\affiliation{%
 Institute for Quantum Computing, University of Waterloo, 200 University Avenue West, Waterloo,
Ontario N2L 3G1, Canada}

\author{Sahel Ashhab}

\affiliation{Advanced ICT Research Institute, National Institute of Information and
Communications Technology, 4-2-1, Nukui-Kitamachi, Koganei, Tokyo 184-8795, Japan}
\affiliation{Research Institute for Science and Technology, Tokyo University of Science, 1-3 Kagurazaka, Shinjuku-ku, Tokyo 162-8601, Japan}

\date{\today}

\begin{abstract}
The dispersive regime of $n$-photon qubit-oscillator interactions is analyzed using Schrieffer-Wolff perturbation theory. Effective Hamiltonians are derived up to the second order in the perturbation parameters. These effective descriptions reveal higher-order qubit-oscillator cross-Kerr and oscillator self-Kerr terms. The cross-Kerr term combines a qubit Pauli operator with an $n$-degree polynomial in the oscillator photon number operator, while the self-Kerr term is an $(n-1)$-degree polynomial in the oscillator photon number operator. In addition to the higher-order Kerr terms, a qubit-conditional $2n$-photon squeezing term appears in the effective non-rotating-wave-approximation Hamiltonian. Furthermore, perturbation theory is applied to the case of multiple qubits coupled to a shared oscillator. A photon-number-dependent qubit-qubit interaction emerges in this case, which can be leveraged to tune the effective multiqubit system parameters using the oscillator state. Results for the converse setup of multiple oscillators and a single qubit are also derived. In this case, a qubit-conditional oscillator-oscillator nonlinear interaction is found. {The spectral instabilities plaguing multiphoton qubit-oscillator models are carefully treated by introducing stabilizing higher-order terms in the Hamiltonian. The stabilizing terms preserve low-photon subspaces, avoid negative infinite energies and facilitate reliable numerical calculations used to validate analytical predictions.} The effective descriptions developed here offer a simple and intuitive physical picture of dispersive multiphoton qubit-oscillator interactions that can aid in the design of implementations harnessing their various nonlinear effects.
\end{abstract}

\maketitle


\section{\label{sec:level1} Introduction}

The simplest picture of a quantum-mechanical light-matter interaction is captured by a single qubit coupled to a single harmonic oscillator. The prototypical model is known as the (quantum) Rabi model and was first introduced by Jaynes and Cummings in 1963 \cite{JaynesCummings} as a quantum-mechanical generalization of the pioneering semiclassical model formulated in 1936 by Rabi \cite{Rabi_PhysRev.49.324,Rabi2_PhysRev.51.652}. Jaynes and Cummings  simplified the Rabi model to the widely-used model that now bears their names, the Jaynes-Cummings (JC) model, which has a closed-form solution and applies under the rotating-wave-approximation (RWA) where the counter-rotating and energy-non-conserving terms are neglected. These qubit-oscillator models form the theoretical underpinning of various disciplines in quantum physics such as cavity quantum electrodynamics (QED) \cite{Haroche_Raimond}, circuit QED \cite{CircuitQEDReview} and trapped ions \cite{TrappedIonsReview}. 

The qubit-oscillator interaction described in the Rabi and JC models is one of a linear nature where single photons (or, more generally, excitations) are created and annihilated in the oscillator. Generalizations were later formulated with the qubit-oscillator interactions being intensity or photon-number dependent \cite{IntensityDep_BUCK1981132,IntensityDep_Multiph_Singh_PhysRevA.25.3206,IntensityDep_Buzek_PhysRevA.39.3196}. Additionally, multiphoton generalizations were proposed where more than one photon is created or annihilated in each emission or absorption event \cite{IntensityDep_BUCK1981132,Multiphoton_Seb,Multiphoton_Aliskenderov_1987,Multiphoton_Ashraf_PhysRevA.42.6704}. For a review on various generalizations of the Rabi model, see Ref.\cite{JCM_Descendants} and references therein. 

The aforementioned studies spurred an active area of research focused on the spectral and dynamical features of the generalized Rabi and JC models. One particularly peculiar phenomenon occurs in some variants of the $n$-photon Rabi models; the feature known as the ``spectral collapse'' where the joint qubit-oscillator spectra and eigenstates transition from a discretely indexed set to a degenerate continuum above a critical qubit-oscillator coupling strength \cite{SpecCollapseNg1999,SpecCollapse_CEmary_2002}. For the two-photon Rabi model, a quasi-analytical solution was found in the studies of Refs.~\cite{TwoPh_Travenec_PhysRevA.85.043805,SpecCollapse_Duan_2016} based on the same method used by Braak \cite{Braak2011_PhysRevLett.107.100401} to solve the one-photon case. Other works used generalized Bogoliubov transformations to arrive at the same solution \cite{TwoPh_Chen_PhysRevA.86.023822,TwoPh_Cui_2017}. In parallel with the spectral and analytical studies of multiphoton models, there have been a number of quantum information applications developed that leverage nonlinear qubit-oscillator interactions \cite{StateDepSqz_PhysRevA.101.052331,ImplementationSC0,SanerNonclassicalHOStates}. Experimentally feasible implementations were proposed for superconducting circuits \cite{ImplementationSC0,ImplementationSC1,ImplementationSC2,ImplementationSC3} and trapped ions \cite{StateDepSqz_PhysRevA.101.052331,ImplementationTI2}. Generally, the engineering of nonlinear interactions between a qubit and an oscillator remains an active area of research with novel nonlinear interactions realized in various platforms such as opto/electromechanics \cite{vonLüpke2024_multimode,Marti2024_mechSqz}  and trapped ions \cite{SanerNonclassicalHOStates}.

In this paper, we aim to add more insight to the study of multiphoton qubit-oscillator interactions. While quasi-analytical solutions for the two-photon Rabi model were found \cite{TwoPh_Travenec_PhysRevA.85.043805,SpecCollapse_Duan_2016,TwoPh_Chen_PhysRevA.86.023822,TwoPh_Cui_2017}, they lack an intuitive and physically-motivated picture.  We seek to uncover a more intuitive picture than the rigorous mathematical solutions of the spectra of $n$-photon qubit-oscillator models in the dispersive regime, where the qubit-oscillator $n$-photon coupling is much smaller than the $n$-photon qubit-oscillator detuning as we will define later. We resort to second-order Schrieffer-Wolff (SW) perturbation theory to obtain analytic expressions for the qubit-oscillator spectra in the dispersive regime by generalizing techniques employed with the Rabi and JC models \cite{Blais_QIP_cQED,Zueco_DispersiveRegime,DissipAndUSC}. We then consider two further multipartite generalizations of this multiphoton dispersive perturbation theory; the case of many qubits nonlinearly coupled to a shared oscillator and the converse case of many oscillators nonlinearly coupled to a shared qubit. {Furthermore, we discuss the spectral instabilities arising in multiphoton qubit-oscillator interaction models and remedy them using stabilizing higher-order terms in the Hamiltonian.}

The paper is structured as follows. In Sec.~\ref{sec:BriefQOInt}, we review relevant previous work. In particular, we recall the typically used SW perturbation theory for the dispersive regime of linear qubit-oscillator interactions that we seek to generalize. {The reader who is familiar with SW perturbation theory applied to the linear interaction can skip Sec.~\ref{sec:BriefQOInt}.} Section~\ref{sec:SingleQO} defines the multiphoton dispersive regime and lays out the generalized SW transformations for the RWA and non-RWA regimes. The analytically-obtained spectra are compared to numerical results to establish a regime of validity. The transformations are then used in the multiqubit-single-oscillator scenario in Sec.~\ref{sec:Multiqubit}. The multioscillator-single-qubit converse case is discussed in Sec.~\ref{sec:Multimode}. The results are summarized and conclusions are given in Sec.~\ref{Sec:Conc}. Appendix~\ref{app:ComComb} outlines the details of the combinatorics and commutators used in the SW perturbation theory developed in the main text. {Appendix \ref{app:Stab} explains the spectrum stabilization relying on higher-order terms and its details.}

\section{Background}\label{sec:BriefQOInt}

In this section, we review the Rabi model and its subsequent approximations and relevant generalizations that are instrumental for the results to be presented. We begin with models relying on linear interactions in the case of a single qubit-oscillator system followed by the multipartite cases of many qubits linearly coupled to a single oscillator and many oscillators linearly coupled to a single qubit. In each case, we derive an effective RWA and non-RWA dispersive Hamiltonian using second-order perturbation theory. 

\subsection{Single qubit-oscillator system}

The Hamiltonian of the Rabi model reads ($\hbar=1$)
\begin{align}\label{eq:RabiHam}
    \hat{H}_{R}= \omega_o \adag\aop+ \frac{\omega_q}{2}\sigZ + g\sigX(\adag +\aop),
\end{align}
where $\omega_q$ is the qubit transition frequency, $\omega_o$ is the oscillator resonance frequency and $g$ is the qubit-oscillator coupling strength. Here, $\sigZ=\dyad{e}-\dyad{g}$ describes the population difference between the qubit's ground, $\ket{g}$, and excited, $\ket{e}$, states, $\sigX=\sigP+\sigM$ is the dipole operator with $\sigP=\dyad{e}{g}$ and $\sigM=\sigP^\dagger$, and $\adag$ and $\aop$ are the oscillator's creation and annihilation operators obeying $[\aop,\adag]=\hat{\mathbb{I}}$. When the coupling strength and qubit and oscillator frequencies satisfy
\begin{align}\label{eq:JCRWA}
 |\Delta|\ll \Sigma \text{ and } g\ll \min(\omega_q,\omega_o)
\end{align}
with $\Delta=\omega_q-\omega_o$ and $\Sigma=\omega_q+\omega_o$,
the Hamiltonian of Eq.~\eqref{eq:RabiHam} can be simplified using an RWA to read as
\begin{align}
    \hat{H}_R\simeq \hat{H}_{JC}= \omega_o \adag\aop+ \frac{\omega_q}{2}\sigZ + g(\sigP\aop +\sigM\adag).
\end{align}
The spectra and eigenstates of the JC Hamiltonian are exactly solvable. Within the JC regime of validity defined in Eq.~\eqref{eq:JCRWA} lies the dispersive regime where
\begin{align}\label{eq:RWADispersiveCond}
    g \ll |\Delta|\ll\Sigma.
\end{align}
In the dispersive regime, the JC eigenstates are well approximated by the bare states of $\hat{H}_0=\omega_o\adag\aop +\omega_q\sigZ/2$, $\{\ket{g,n},\ket{e,n}\}$, and, as such, the JC interaction term can be treated as a perturbation. Thus, it is helpful to apply a SW transformation \cite{OriginalSWPaper} that perturbatively diagonalizes the Hamiltonian in the bare basis. The transformation is defined as \cite{Haroche_Raimond,Blais_QIP_cQED}
\begin{subequations}\label{eq:SchriefferWolf1Ph}
 \begin{align}
     \hat{U}_{\text{Disp,RWA}}=\exp(\lambda \hat{X}_-),
 \end{align}
 where
 \begin{align}
     \lambda=\frac{g}{\Delta}
 \end{align}
 and
 \begin{align}
     \hat{X}_\pm =\sigM\adag\pm \sigP\aop. 
 \end{align}
\end{subequations}
The interaction term in the JC model corresponds to $g\hat{X}_+$. Then, we transform the Hamiltonian using this unitary operator and expand using the Baker-Campbell-Haussdorf (BCH) identity, up to second order in $\lambda$, to obtain the dispersive RWA Hamiltonian
\begin{align}\label{eq:RWADispHam}
    \hat{H}_{\text{Disp,RWA}}&=\hat{U}_{\text{Disp,RWA}}^\dagger \hat{H}_{\text{JC}}\hat{U}_{\text{Disp,RWA}}\nonumber\\ &\simeq\omega_o\adag\aop + \frac{\omega_q}{2}\sigZ +\chi \sigZ\left(\frac{1}{2}+\adag\aop\right),
\end{align}
where $\chi=g^2/\Delta$ is the dispersive shift. Here, we employ second-order SW perturbation theory since the first-order term in the BCH expansion, $\lambda[\hat{H}_{\text{JC}},\hat{X}_-]$, yields terms quadratic in $\lambda$, and the second-order term, $\lambda^2[[\hat{H}_{\text{JC}},\hat{X}_-],\hat{X}_-]/2!$, leads to terms that are quadratic and quartic in $\lambda$. Thus,  we need to go to second order in SW perturbation theory to
account for all the terms that are quadratic in $\lambda$. This will be the case for all perturbative expansions in this paper. This Hamiltonian is typically employed for quantum-nondemolition measurements of a qubit using an oscillator. The perturbative second-order expansion is typically considered valid when $\lambda \ll 1$. {However, strictly speaking, there is an additional constraint; the oscillator average occupation must remain lower than the critical photon number $n_{\text{ph},c}=1 /(4\lambda^2) = \Delta^2/4g^2$ \cite{Blais_QIP_cQED,ReadoutQubitInducedNonlinearity}. The critical photon number stems from the JC energy eigenvalues,
\begin{align}\label{eq:JCSpectra}
    E_{\pm,n_{\text{ph}}}^{\text{JC}}=\left(n_{\text{ph}}+\frac{1}{2}\right)\omega_o \pm\sqrt{g^2(n_{\text{ph}}+1) +\Delta^2},
\end{align}
with $n_{\text{ph}}$ being the photon occupation. The dispersive description makes the approximation
\begin{align}
        \sqrt{g^2(n_{\text{ph}}+1) +\Delta^2}&=\Delta\sqrt{\lambda^2(n_{\text{ph}}+1)+1}\nonumber\\ &\simeq\Delta\left(\frac{\lambda^2}{2}(n_{\text{ph}}+1)+1\right)\nonumber\\ &=\frac{\chi}{2}(n_{\text{ph}}+1)+\Delta.
\end{align}
Thus, when the energy difference between adjacent dressed states becomes comparable to the coupling strength, i.e. $g\sqrt{n_{\text{ph}}}\sim|\Delta|$, the linearization of the square root term, $\sqrt{g^2(n_{\text{ph}}+1)+\Delta^2}$, is no longer a valid approximation. Equivalently, this occurs when the photon occupation $n_{\text{ph}}\sim|\Delta|^2/g^2.$ Thus, the approximate description becomes less accurate for larger $\lambda$ and/or higher oscillator photon occupation.} In either of these cases, higher-order corrections in the expansion are required \cite{Blais_QIP_cQED}. The critical photon number places a bound on the linear dispersive regime where the oscillator does not exhibit any nonlinearity, e.g. $(\adag\aop)^2$, as it would from higher-order corrections. Higher-order corrections and the effect of qubit-induced oscillator nonlinearity have been studied in work on the efficacy of qubit readout \cite{ReadoutQubitInducedNonlinearity}.

Due to the accessible strong nonlinearities in circuit QED implementations, it became possible to reach the ultrastrong coupling regime where the hierarchy of energy scales in Eq.~\eqref{eq:JCRWA} no longer holds. In the ultrastrong coupling regime, the effects of counter-rotating terms neglected from Eq.~\eqref{eq:RabiHam} become considerable and have been experimentally verified \cite{USC_FornDiaz2010,USC_Niemczyk2010,BeyondUSC_Yoshihara2017}. It is then possible to be simultaneously in the dispersive and ultrastrong coupling regimes where $|\Delta|\ll \Sigma$ in Eq.~\eqref{eq:RWADispersiveCond} is no longer necessarily true. We refer to this as the non-RWA dispersive regime \cite{Zueco_DispersiveRegime}. This defines a hierarchy where the non-RWA dispersive regime contains its RWA counterpart within it. When the RWA no longer applies, we need to extend the SW transformation of Eq.~\eqref{eq:SchriefferWolf1Ph} to account for the counter-rotating terms present in Eq.~\eqref{eq:RabiHam}. The new transformation reads \cite{Zueco_DispersiveRegime}
\begin{subequations}\label{eq:SchriefferWolf1Ph_nonRWA}
 \begin{align}
     \hat{U}_{\text{Disp}}=\exp(\lambda \hat{X}_-+\lamBar \hat{Y}_-),
 \end{align}
 where
 \begin{align}
     \lamBar=\frac{g}{\Sigma}
 \end{align}
 and
 \begin{align}
     \hat{Y}_\pm =\sigM\aop \pm \sigP\adag. 
 \end{align}
\end{subequations}
Note that $\hat{Y}_{\pm}$ and $\hat{X}_{\pm}$ do not commute. In the Rabi model, the interaction part can be expressed as $g(\hat{X}_++\hat{Y}_+).$ Applying the same procedure of transforming the Hamiltonian and using the BCH expansion to second order in both $\lambda$ and $\lamBar$, we obtain the non-RWA dispersive Hamiltonian
\begin{align}\label{eq:nonRWADispHam}
    \hat{H}_{\text{Disp}}&=\hat{U}_{\text{Disp}}^\dagger \hat{H}_{\text{R}}\hat{U}_{\text{Disp}}\nonumber\\ &\simeq
    \omega_o\adag\aop + \frac{\omega_q}{2}\sigZ + (\chi+\xi)\sigZ\left(\frac{1}{2}+\adag\aop\right)\nonumber\\ &\,\,\,\,\, +\frac{(\chi+\xi)}{2}\sigZ(\adagT +\aop^2) ,
\end{align}
where $\xi=g^2/\Sigma$ is the Bloch-Siegert shift coefficient. Note that the last two terms in the Hamiltonian can be combined into one term, $(\chi+\xi)\sigZ(\adag+\aop)^2/2$. However, we choose to keep these terms separated since the generalized multiphoton we present later are in the same form. Interestingly, the non-RWA dispersive Hamiltonian exhibits qubit-conditional squeezing. This resulting interaction was proposed as a mechanism for generating squeezed states in the oscillator \cite{DispSqueezing}.

It is worth noting that one can absorb the qubit-conditional squeezing into the transformation by introducing an appropriate generator in the non-RWA Schrieffer-Wolff transformation of Eq.~\eqref{eq:SchriefferWolf1Ph_nonRWA}. This ensures that the Hamiltonian is diagonal in its bare basis which is not the case in the presence of $\sigZ(\adagT +\aop^2)$. The new transformation reads \cite{DissipAndUSC}
\begin{subequations}
\begin{align}
     \hat{\widetilde{U}}_{\text{Disp}}=\exp(\lambda \hat{X}_-+\lamBar \hat{Y}_- + \zeta \hat{Z}  ),
 \end{align}
 where
 \begin{align}
     \zeta=\frac{g\lamBar}{2\omega_o}
 \end{align}
 and
 \begin{align}
     \hat{Z} =\sigZ(\aop^2 - \adagT). 
 \end{align}
\end{subequations}
 Then, using this transformation, we obtain a diagonal non-RWA dispersive Hamiltonian in the bare basis up to second order in all perturbation parameters. 

\subsection{Multiple qubits coupled to a single oscillator}

We now turn our attention to the case of multiple qubits linearly coupled to a single oscillator known as the Dicke model \cite{DickePaper}. The Dicke Hamiltonian is
\begin{align}
    \hat{H}_{\text{D}}^N=&\omega_o \adag\aop+\sum_{l=1}^N \frac{\omega_{q}^{(l)}}{2}\sigZ^{(l)}   \nonumber\\ &+ \sum_{l=1}^N g^{(l)}\sigX^{(l)}(\adag +\aop),
\end{align}
where the superscript $(l)$ signifies the $l$th qubit.  The origins of this model lie in the pioneering work of Dicke on superradiance in ensembles of two-level atoms \cite{DickePaper}. However, the above form of this model was used in the study of superradiant phase transitions of a radiation field mode coupled to an ensemble of two-level systems past a critical coupling strength \cite{Hepp_Lieb_DickeModel,Wang_Hioe_DickeModel}.

When the qubits are individually in the JC-RWA regime defined by Eq.~\eqref{eq:JCRWA}, i.e. $|\Delta^{(l)}|\ll \Sigma^{(l)} \text{ and } g^{(l)}\ll \min(\omega_q^{(l)},\omega_o)$ holds for all $\omega_q^{(l)}$ and $g^{(l)}$ with $\Delta^{(j)}=\omega_q^{(j)}-\omega_o$ and $\Sigma^{(j)}=\omega_q^{(j)}+\omega_o$, the counter-rotating terms can be neglected to obtain the Tavis-Cummings (TC) Hamiltonian \cite{TavisCummingsModel}
\begin{align}
    \hat{H}_{\text{D}}^N\simeq \hat{H}_{\text{TC}}^N=&\omega_o \adag\aop+\sum_{l=1}^N \frac{\omega_{q}^{(l)}}{2}\sigZ^{(l)}   \nonumber\\ &+ \sum_{l=1}^N g^{(l)}(\sigP^{(l)}\aop +\sigM^{(l)}\adag).
\end{align}
Very much like the JC model, the TC model is exactly solvable and its spectra can be found analytically. This model serves as an important basis for many quantum computing implementations such as circuit QED \cite{Blais_QIP_cQED}. In such a scenario, the oscillator can mediate interactions between qubits that are not directly coupled to each other. This is typically performed in the dispersive regime where Eq.~\eqref{eq:RWADispersiveCond} holds for each qubit, i.e. $ g^{(j)}\ll|\Delta^{(j)}|\ll\Sigma^{(j)}$. In such a regime, we can apply a multipartite SW transformation that straightforwardly follows from the single qubit-oscillator case. Following the same procedure as before, the multiqubit RWA dispersive Hamiltonian reads as \cite{Blais_QIP_cQED}
\begin{align}
    \hat{H}_{\text{Disp,RWA}}^N &\simeq \omega_o \adag\aop + \sum_{l=1}^N 
 \frac{\omega_q^{(l)}}{2}\sigZ^{(l)}\nonumber\\&\,\,\,\,\,\,\,\,+\sum_{l=1}^{N}\chi^{(l)} \sigZ^{(l)}\left(\frac{1}{2}+\adag\aop\right)\nonumber\\&\,\,\,\,\,\,\,\,+\sum_{m>l}\widetilde{\chi}^{(l,m)}(\sigP^{(l)}\sigM^{(m)}+\sigM^{(l)}\sigP^{(m)})
\end{align}
where $\chi^{(l)}=(g^{(l)})^2/|\Delta^{(l)}|$, and
$
\widetilde{\chi}^{(l,m)}=g^{(l)}g^{(m)} \left(\frac{1}{\Delta^{(l)}}+ \frac{1}{\Delta^{(m)}}\right).$ In this setup, the oscillator mediates a qubit-qubit interaction of the isotropic XY type which is an  excitation-conserving interaction. This type of mediated interaction has already been extensively used for implementing two-qubit gates and state tomography. The collective qubit-induced dispersive shift on the oscillator's frequency has been experimentally observed with up to 4000 qubits coupled to a shared oscillator \cite{TC_Dicke_ExpKakuyanagi}.

Naturally, as the qubits' coupling to the shared oscillator becomes stronger, the hierarchy of energy scales is different than that of the RWA regime. The new parameter regime is then defined by $g^{(j)}\ll |\Delta^{(j)}|$. Here, $|\Delta^{(j)}|\ll\Sigma^{(j)}$ is no longer necessarily true, and, once again, there is a hierarchy where the non-RWA dispersive regime contains within it the RWA dispersive regime. In this case,
 we may now perturbatively expand the Dicke Hamiltonian where the counter-rotating terms are included as we did for the Rabi Hamiltonian.
 Then, the multiqubit non-RWA dispersive Hamiltonian becomes \cite{Zueco_DispersiveRegime}
\begin{align}
    \hat{H}_{\text{Disp}}^N &=\hat{U}_{\text{Disp}}^{N\dagger }\hat{H}_{\text{D}}^N \hat{U}_{\text{Disp}}^{N}\nonumber\\ &\simeq \omega_o \adag\aop + \sum_{l=1}^N 
 \frac{\omega_q^{(l)}}{2}\sigZ^{(l)}\nonumber\\&\,\,\,\,\,\,\,\,+\sum_{l=1}^{N}(\chi^{(l)}+\xi^{(l)}) \sigZ^{(l)}\left(\frac{1}{2}+\adag\aop\right)\nonumber\\&\,\,\,\,\,\,\,\,+\sum_{m>l}(\widetilde{\chi}^{(l,m)}-\widetilde{\xi}^{(l,m)})(\sigX^{(l)}\sigX^{(m)}),
\end{align}
where
$\xi^{(l)}=(g^{(l)})^2/\Sigma^{(l)}$, and
$
\widetilde{\xi}^{(l,m)}=g^{(l)}g^{(m)} \left(\frac{1}{\Sigma^{(l)}}+ \frac{1}{\Sigma^{(m)}}\right).$ The presence of the counter-rotating terms gives rise to an Ising-type mediated qubit-qubit interaction, $\sigX^{(l)}\sigX^{(m)}$, which is not excitation preserving. {A related study, in which the coupling strength was pushed to the deep-strong coupling regime, was performed in Ref.~\cite{Jaako_BeyondUSCDicke2016}.}

\subsection{Multiple oscillators coupled to a single qubit}
We now briefly highlight the converse multipartite case where multiple oscillators are coupled to a shared qubit. In relation to the quantum computational applications of the multiqubit-oscillator models, the multioscillator-qubit models can be thought of as an alternative paradigm; the computational or logical qubits are bosonically encoded in the oscillator, and the physical qubit is used to control and readout the bosonic modes \cite{BosonicControlReview,HybridQubitOscillatorReview}.

We start with the multimode Rabi  (MR) model and its Hamiltonian that reads
\begin{align}
    \hat{H}_{\text{MR}}=&\sum_{k} \omega_{k} \adagk\aopk + \frac{\omega_q}{2}\sigZ \nonumber\\&+ \sum_{k}g_{k}\sigX( \aopk^{ \dagger} +\aopk ),
\end{align}
where we label each mode and its frequencies with a subscript $k$ (contrasted with a superscript for the case of multiple qubits). It is worth noting
that a transmission line resonator contains an infinite, or at least a
large, number of modes, such that a qubit coupled to a transmission
line resonator is one case of multiple oscillators coupled to a single
qubit \cite{Niemczyk2010,Sundaresan_Houck_nonRWAMR,Ao2023}. Similarly to the relation between the Dicke and TC model, when the oscillators are within the RWA regime of Eq.~\eqref{eq:JCRWA}, we can simplify the MR model by neglecting the counter-rotating terms. Thus, we perform an RWA and obtain the multimode JC (MJC) Hamiltonian
\begin{align}
    \hat{H}_{\text{MJC}}=&\sum_{k} \omega_{k} \adagk\aopk + \frac{\omega_q}{2}\sigZ \nonumber\\&+ \sum_{k}g_{k}(\sigP\aopk +\sigM\aopk^{\dagger} ),
\end{align}

Now that we have explictly derived the single qubit-oscillator and multiqubit-oscillator cases, we simply state the results for the different regimes, and the derivations follow exactly from the multiqubit case with the exception that it is now a multioscillator system. We do not redefine all the parameters, and instead, we assume the indices now run over the oscillators rather than the qubits. In the RWA regime, the multioscillator RWA dispersive regime Hamiltonian reads as
\begin{align}
    \hat{H}_{\text{Disp,RWA}}^{\text{MO}}&\simeq \sum_k \omega_k \adagk\aopk +  
 \frac{\omega_q}{2}\sigZ\nonumber\\&\,\,\,\,\,\,\,\,+\sum_{k}\chi_k \sigZ\left(\frac{1}{2}+\adag_k\aop_k\right)\nonumber\\&\,\,\,\,\,\,\,\,+\sum_{j>k}\frac{\widetilde{\chi}_{j,k}}{2}\sigZ(\aopj^\dagger\aopk+\aopj\aopk^{\dagger}).
\end{align}
Here, a qubit-conditional oscillator-oscillator beamsplitter interaction emerges between the uncoupled oscillators. This type of interaction has been leveraged for creating nonclassical states entangled between the oscillators and more generally, for implementing gates for bosonic encodings \cite{MariantoniQSwitch,ConditionalBS}. As expected, in the RWA regimes where the counter-rotating terms are neglected, the qubit-mediated oscillator-oscillator interaction is of the excitation-conserving type (beamsplitter interaction). Similarly to the previous section, we now consider the case where the RWA begins to breakdown, and the counter-rotating terms must be taken into account. In this scenario, the multioscillator Hamiltonian in the non-RWA dispersive regime becomes
\begin{align}
    \hat{H}_{\text{Disp}}^{\text{MO}}&\simeq \sum_k \omega_k \adagk\aopk +  
 \frac{\omega_q}{2}\sigZ\nonumber\\&\,\,\,\,\,\,\,\,+\sum_{k}(\chi_k+\xi_k) \sigZ\left(\frac{1}{2}+\adag_k\aop_k\right)\nonumber\\&\,\,\,\,\,\,\,\,+\sum_{j>k}\frac{(\widetilde{\chi}_{j,k}+\widetilde{\xi}_{j,k})}{2}\sigZ(\aopj^\dagger+\aopj)(\aopk^\dagger+\aopk).
\end{align}
In this regime, the qubit-conditional interactions include both a beamsplitter and two-mode squeezing interaction (non-excitation preserving). The presence of both terms allows for different continous variable gates such as a qubit-controlled SUM gate or a qubit-controlled-control-$Z$ gate \cite{UniversalQCCV,HybridQubitOscillatorReview}. {The MR model was leveraged in the dispersive regime to generate entangled cat states in two three-dimensional superconducting cavities mediated by a transmon qubit \cite{Wang_CatTwoBoxes2016}.}

\section{ Single Qubit-Oscillator Multiphoton Interaction Dispersive Regime}\label{sec:SingleQO}

In the previous section, we highlighted the essence of the RWA and non-RWA dispersive regimes when linear interactions are at play between a single qubit and an oscillator, multiple qubits and an oscillator, and finally, multiple oscillators and a qubit. In this section, we begin by introducing the single qubit-oscillator multiphoton generalizations of interest: the $n$-Rabi ($n$R) model and its sister model, the $n$-photon Jaynes-Cummings ($n$JC) model. We also discuss some of the mathematical complications arising for the $n$R and $n$JC Hamiltonians in the case of $n>2$ as well as their remedies. Then, we proceed to develop the multiphoton generalization of the SW transformations presented in the previous section. 

The Hamiltonian of the $n$R model reads
\begin{align}\label{eq:nRHam}
    \hat{H}_{n\text{R}}=\omega_o\adag\aop + \frac{\omega_q}{2}\sigZ + g_n \sigX(\adagn +\aop^n),
\end{align}
where $g_n$ is the $n$-photon coupling strength. It is worth noting that this model is highly relevant for quantum information and quantum optics as its interaction is in the form of a qubit-conditional (generalized) squeezing generator, $g_n\sigX(\adagn+\aop^n)$, which is of importance for synthesizing gates and for interferometry. There is a similar model and its interaction Hamiltonian is $g_n\sigX(\adag+\aop)^n$, which we refer to as the full $n$R model. The full $n$R model naturally arises when studying physical implementations of non-dipolar qubit-oscillator interactions \cite{ImplementationSC1}. The $n$R and $n$JC interaction terms are more commonly studied than the full $n$R model due to their relevance for quantum information applications. It is worth noting that for weak coupling strengths, all three models are equivalent descriptions \footnote{Explicitly, the $n$JC model serves as an excellent approximation for the full $n$R and $n$R models in the weak coupling regime.}. Apart from some commentary in the introductory part of this section, we do not concern ourselves with the full $n$R model in this paper.

When the qubit and oscillator are near $n$-photon resonance, $\omega_q\simeq n\omega_o $, and the coupling strength is small compared to the qubit and oscillator frequencies,  the effects of the counter-rotating terms, $\sigP\adagn$ and $\sigM\aop^n$, become negligible. Explicitly, when the coupling strength and qubit and oscillator frequencies satisfy
\begin{align}\label{eq:nJCRWA}
 |\Delta_n|\ll \Sigma_n \text{ and } g_n\ll \omega_o,
\end{align}
where $\Delta_n=\omega_q-n\omega_o$ and $\Sigma_n=\omega_q+n\omega_o$, the $n$R Hamiltonian can be simplified to the $n$JC Hamiltonian,
\begin{align}\label{eq:nJCHam}
    \hat{H}_{n\text{R}}\simeq\hat{H}_{n\text{JC}}=\omega_o\adag\aop + \frac{\omega_q}{2}\sigZ + g_n (\sigP\aop^n +\sigM\adagn).
\end{align}
As in the linear interaction $(n=1)$ case, the $n$JC model is exactly solvable and its spectra can be analytically computed \cite{IntensityDep_BUCK1981132}. The $n$JC model predicts singlets for the states $\ket{g,k}$, where $k<n$. The singlets are followed by doublets that separate and eventually cross \cite{Multiphoton_Seb,Multiphoton_Aliskenderov_1987,Multiphoton_Ashraf_PhysRevA.42.6704}.

A key phenomenon occurring in the $n$R model for $n=2$ is that of spectral collapse, where the discretely indexed states and spectra collapse to a continuum. This collapse happens at a critical coupling ratio $r_{c}=g_{2}/\omega_o=1/2$. There have been some explanations such as the qubit inducing an inversion of the oscillator's potential which admits no bound states \cite{SpecCollapseTwoPh}. It is worth noting that the full $n$R model for $n=2$ exhibits the same type of spectral collapse at $r_c\simeq1/4$ \cite{ImplementationSC1}. Interestingly, in the two-photon JC model, a spectral `bunching' of one energy level out of each doublet (those with a negative sign; see Eq.~\eqref{eq:nJCSpectra}) where all these energy levels cross each other at $g_2/\omega_o\simeq1$ and deviate again. While this is an interesting mathematical feature, it is important to note that this spectral bunching in the two-photon JC model occurs well beyond its regime of validity (see Eq.~\eqref{eq:nJCRWA}).

The $n$R model for $n>2$ suffers from spectral instabilities due to the unboundedness of the Hamiltonian from below which leads to states with infinite negative energies at any nonzero coupling ($g_n>0$). This means that calculations even for low coupling strengths will be plagued by the infinite negative energies affecting the Hamiltonian spectrum. Other studies have shown that the $n$R model for $n>2$ is, strictly speaking, not self adjoint and, thus, they concluded it is unphysical \cite{Braak_kPhotonRabiModel}. It is important to note that the $n$JC Hamiltonian, while exactly solvable, also suffers from negative infinite energies. This becomes obvious when expressing the $n$JC Hamiltonian in the following form:
\begin{align}
    &\hat{H}_{n\text{JC}}=\sum_{k=0}^{n-1} (E_{g,k}\ket{g,k}\bra{g,k}) \nonumber\\&+ \bigoplus_{l=0}^\infty \begin{pmatrix} \omega_q/2 + l\omega_o & g_n\sqrt{(l+1)...(l+n)} \\ g_n\sqrt{(l+1)...(l+n)}& -\omega_q/2 + (l+n)\omega_o  \end{pmatrix}, 
\end{align}
where the $2\times 2$ matrices act on the subspaces spanned by $\{\ket{e,l},\ket{g,l+n}\}$ for nonnegative integers $l.$
This form explicitly shows the aforementioned singlets and doublets which are represented by the first and second terms, respectively. To show the negative infinite energies, we can simply obtain the doublet energies by diagonalizing the $2\times 2$ matrix in each fixed $l$ subspace;
\begin{align}\label{eq:nJCSpectra}
    E_{\pm,l}^{(n)}=\left(l+\frac{n}{2}\right)\omega_o \pm\sqrt{g_n^2(l+n)!/l! +\Delta_n^2}.
\end{align}
For very large $l$, the second term (under the square root) scales as $l^{n/2}$ while the first term which is strictly postive, scales as $l$. This means that there are infinitely many states with infinite negative energies for all $n>2$. Even though the $n$JC Hamiltonian is exactly solvable, it is also problematic because it is unbounded from below. However, when studying the $n$R or $n$JC Hamiltonian, one must keep in mind that, in practice, these models never arise as the sole terms in a realistic system Hamiltonian. Rather, they are arrived at as an effective description in a weak-coupling and low-energy regime. These effective descriptions typically neglect other terms that bound these spectral instabilities such as unbounded states (see, for example, the derivations of effective multiphoton Hamiltonians in Refs.~\cite{ImplementationSC0} and~\cite{CWilsonPRX}, where spurious terms that make the total system Hamiltonian bounded are neglected to obtain similar effective descriptions).

In this paper, we are interested in the study of the dispersive regime of multiphoton qubit-oscillator interactions by means of SW second-order perturbation theory. As with typical use cases of SW-type perturbation theory, we seek to obtain effective low-energy descriptions decoupled from the high-energy subspaces of the system \cite{BRAVYI20112793}. Thus, for completeness, we can remedy the aforementioned spectral instabilities caused by the unboundedness of the $n$R and $n$JC Hamiltonians by considering them as an approximation of more realistic Hamiltonians that contain additional bounding terms that stabilize the high-photon-number part of the Hilbert space. Such terms avoid the issues of infinite negative energies by bounding the Hamiltonians from below. These terms will have a very small coupling strength and will only affect the high-photon-number states of the system and will not alter the low-energy effective descriptions we seek. Thus, we add a bounding term to the $n$R Hamiltonian such that it reads
\begin{align}\label{eq:Mod_nRHam}
    \hat{\widetilde{H}}_{n\text{R}}=&\omega_o\adag\aop + \frac{\omega_q}{2}\sigZ + g_n \sigX(\adagn +\aop^n) \nonumber\\&+ \eta g_na^{\dagger m_n}\aop^{m_n},
\end{align}
where $m_n=\lfloor{n/2}\rfloor+1$ with $\eta$ being a positive real number that determines strength of the perturbation and $\lfloor x\rfloor$ denoting the floor function; we only consider cases that obey $\eta\ll 1$ such that the low-energy subspaces are barely affected by the term $\aop^{\dagger m_n}\aop^{m_n}$. With this perturbation, we ensure that the Hamiltonian is bounded from below. We can now study the dispersive regime of the $n$R (and $n$JC) Hamiltonian for $n>2$ while being assured that the spectral instabilities of the high-photon-number subspaces do not affect the low-energy subspaces since, in fact, for low-energy regimes, the bounding term can be ignored. 

We now proceed to develop the multiphoton generalization of the SW transformations presented in Sec.~\ref{sec:BriefQOInt} and apply it to the $n$JC and $n$R model. We aim to use second-order SW perturbation theory to better understand the system spectra in the multiphoton dispersive regime.

\subsection{RWA dispersive regime}
We define the multiphoton RWA dispersive regime's  hierarchy of energy scales as that of Eq.~\eqref{eq:nJCRWA} along with
\begin{align}
    g_n\ll |\Delta_{n}|.
\end{align}
This allows us to define a generalized multiphoton SW transformation along with a dimensionless perturbation parameter:
\begin{align}
    \hat{U}_{\text{Disp,RWA}}^{(n)}:=\exp(\lambda_n \hat{X}_{-n}),
\end{align}
where
\begin{align}
    \hat{X}_{\pm n}= \sigM \adagn \pm \sigP \aop^n,
\end{align}
and $\lambda_n=g_n/\Delta_n$. The $n$JC interaction term can be expressed as $g_n\xpn$. Due to the parameter regime we are working in, it follows that $\lambda_n\ll 1$. We now transform the $n$JC Hamiltonian and expand to second order in $\lambda_n$ such that the RWA dispersive Hamiltonian reads

\begin{align}\label{eq:RWA_DispHam}
    \hat{H}_{\text{Disp,RWA}}^{(n)}&= \hat{U}_{\text{Disp,RWA}}^{(n)\dagger}\hat{H}_{n\text{JC}}\hat{U}_{\text{Disp,RWA}}^{(n)}\nonumber\\ &= \hat{H}_{n\text{JC}} + \lambda_n[\hat{H}_{n\text{JC}},\xmn] \nonumber\\ &\,\,\,\,\,\,\,+\frac{\lambda_n^2}{2}[[\hat{H}_{n\text{JC}},\xmn],\xmn] +...\nonumber\\ &\simeq\omega_o\adag\aop + \frac{\omega_q}{2}\sigZ +\frac{\chi_n}{2}\sum_{k=0}^n C_{n,k}^{+}\sigZ(\adag\aop)^k \nonumber\\ &\,\,\,\,\,\,\, +\frac{\chi_n}{2}\sum_{k=1}^{n-1} C_{n,k}^{-}(\adag\aop)^k, 
\end{align}
where 
\begin{align}
    C_{n,k}^{\pm}=(-1)^{n+k} s_1(n+1,k+1)\pm s_1(n,k).
\end{align}
Here, $s_1(n,k)$ is the Stirling number of the first kind which denotes the number of permutations of $n$ elements which contain $k$ permutation cycles, and $\chi_n=g_n^2/\Delta_n$ is the $n$-photon dispersive shift. The coefficients $C_{n,k}^\pm$ arise due to the combinatorics of the commutator expansion, and they have a concrete interpretation; $C_{n,k}^+$ are the commutator coefficients arising due to the qubit-state-dependent ($\sigZ$) part of the commutator, and $C_{n,k}^-$ arise due to the qubit-state-independent terms. Explicitly, the commutator $[\xpn,\xmn]$ contains $[\aop^n,\adagn]$ which requires multiple reductions of one side of the entries to be able to evaluate it using, e.g., $[\aop,f(\adag,\aop)]=\partial f(\adag,\aop)/\partial \adag$. The full derivation and simplification of the commutator into a diagonal form is found in App.~\ref{app:ComComb}. 

\begin{table}[t]
\caption{\label{tab:CoefficientsTable} Table of combinatorial commutator coefficients $C_{n,k}^{\pm}$ for dispersive $n$-photon interactions up to $n=4$.}
\begin{ruledtabular}
\begin{tabular}{c|ccccc|ccccc}
 &\multicolumn{5}{c}{$C_{n,k}^{+}$}\vline &\multicolumn{5}{c}{$C_{n,k}^{-}$}\\
 \hline
 $n\backslash k$&0&1&2&3
&4 &0& 1&2&3&4\\ \hline
1& 1& 2 & - &- &- &1&0 & - &- &-\\
2& 2& 2 & 2 &- &- &2&4 & 0 &- &-\\
3& 6&13 & 3 &2 &- &6&9 & 9 &0 &-\\
4& 24&44 & 46 &4 &2 &24&56 & 24 &16 &0\\
 
\end{tabular}
\end{ruledtabular}
\end{table}

The Hamiltonian contains two polynomials in the photon number operator, $\chi_n\sum_{k=0}^n C_{n,k}^{+}\sigZ(\adag\aop)^k$ and $\chi_n\sum_{k=0}^{n-1} C_{n,k}^{-}(\adag\aop)^k$, with the qubit-state-dependent polynomial truncating at order $n$ and the other truncating at order $n-1$. The qubit-state-independent polynomial truncating at order $n-1$ has to do with the coefficients $C_{n,k}^-$; they are zero when $n=k$ for all $n$. Due to this fact, the qubit-state-dependent series causes a larger shift between energy levels. The qubit-state dependence is reflected in the sign of the shift; the shift is negative when the qubit is in $\ket{g}$, and it is positive when the qubit is in $\ket{e}$. Interestingly, for interaction orders greater than $n=2$, the qubit-oscillator interaction induces a dispersive anharmonicity in the oscillator. This usually occurs for the dispersive linear interaction ($n=1$) when accounting for expansion terms beyond second order, and due to their  higher-order nature, the strength of such terms is usually very small~\cite{NonlinearReadout}. For a given order $n$, the qubit-state-dependent $n$th order shift, as well as the lower-order dispersive shifts ($n-1$, $n-2$, ...), are all present in the Hamiltonian. Interestingly, the lower-order shifts are always larger than the $n$th order shift as seen in the table of $C_{n,k}^+$ values; $C_{n,n}^+=2$ for all $n$. 

The resulting higher-order self-Kerr terms induce multiphoton blockades in the presence of a drive on the oscillator \cite{Miranowicz_PhotonBlockade}. Additionally, when the oscillator is initialized in a coherent state, these higher-order self-Kerr terms can be used to make multi-component cat states \cite{Yurke_AmplitudeDispersion}. It is interesting to note that the one-photon JC Hamiltonian under exact diagonalization reveals higher-order Kerr terms with similar combinatorial factors \cite{KCSmith_ExactKBodyJC}.

\begin{figure}[t]
        \includegraphics[scale=.9,trim={0.1cm 0.25cm 0cm 0cm}]{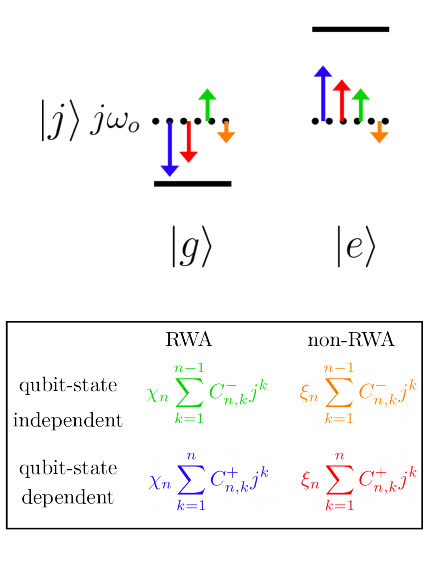}
        \caption{Energy-level dispersive shifts for the $j$th Fock state. {The bare frequency of the $j$th Fock is $j\omega_o$, and in the presence of a dispersive multiphoton interaction with the qubit, second-order SW perturbation theory predicts that the Fock state experiences a shift with four contributions; two qubit-state-dependent shifts and two qubit-state-independent shifts. The shifts are also classified according the interaction terms giving rise to them; RWA, $\sigP \aop^n$ and $\sigM\adagn$, and non-RWA, $\sigP \adagn$ and $\sigM \aop^n$, terms.} The qubit-state-independent shifts in the RWA (green) and non-RWA (orange) cases are always weaker than the qubit-state-dependent RWA (blue) and non-RWA (red) shifts. This is due to the fact that the former's polynomial is of degree $n-1$, while the polynomial of the latter is of degree $n$. Additionally, both RWA shifts are larger than the non-RWA shifts since $\chi_n > \xi_n$.}
        \label{fig:EnergyLevels}
\end{figure}
\begin{figure}[t]
        \includegraphics[scale=1,trim={0cm 0cm 0cm 0cm}]{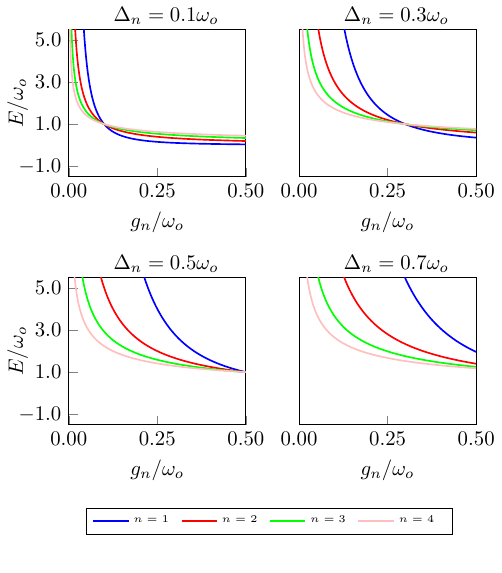}
        \caption{ {Critical photon number scaling for different interaction orders. The plots show the function $(\Delta_n / g_n)^{2/n}$ for different $\Delta_n$ values and orders $n.$ Energies below the curve are considered well-approximated by the RWA dispersive description. Overall, the permissible region for states with larger energy content shrinks for higher order $n.$ As expected, increasing $\Delta_n$ improves the approximation (as $\lambda_n$ becomes smaller), and, thus, increases the area under the curve.}}
        \label{fig:CriticalPhotonNumber}
\end{figure}

\subsection{Non-RWA dispersive regime}

When the first inequality of Eq.~\eqref{eq:nJCRWA} no longer holds, it is possible to extend the usefulness of the SW transformation by adding another generator which accounts for the counter-rotating terms. Thus, as performed in Sec.~\ref{sec:BriefQOInt}, we resort to an additional generator in the SW transformation and apply it to the $n$R model which includes the counter-rotating terms. We now define the multiphoton non-RWA dispersive regime in a similar fashion to the case of the linear interaction \cite{Zueco_DispersiveRegime}. The non-RWA regime obeys
\begin{align}\label{eq:nonRWARegime}
    g_n\ll |\Delta_n|,
\end{align}
where $|\Delta_n|\ll \Sigma_n$ is no longer necessarily true. Thus, this paves the way for the generalized non-RWA SW transformation:
\begin{align}
    \hat{U}_{\text{Disp}}^{(n)}:=\exp(\lambda_n\xmn +\lamBar_n \hat{Y}_{-n}),
\end{align}
where
\begin{align}
    \hat{Y}_{\pm n}= \sigM \aop^n \pm \sigP \adagn
\end{align}
and $\lamBar_n=g_n/\Sigma_n$. Here, $\xpmn$ and $\ypmn$ do not commute, just as in the linear case. Equation~\eqref{eq:nonRWARegime} ensures that $\lamBar_n,\lambda_n\ll 1$ and allows for the perturbative expansion using $\hat{U}_{\text{Disp}}^{(n)}$ to second order in $\lamBar_n$ and $\lambda_n$. Then, the non-RWA dispersive Hamiltonian reads

\begin{align}\label{eq:nonRWA_DispHam}
    \hat{H}_{\text{Disp}}^{(n)}&= \hat{U}_{\text{Disp}}^{(n)\dagger}\hat{H}_{n\text{R}}\hat{U}_{\text{Disp}}^{(n)}\nonumber\\ &= \hat{H}_{n\text{R}} + [\hat{H}_{n\text{R}},\lambda_n\xmn+\lamBar_n\ymn]   \nonumber\\&\,\,\,\,\,\,\,+\frac{1}{2}\Big[[\hat{H}_{n\text{R}},\lambda_n\xmn+\lamBar_n\ymn],\nonumber\\ &\,\,\,\,\,\,\,\,\,\,\,\,\,\,\,\,\,\,\,\,\,\,\,\,\lambda_n\xmn+\lamBar_n\ymn\Big] +...\nonumber\\&\simeq\omega_o \adag\aop + \frac{\omega_q}{2}\sigZ + \frac{(\chi_n+\xi_n)}{2}\sum_{k=0}^n C_{n,k}^{+}\sigZ(\adag\aop)^k \nonumber\\ &\,\,\,\,\,\,\, +\frac{(\chi_n-\xi_n)}{2}\sum_{k=1}^{n-1} C_{n,k}^{-}(\adag\aop)^k\nonumber\\ &\,\,\,\,\,\,\, +\frac{(\chi_n+\xi_n)}{2}\sigZ(\hat{a}^{\dagger 2n} +\aop^{2n}),
\end{align}
where $\xi_n=g_n^2 /\Sigma_n$ is the $n$-photon Bloch-Siegert shift coefficient. The non-RWA regime transformation yields an additional shift, $\xi_n$, dependent on the counter-rotating terms which oscillate with $\omega_q + n\omega_o$. The commutators $[\xpn,\ymn],\,[\ypn,\xmn],$ and $[\ymn,\ypn]$ give rise to the same combinatorial coefficients as $[\xpn,\xmn]$ and are also discussed in App.~\ref{app:ComComb}. The non-RWA regime Hamiltonian exhibits a generalized $2n$-photon squeezing term, $\sigZ(\aop^{\dagger 2n}+\aop^{2n})$. This is consistent with the one-photon dispersive calculations presented in Sec~\ref{sec:BriefQOInt}. Unlike in the RWA regime, here, the qubit-state-independent oscillator anharmonicity is negative and pushes the energy levels down.

The different shifts arising in the RWA and non-RWA dispersive regime are summarized in Fig.~\ref{fig:EnergyLevels}. Generally, the RWA effects are stronger than the non-RWA effects since $\chi_n>\xi_n$. 

Note that the non-RWA Hamiltonian is not diagonal in the bare basis. An easy remedy is to add the generator of the qubit-conditional $2n$-photon squeezing term \cite{DissipAndUSC},
\begin{align}\label{eq:SqSWT}
    \hat{\widetilde{U}}_{\text{Disp}}^{(n)}:=\exp(\lambda_n\xmn +\lamBar_n \hat{Y}_{-n}+ \zeta_n \hat{Z}_n  )
\end{align}

where $\zeta_n=g_n \lamBar_n / 2n\omega_o$ and 
\begin{align}
    \hat{Z}_n=\sigZ (\aop^{2n} - \aop^{\dagger 2n}).
\end{align}
With this transformation, the generalized $2n$-photon squeezing term vanishes, and the non-RWA Hamiltonian becomes diagonal to second order in $\lamBar_n$ and $\lambda_n$. Then, we may analytically find the spectra of the system as
\begin{figure*}[t]
\includegraphics[scale=1.07,trim={.4cm .5cm 0 0}]{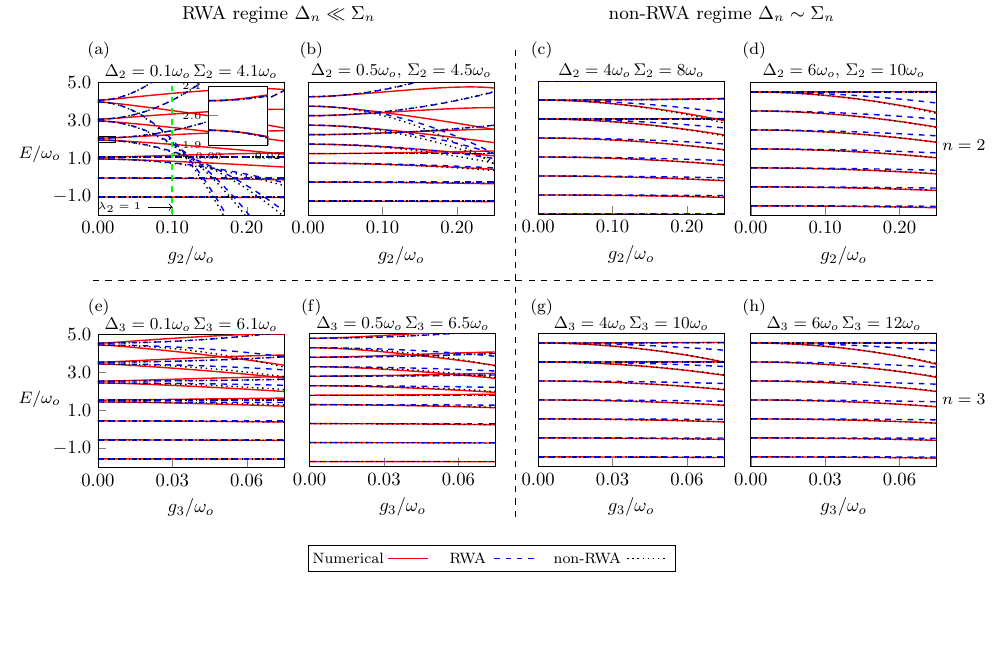}
\caption{\label{fig:QubitOscillatorSpectra} {Multiphoton qubit-oscillator interaction spectra in the dispersive regime. The analytic spectra from the RWA and
non-RWA perturbative expansions are compared with the numerically-obtained spectra. For each
plot, $\Delta_n$ is fixed (which also fixes $\Sigma_n$), and, as such, the x-axis ($g_n$) can be thought of as tuning the perturbation parameters $\lambda_n$ and $\lamBar_n$. The top row ((a)-(d)) corresponds to $n=2$, and the bottom row ((e)-(h)) corresponds to $n=3$. Plots (a), (b), (e) and (f) are in the RWA regime where $\Delta_n\ll \Sigma_n$. Similarly, plots (c), (d), (g) and (h) are in the non-RWA regime where $\Delta_n\sim \Sigma_n.$  For smaller values of $\Delta_n$, the perturbative spectra deviate from the numerical ones at smaller coupling strengths. This is because smaller $\Delta_n$ values correspond to larger $\lambda_n$ values. When $\Delta_n\sim\Sigma_n$, the non-RWA spectra become a better approximation of the numerical results than the RWA spectra.
}  } 
\end{figure*}
\begin{align}
    E^{e/g}_{j}(n)=& \bra{e/g}\bra{j}\hat{\widetilde{H}}_{\text{Disp}}^{(n)}\ket{e/g}\ket{j}\nonumber\\=&\omega_o j +\frac{(\chi_n-\xi_n)}{2}\sum_{k=1}^{n-1}C_{n,k}^- j^k \nonumber\\&\pm \frac{(\chi_n+\xi_n)}{2}\sum_{k=1}^{n}C_{n,k}^+ j^k \pm \frac{\omega_q}{2},
\end{align}
where $\hat{\widetilde{H}}_{\text{Disp}}^{(n)}$ is the non-RWA Hamiltonian without the $2n$-photon squeezing term. From this last equation, we can set $\xi_n=0$ to obtain the RWA spectrum. Generally, if we transform to the usual interaction picture via $\exp(-i\omega_q t \sigZ/2 -i \omega_o t\adag\aop  )$, when the energy scales follows Eq.~\eqref{eq:nJCRWA}, $\xi_n$ will be negligibly small and we can drop the fast-oscillating $2n$-photon squeezing terms, $\aop^{\dagger 2n}$ and $\aop^{2n}$, which allows us to recover the RWA results.

\subsection{Comparison with numerical results}

We now compare the analytically derived results of the RWA and non-RWA spectra with the numerically-obtained spectra of the $n$R model (for details on the numerical spectra and stabilization, see App.~\ref{app:Stab}).

We prelude the discussion with a reminder of the looming spectral instabilities for higher-order interactions. The spectral instabilities of the nR model for $n>2$ restrict the low-energy regime in a smaller set of parameters than $n=2$. Thus, the low-energy stable regime shrinks for higher-order interactions. This means that, in general, the regime of validity of perturbation theory becomes smaller for larger $n$. Additionally, the dispersive regime where the qubit-oscillator system can be approximately described in its bare basis comprises a smaller range of parameters for higher order $n$. {In addition to the regime of validity set by the size of the perturbation parameters, one can derive a heuristic bound based on a critical photon number in a similar fashion to the one-photon critical photon number discussed in Sec.\ref{sec:BriefQOInt}. The effect of SW perturbation theory in the RWA regime can be described as a linearization of the square root term in the $n$JC energy eigenvalues of Eq.~\eqref{eq:nJCSpectra}. Let $n_{\text{ph}}$ be the oscillator occupation. When $g_n\sqrt{(n_\text{ph}+n)!/n_{\text{ph}}!}\sim \Delta_n$, the approximate linearization of the square root begins to break down. Thus, using the fact that $\sqrt{(n_\text{ph}+n)!/n_{\text{ph}}!}$ scales as $ n_{\text{ph}}^{n/2}$, we can arrive at a heuristic critical photon number for the $n$JC dispersive regime; $n_{\text{ph},c}^{(n)}\propto (\Delta_n / g_n)^{2/n}.$ This heuristic bound predicts a shrinking regime of validity for the RWA dispersive spectra for higher order $n.$ Figure \ref{fig:CriticalPhotonNumber} shows the behaviour of the critical photon number in the cases of $n=1,2,3$ and $4$ as a function of $g_n$ for different values of $\Delta_n.$}

{In Fig.~\ref{fig:QubitOscillatorSpectra}, we compare the analytical results from second-order SW perturbation theory with numerical diagonalization of the $n$R Hamiltonian\footnote{In the case of $n=2$, the numerically-obtained spectra match the quasi-analytic solutions of Refs.~\cite{TwoPh_Travenec_PhysRevA.85.043805,SpecCollapse_Duan_2016}.}. Here, we use an oscillator truncation of $N_T=300$ and focus on the states that correspond to the first few bare qubit-oscillator states at $g_n=0$, namely, the first few states $\ket{g/e}\ket{l}$. The details regarding stabilization and justification for using $N_T=300$ are found in App.~\ref{app:Stab}.}

    {We start by noting general trends. The description provided by perturbation theory is most accurate when $\lambda_n$ is small for the RWA regime (and, additionally, when $\lamBar_n$ is small for the non-RWA regime), as expected. This is seen in the improvement of overlap between the numerical and RWA analytical spectra between Fig.~\ref{fig:QubitOscillatorSpectra}(a) and (b) for $n=2$, and (e) and (f) for $n=3$. For each plot $\Delta_n$ is fixed which makes the x-axis, $g_n/\omega_o$ proportional to $\lambda_n$. Thus, for plots with larger $\Delta_n$, the range of $\lambda_n$ values used in the plot is smaller. The deviation from the analytic description occurs in a similar fashion to the deviation of the general $n$R spectra from the $n$JC spectra in two ways. The first trend is one within a fixed interaction order; the higher energy levels deviate at smaller ratios of $g_n/\omega_o$. The second trend is that the higher-order interactions deviate at smaller ratios of $g_n/\omega$ --- this can be immediately attributed to the smaller dispersive regime for higher-order interactions as discussed earlier. Figures~\ref{fig:QubitOscillatorSpectra}(c) and (d) show the non-RWA dispersive regime for $n=2$. We observe that as $|\Delta_n|$ becomes of comparable magnitude to $\Sigma_n$, the non-RWA analytic spectra (black dotted lines) serve as better approximations for the numerical spectra than their RWA counterparts. This is also the case for $n=3$ as seen in Fig.~\ref{fig:QubitOscillatorSpectra}(g) and (h).}

\begin{figure}[t]
        \includegraphics[scale=1.1,trim={.5cm 0cm 0cm 0cm}]{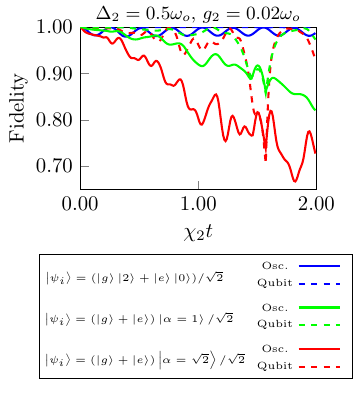}
        \caption{{Fidelity of time-evolved states under the dispersive approximation. The qubit and oscillator subsystem fidelities are obtained by time-evolving the initial states under Eq.~\eqref{eq:RWA_DispHam} and Eq.~\eqref{eq:nRHam} for the case of $n=2$ then tracing out the oscillator and qubit, respectively.}}
        \label{fig:Dynamics}
\end{figure}

    {Thus far, we were concerned with the accuracy of the perturbation theory with regards to predicting spectra. In the dispersive regime, the qubit and oscillator do not exchange excitations, and, thus, it is of interest to examine the accuracy of the dispersive description's dynamics. Here, we consider the case of $n=2$ in the RWA regime. We prepare an initial qubit-oscillator state, $\ket{\psi_i}$, and time-evolve it under the dispersive RWA Hamiltonian in Eq.~\eqref{eq:RWA_DispHam} \footnote{This is because we choose to use parameters in the RWA regime and, but the discussion applies to the non-RWA Hamiltonian as well.} and the $n$R Hamiltonian in Eq.~\eqref{eq:nRHam}. Then, we find the qubit subystem fidelity by tracing out the oscillator and comparing the reduced qubit density matrices time-evolved under the two Hamiltonians. Similarly, we obtain the oscillator subsystem fidelity by tracing out the qubit instead.  Figure~\ref{fig:Dynamics} shows the qubit and oscillator subsystem fidelities for different initial states. When there is a small and well-defined number of excitations in the oscillator such as in the entangled initial state $\ket{\psi_i}=(\ket{g}\ket{2}+\ket{e}\ket{0})/\sqrt{2}$, the dispersive approximation provides an accurate dynamical description as shown by the high fidelities of the blue lines in Fig.~\ref{fig:Dynamics}. When the system is initialized in $\ket{\psi_i}=(\ket{g}+\ket{e})\ket{\alpha}/\sqrt{2}$ with $|\alpha|^2=1$, the subsystem fidelities are lower than the previous case, but they remain above $90\%$ up to $\chi_2t\simeq1.$ In this case, the oscillator has a worse subsystem fidelity than the qubit. When the coherent state has an average of two photons, $|\alpha|^2=2$, the dynamics generated by the dispersive Hamiltonian are consistently accurate for the qubit (mostly maintaining above $90\%$ subsystem fidelity), but the accuracy for the oscillator is only high while $\chi_2t\ll1$, as shown by the decay of fidelities of the red solid line in Fig.~\ref{fig:Dynamics}. The lower fidelities for coherent states are rooted in the presence of large Fock states in a coherent state, which might have small amplitudes, e.g. $\expval{6|\alpha=\sqrt{2}}\approx0.11$, but contribute to large shifts in the dispersive approximation, which causes a discrepancy in the dynamics.}

\subsection{Experimental signature of dispersive multiphoton qubit-oscillator interactions}\label{sec:ExpSignatures}

In the previous sections, we presented the analytical perturbation theory results and established a regime of validity in comparison with numerical results. Here, we seek to highlight a signature of the multiphoton dispersive interactions to test the differing interaction orders. Furthermore, we provide estimates for the dispersive regime couplings and shifts based on realistic circuit QED implementation proposals \cite{ImplementationSC0,ImplementationSC1}. 

\begin{figure}[t]
        \includegraphics[scale=1,trim={0.1cm 0.25cm 0cm 0cm}]{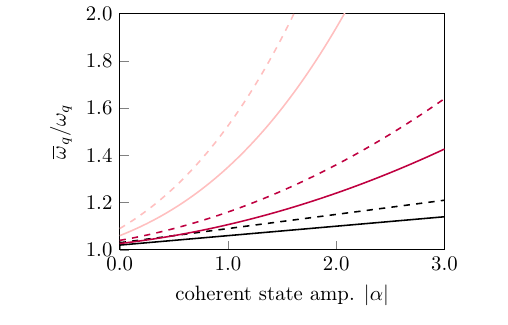}
        \caption{Effective qubit frequency due differing interaction orders as a function of oscillator coherent state amplitude. The ratio of effective qubit frequency to bare qubit frequency is plotted as a function of the amplitude of the coherent state populating the oscillator. The black, purple and pink lines correspond to $n=1$, $n=2$ and $n=3$ dispersive interactions, respectively; solid lines represent $g_n=0.02\omega_o$ and dashed lines represent $g_n=0.03\omega_o$.}
        \label{fig:FreqShiftCoherentState}
\end{figure}

Situating ourselves in the RWA regime (what follows works just as well in the non-RWA regime), we rely on the Hamiltonian of Eq.~\eqref{eq:RWA_DispHam}. We consider the effective (dressed) qubit frequency as a function of the oscillator population and its powers,
\begin{align}
    \overline{\omega}_q^{(n)}=\omega_q+\chi_n\sum_{k=0}^n C_{n,k}^+ \expval{(\adag\aop)^k},
\end{align}
where the oscillator occupation dresses the qubit frequency in a manner dependent on the interaction order. For an experimentally-motivated illustration, let us consider the oscillator to be occupied by a coherent state $\ket{\alpha}$, where $\expval{\adag\aop}=|\alpha|^2.$ Here, we can imagine the oscillator being populated by some external linear drive $\propto(\adag\pm\aop)$. Using the identity  $(\adag\aop)^k=\sum_{l=0}^k s_2(k,l) \aop^{\dagger l}\aop^l$ \cite{KATRIEL2000159}, we can now easily evaluate $\expval{(\adag\aop)^k}{\alpha}=\sum_{l=0}^k s_2(k,l)|\alpha|^l.$ We may now observe the growth in th effective qubit frequency as a function of the amplitude of the coherent state, $|\alpha|$, populating the oscillator. Figure~\ref{fig:FreqShiftCoherentState} shows the trends in effective qubit frequency for different interaction orders, $n$, for two different $g_n$ values at each order. Notably, the growth of the shift is a polynomial of degree $n$. The difference in growth behavior of the effective qubit shifts as a function of coherent state amplitude serves as a potential experimental signature for distinguishing $n=1,\,n=2$ and $n=3$ (and beyond) dispersive interactions.

In the case of $n=2$, for a superconducting circuit implementation, we can provide realistic estimates of the qubit-oscillator system parameters in the dispersive regime and the corresponding shifted effective qubit frequency. For a qubit frequency $\omega_q=2\pi\times13.5\text{ GHz}$ and oscillator frequency $\omega_o=2\pi\times4.50\text{ GHz}$, i.e. $\Delta_2=\omega_o$, there are circuit design proposals with $g_2$ anywhere from $25 \text{ MHz}$ \cite{ImplementationSC0} up to $g_2\sim 1\text{ GHz}$ \cite{ImplementationSC2}. For a simple estimate in the dispersive regime, we can choose $g_2=\text{90 MHz}$ ($=0.02\omega_o$) with $\chi_2=3.6 \text{ MHz}.$ We can extract the qubit shift, $\delta\omega_q=\overline{\omega}_q-\omega_q$, as a function oscillator coherent state population using the solid purple line in Fig.~\ref{fig:FreqShiftCoherentState}. As an example, when the oscillator is populated with a coherent state with amplitude $|\alpha|=1$, the qubit shift will be $\delta\omega_q\simeq 1.45\text{ GHz}.$ This shift is measurable and can be used to verify our predictions of the mulitphoton dispersive regime.

\section{Multiple qubits coupled to a single oscillator \label{sec:Multiqubit}}
With the established technique from the single qubit-oscillator multiphoton dispersive regime, we now move on to the multipartite scenario where $N$ qubits are coupled to a shared oscillator through an $n$-photon interaction.

We begin with the $n$-photon Dicke ($n$D) Hamiltonian (generalizing from Sec.~\ref{sec:BriefQOInt}) whose Hamiltonian reads
\begin{align}
    \hat{H}_{n-\text{D}}^N=&\omega_o \adag\aop+\sum_{l=1}^N \frac{\omega_{q}^{(l)}}{2}\sigZ^{(l)}   \nonumber\\ &+ \sum_{l=1}^N g_n^{(l)}(\sigP^{(l)}+\sigM^{(l)})(\aop^n +\adagn),
\end{align}
where we recall that the superscript $(l)$ denotes the $l$th qubit. Of course, just as in the linear case, when each of the qubits is individually in the $n$JC energy-scale regime, we can simplify the $n$D Hamiltonian by applying an RWA, as done in the single qubit-oscillator case. This leads us to the $n$-photon Tavis-Cummings Hamiltonian ($n$TC),

\begin{align}
    \hat{H}_{n-\text{TC}}^N=&\omega_o \adag\aop+\sum_{l=1}^N \frac{\omega_{q}^{(l)}}{2}\sigZ^{(l)}   \nonumber\\ &+ \sum_{l=1}^N g_n^{(l)}(\sigP^{(l)}\aop^n +\sigM^{(l)}\adagn).
\end{align}
The $n$TC model is exactly solvable, and just like the $n$JC model, the $n$TC Hamiltonian predicts $2N(n-1)$ singlets followed by doublets. Here, there are $N$-times the singlets and doublets of the single-qubit model due to the additional qubits.

The spectral collapse for $n=2$ persists in the multiqubit case. The presence of multiple qubits changes the critical coupling at which the collapse occurs. It was shown for the case of $n=2$ that the collapse occurs at a smaller collective critical collapse coupling \cite{ImplementationSC1}. Explicitly, for a fixed oscillator frequency, if we take a single qubit-oscillator system where the critical collapse coupling is $g_{2,c}$, the individual qubit couplings at which the multipartite system's spectral collapse occurs are all less than the single qubit case, $g_{2}^{(1)},...,g_{2}^{(N)} < g_{2,c}$; the shifted multipartite ($N$ qubit) critical collapse coupling $\widetilde{g}_{2,c}^{N}$ is less than $g_{2,c}$ due to coupling enhancement in Dicke-type models. When all individual couplings are the same, the enhancement scales as $\sqrt{N}$. For varying individual coupling strengths, there is no simple
scaling factor that describes the enhancement of individual coupling
strengths. However, some approximate relations can be derived, for
example using a mean-field treatment of the interactions between the
different subsystems \cite{Ashhab_SuperradianceParameterFluctuation}. Additionally, the discussion on the spectral instabilities for $n>2$ follows from the single qubit-oscillator case and applies just as well. {Similarly to the one-photon Dicke model, the multiphoton Dicke model exhibits normal and superradiant phases. The dispersive regime is concerned with qubit-oscillator coupling strengths that are well within the normal phase \cite{Garbe_TwoPhDicke_SPT,Chen_TwoPhDicke_SPT}.}

\subsection{RWA dispersive regime}

We may now generalize the multiphoton dispersive regime and its SW transformation of the single qubit-oscillator case to the case of multiple qubits. We start with the RWA regime where the qubits all obey $g_n^{(j)}\ll |\Delta_n^{(j)}|.$ Next, we straightforwardly generalize the generators to be a sum over the qubit indices, and the parameters are now also labelled with the same indices. Then, the multiqubit SW transformation reads as
\begin{align}
    \hat{U}_{\text{Disp,RWA}}^{(n)}:=\exp(\sum_{j}\lambda_n^{(j)} \hat{X}_{-n}^{(j)}).
\end{align}
Similarly, we transform the Hamiltonian and expand to first order in all the perturbation parameters to obtain

\begin{align}\label{eq:RWAMultiqubitDispHam}
    &\hat{H}_{\text{Disp,RWA}}^{N,(n)}=\hat{U}_{\text{Disp,RWA}}^{(n)\dagger}\hat{H}_{n-\text{TC}}^N\hat{U}_{\text{Disp,RWA}}^{(n)}\nonumber\\ &=\hat{H}_{n-\text{TC}}^N + \sum_{l}\lambda_l[\hat{H}_{n-\text{TC}}^N,\xmn^{(l)}] \nonumber\\&\,\,\,\,\,\,+ \sum_{l,m}\frac{\lambda_l\lambda_m}{2}[[\hat{H}_{n-\text{TC}}^N,\xmn^{(l)}],\xmn^{(m)}] +...
\nonumber\\&\simeq \sum_{l} \hat{H}_{\text{Disp,RWA}}^{(l),(n)}\nonumber\\ & \,\,\,\,\,\,+ \sum_{l>m}\sum_{k=0}^{n-1}\bigg[\widetilde{\chi}_n^{(l,m)}C_{n,k}^-(\sigP^{(l)}\sigM^{(m)}+\sigM^{(l)}\sigP^{(m)})(\adag\aop)^k\bigg]
\end{align}
where $$\widetilde{\chi}_n^{(l,m)}=g_n^{(l)}g_n^{(m)} \left(\frac{1}{\Delta_n^{(l)}}+ \frac{1}{\Delta_n^{(m)}}\right),$$ and $\hat{H}_{\text{Disp,RWA}}^{(l),(n)}$ is the $l^{\text{th}}$ qubit RWA dispersive Hamiltonian of Eq.~\eqref{eq:RWA_DispHam}. In this case, a photon-number dependent oscillator-mediated qubit-qubit interaction emerges. The qubit-qubit interaction between the $l$th and $m$th qubits vanishes if $\Delta_n^{(l)}=-\Delta_n^{(m)}$ since in this case $\widetilde{\chi}^{(j,k)}_n=0$. Similarly to the linear interaction, the RWA regime gives rise to an anisotropic XY excitation-preserving qubit-qubit interaction of the form $\sigP^{(l)}\sigM^{(m)} + \text{H.c.}.$
\\

\subsection{Non-RWA dispersive regime}
We now proceed to generalize the single qubit-oscillator non-RWA dispersive regime to the case of multiple qubits. Here, the energy scale in effect obeys $g_n^{(j)}\ll |\Delta_n^{(j)}|,\,\Sigma_n^{(j)}$ for each qubit. To this extent, we use the non-RWA generalized SW transformation with a sum of generators over the qubit indices such that the unitary reads
\begin{align}
    \hat{U}_{\text{Disp}}^{(n)}:=\exp(\sum_{j}\lambda_n^{(j)} \hat{X}_{-n}^{(j)} +\sum_{j}\lamBar_n^{(j)} \hat{Y}_{-n}^{(j)}).
\end{align}
Then, the multiqubit non-RWA dispersive Hamiltonian, up to first order in all perturbation parameters, is expanded as  
\begin{align}\label{eq:nonRWAMultiqubitDispHam}
    &\hat{H}_{\text{Disp}}^{N,(n)}=\hat{U}_{\text{Disp}}^{(n)\dagger}\hat{H}_{n-\text{D}}^N\hat{U}_{\text{Disp}}^{(n)}\nonumber\\ &=\hat{H}_{n-\text{D}}^N + \sum_{l}[\hat{H}_{n-\text{D}}^N,\lambda_l\xmn^{(l)} +\lamBar_l\ymn^{(l)}]\nonumber\\ &\,\,\,\,\,\,\,+ \sum_{l,m} \Big[[\hat{H}_{n-\text{D}}^N,\lambda_l\xmn^{(l)} +\lamBar_l\ymn^{(l)}],\nonumber\\&\,\,\,\,\,\,\,\,\,\,\,\,\,\,\,\,\,\,\,\,\,\,\,\,\,\,\,\,\,\,\lambda_m\xmn^{(m)} +\lamBar_m\ymn^{(m)}\Big]+...\nonumber\\ &\simeq\sum_{l=1}^N \hat{H}_{\text{Disp}}^{(l),(n)}\nonumber\\ &\,\,\,\,\,\,\,+ \sum_{l>m}\sum_{k=0}^{n-1}\bigg[(\widetilde{\chi}_n^{(l,m)}-\widetilde{\xi}_n^{(l,m)})C_{n,k}^-(\sigX^{(l)}\sigX^{(m)})(\adag\aop)^k\bigg]
    \end{align}
    where $$\widetilde{\xi}_n^{(l,m)}=g_n^{(l)}g_n^{(m)}\left(\frac{1}{\Sigma_n^{(l)}}+ \frac{1}{\Sigma_n^{(m)}}\right),$$and $\hat{H}_{\text{Disp}}^{(l),(n)}$ is the $l^{\text{th}}$ qubit non-RWA dispersive Hamiltonian of Eq.~\eqref{eq:nonRWA_DispHam}. 
    As expected, the non-RWA dispersive regimes gives rise to an Ising-type oscillator-mediated qubit-qubit interaction. Here, the qubit-qubit interaction is also enhanced by the photon-number coupling, but the coupling strength diminishes due to the $-\widetilde{\xi}_n^{(l,m)}$ non-RWA coefficient. One can also eliminate the $2n$-photon squeezing terms by applying the extended transformation of Eq.~\eqref{eq:SqSWT}. 
   
    \subsection{Comparison with numerical results}
     
    We now compare the perturbative expansion calculations with the numerical results. Specifically, we focus on the two-qubit $n$D model for $n=2$.
    
    It is possible to exactly diagonalize the RWA Hamiltonian of Eq.~\eqref{eq:RWAMultiqubitDispHam} by simply diagonalizing the two-qubit system. Essentially, the Hamiltonian becomes a block-diagonal matrix of $4\times4$ blocks. Similarly, in the non-RWA case, we can first apply the extended transformation accounting for the $2n$-photon squeezing term presented in Eq.~\eqref{eq:SqSWT}, and then diagonalize the two-qubit part of the Hamiltonian. { Explicitly, the block-diagonal Hamiltonian matrix in the joint qubit basis $\{\ket{ee},\ket{eg},\ket{ge},\ket{gg}\}$ for a fixed oscillator state $\ket{j}$ reads}
    \begin{widetext}
        {
        \begin{align}
\hat{\widetilde{H}}^{(n)}_j=\frac{1}{2}\begin{pmatrix}
                \Sigma^{(1,2)} + \Upsilon^{(1,2)}_{+j,n}+\mu_{j,n}^{(1,2)}&0 &0&\Xi^{(1,2)}_{j,n}\\ 0&\Delta^{(1,2)} + \Upsilon^{(1,2)}_{-j,n}+\mu_{j,n}^{(1,2)}&\Xi^{(1,2)}_{j,n}&0\\ 0&\Xi^{(1,2)}_{j,n}&-\Delta^{(1,2)} -\Upsilon^{(1,2)}_{-j,n}+\mu_{j,n}^{(1,2)}&0\\ \Xi^{(1,2)}_{j,n}&0&0& -\Sigma^{(1,2)} - \Upsilon^{(1,2)}_{+j,n} +\mu_{j,n}^{(1,2)}
            \end{pmatrix},
        \end{align}
        }
    \end{widetext}
    {where $$\Sigma^{(1,2)}=\omega_q^{(1)}+\omega_q^{(2)},$$ $$\Delta^{(1,2)}=\omega_q^{(1)}-\omega_q^{(2)},$$ $$\Xi^{(1,2)}_{j,n}=(\widetilde{\chi}_n^{(1,2)}-\widetilde{\xi}_n^{(1,2)})\sum_{k=1}^{n-1}C_{n,k}^- j^k,$$  $$\Upsilon^{(1,2)}_{\pm j,n}=[(\chi^{(1)}_n+\xi^{(1)}_n)\pm(\chi^{(2)}_n+\xi^{(2)}_n)]\sum_{k=0}^{n}C_{n,k}^+ j^k,$$ and $$\mu_{j,n}=2j\omega_o +(\chi_n^{(1)}+\chi_n^{(2)}-\xi_n^{(1)}-\xi_n^{(2)})\sum_{k=1}^{n-1}C_{n,k}^- j^k.$$ The RWA form can be recovered by setting $\xi^{(j)}=0$ as well as zeroing the matrix entries $\dyad{ee}{gg}$ and $\dyad{gg}{ee}$; this is equivalent to transforming to the appropriate rotating frame and neglecting the fast-oscillating terms.  }

\begin{figure}[t]
        \includegraphics[scale=.9,trim={.5cm 0cm 0cm 0cm}]{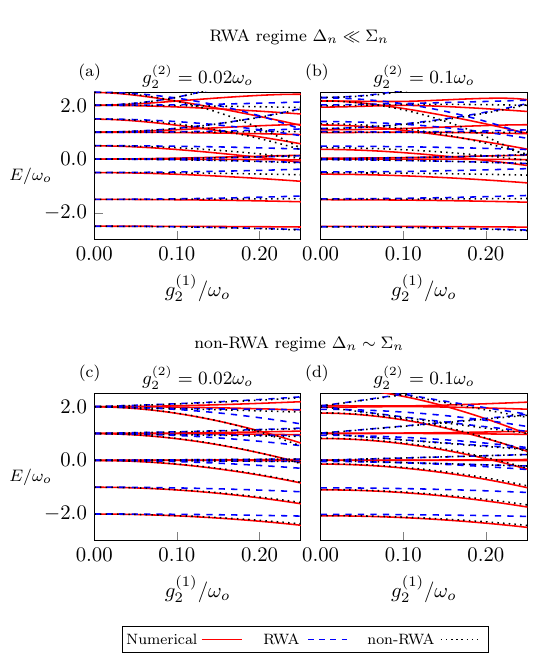}
        \caption{ {Two-qubit two-photon dispersive spectra. For each plot, the first qubit's coupling to the oscillator, $g_2^{(1)}$ is varied along the x-axis while the second qubit's coupling, $g_2^{(2)}$, is fixed. We set $\Delta_2^{(1)}=\Delta_2^{(2)}=0.5\omega_o$ for (a),(b) and $\Delta_2^{(1)}=\Delta_2^{(2)}=6\omega_o$ for (c),(d). The RWA and non-RWA analytics are an excellent approximation of the numerics when the cooperative effects are weak as in (a) and (c). When the cooperative effects are stronger, the analytic spectra deviate at smaller coupling values of the first qubit in (b) and (d).}}
        \label{fig:nDSpectraTwoQubit}
\end{figure}

    {The analytical spectra produced by the RWA and non-RWA expansions are contrasted with the numerical spectra of the two-qubit two-photon Dicke Hamiltonian in Fig.~\ref{fig:nDSpectraTwoQubit}. Here, we use an oscillator truncation $N_T=300$ to produce the numerical spectra. For a single plot, we fix one qubit's coupling to the oscillator and we vary the other qubit's coupling. Here, we assume $\Delta_2^{(1)}=\Delta_2^{(2)}=\Delta_2.$ Figure~\ref{fig:nDSpectraTwoQubit}(a) and (c) show the multiqubit RWA and non-RWA regimes, respectively, where the collective qubit cooperative effects is relatively weak which only slightly shifts the aforementioned collective critical collapse coupling, $\widetilde{g}_{n,c}^N$ ($N=n=2$), slightly lower than $g_{n,c}.$ We find that the perturbative spectra shows very similar trends to the single qubit-oscillator case; the lower-energy states better match the numerical results, the RWA analytical spectra are accurate while $\Delta_2\ll \Sigma_2$, and the non-RWA analytical spectra are accurate for $\Delta_2\sim\Sigma_2.$ Additionally, the impact of stronger cooperative effects on the RWA and non-RWA regimes is shown in Fig.~\ref{fig:nDSpectraTwoQubit}(b) and (d), respectively. In this case the second qubit's coupling to the oscillator is larger and this causes the analytical spectra to deviate at a smaller coupling of the first qubit.}

    \subsection{Oscillator-mediated tunably coupled multiqubit system}
    
    The photon-number dependence of the mediated qubit-qubit interaction allows for the tuning of the effective qubit frequencies and mediated qubit-qubit coupling by carefully selecting the oscillator state. As an example, we can take two qubits interacting with an oscillator via an $n$-photon interaction. For simplicity, we consider the RWA Hamiltonian of Eq.~\eqref{eq:RWAMultiqubitDispHam} and the discussion follows exactly for the non-RWA case. Similarly to the calculations carried out in Sec.~\ref{sec:ExpSignatures}, we consider the oscillator to be populated by a coherent state $\ket{\alpha}.$ We now consider the effective two-qubit Hamiltonian. Then, the effective Hamiltonian (ignoring constant offsets) is
    \begin{subequations}\label{eq:EffParam}
    \begin{align}
        \hat{H}_{\text{Eff}}=&\bra{\alpha}\hat{H}_{\text{Disp,RWA}}^{N=2,(n)}\ket{\alpha}\nonumber\\=& \frac{\overline{\omega}_{q,n}^{(1)}(\alpha)}{2}\sigZ^{(1)}+\frac{\overline{\omega}_{q,n}^{(2)}(\alpha)}{2}\sigZ^{(2)} \nonumber\\&+ \overline{g}_n(\alpha)(\sigP^{(1)}\sigM^{(2)} + \sigM^{(1)}\sigP^{(2)}),
    \end{align}
    where
    \begin{align}
    \overline{\omega}_{q,n}^{(k)}(\alpha)={\omega}_{q}^{(k)}+\chi_n^{(k)}\sum_{k=0}^n \sum_{l=0}^kC_{n,k}^+ s_2(k,l)|\alpha|^l
    \end{align}
    and
    \begin{align}
    \overline{g}_{n}(\alpha)=\widetilde{\chi}_n^{(1,2)}\sum_{k=0}^{n-1}\sum_{l=0}^k C_{n,k}^-  s_2(k,l)|\alpha|^l.
    \end{align}
    \end{subequations}
    We can now tune the effective qubit frequencies and qubit-qubit coupling strength by changing the coherent state amplitude, $|\alpha|$. The effective qubit frequencies follow exactly from the single-qubit-oscillator case in Sec.~\ref{sec:ExpSignatures} (see Fig~\ref{fig:FreqShiftCoherentState}). Interestingly, the effective qubit frequencies are polynomial functions  of $|\alpha|$ of degree $n$, whereas the effective qubit-qubit interactions are polynomial functions of degree $n-1$. The tuning of the effective qubit-qubit interaction can be done \textit{in situ} by adjusting the amplitude of the coherent state using a linear drive on the oscillator ($\propto\adag\pm\aop$). The tunability and different growth behaviour (depending on the interaction order) of $\overline{g}_n(\alpha)$ will significantly affect the effective two-qubit system since this coupling dictates qubit-qubit transitions such as $\ket{ge}\mapsto\ket{eg}$.

    Note that, in principle, we can just easily do the calculations assuming the oscillator to be in some Fock state $\ket{j}$. However, while the preparation of an initial Fock state might be relatively easy using of the two qubits, changing from one Fock state to another while operating the effective two-qubit system will have much more detrimental effects on the two qubits than, e.g., adiabatically changing the coherent state using a linear drive on the oscillator.

\section{Multiple oscillators coupled to a single qubit}\label{sec:Multimode}

We can extend the framework presented in the previous section to the multioscillator scenario. The calculations, straightforwardly, follow from the multiqubit case. We begin with the multiphoton multimode Rabi (MMR) model where we generically assume $N$ oscillators interact with a single qubit via a multiphoton interaction.

We assume that each oscillator can interact with the qubit through a differing interaction order, i.e. the $l$th oscillator interacts with the qubit through an $n_l$-photon interaction where generally $n_l\neq n_m$ for $m\neq l$\footnote{As an example, one can think of a qubit simultaneously interacting with two oscillators where one is couplied via a one-photon interaction and the other via a two-photon interaction.}.  The MMR Hamiltonian reads
\begin{align}
    \hat{H}_{\text{MMR}}=&\sum_{k} \omega_{k} \adagk\aopk + \frac{\omega_q}{2}\sigZ \nonumber\\&+ \sum_{k}g_{k,n_k}\sigX( \aopk^{ \dagger n_k} +\aopk^{n_k} ),
\end{align}
where $g_{k,n_k}$ specifies the $k$th oscillator's coupling strength to the qubit via an $n_k$-photon interaction. Similar to the multiqubit case, when each oscillator is in the $n_k$-photon JC-RWA regime with respect to its interaction with the qubit. The MMR Hamiltonian can be simplified to the multiphoton multimode Jaynes-Cummings (MMJC) Hamiltonian which reads
\begin{align}
    \hat{H}_{\text{MMJC}}=&\sum_{k} \omega_{k} \adagk\aopk + \frac{\omega_q}{2}\sigZ \nonumber\\&+ \sum_{k}g_{k,n_k}(\sigP\aopk^{n_k} +\sigM\aopk^{ \dagger n_k}  ).
\end{align}
We now proceed to apply the SW transformation in the cases of the RWA and non-RWA regime; the RWA regime takes the MMJC Hamiltonian as the starting point whereas the non-RWA regime considers the MMR Hamiltonian.

Similar to the multiqubit case, the RWA energy scale is defined by $g_{k,n_k}\ll |\Delta_{k,n_k}| \ll \Sigma_{k,n_k} $, where $\Delta_{k,n_k}=\omega_q - n_k\omega_k$ and $\Sigma_{k,n_k}=\omega_q + n_k \omega_k$. We now use the same SW transformation as the multqubit RWA case with the indices running over the oscillators instead of the qubits;
\begin{align}
    \hat{U}_{\text{Disp,RWA}}=\exp(\sum_k \lambda_k \hat{X}_{-,k})
\end{align}
where
\begin{align}
\hat{X}_{\pm,k}=\sigM\aopk^{\dagger n_k}\pm \sigP\aopk^{n_k}
\end{align}
and $\lambda_k= g_{k,n}/(\omega_q - n_k \omega_r).$ Then, to second order in each $\lambda_k$, the multiphoton multioscillator RWA dispersive regime Hamiltonian reads
\begin{align}
\hat{H}_{\text{Disp,RWA}}^{\text{MMO}}&=\hat{U}_{\text{Disp,RWA}}^\dagger \hat{H}_{\text{MMJC}}\hat{U}_{\text{Disp,RWA}}   \nonumber\\ &\simeq \sum_{k} \hat{H}_{\text{Disp,RWA}}^{k}\nonumber\\ &\,\,\,\,\,\,\, +\sum_{l> k}\widetilde{\chi}_{k,l}\sigZ(\aop_{k}^{\dagger n_k} \aop_{l}^{n_l} + \aop_{k}^{n_k} \aop_{l}^{\dagger n_l}), 
\end{align}
where $\hat{H}_{\text{Disp,RWA}}^{k}$ is the $k$th oscillator RWA dispersive Hamiltonian of Eq.~\eqref{eq:RWA_DispHam} with $n_k$ replacing $n$. Here, a qubit-conditional multiphoton $n_k$-to-$n_l$-downconversion interaction emerges.

Similarly to the multiqubit case, the non-RWA extended generator can also be generalized from the multiqubit case where it runs over oscillator indices;
\begin{align}
\hat{Y}_{\pm,k}=\sigM\aopk^{ n_k}\pm \sigP\aopk^{\dagger n_k}.
\end{align}
Thus, the non-RWA transformation becomes
\begin{align}
    \hat{U}_{\text{Disp}}=\exp(\sum_k \lambda_k \hat{X}_{-\,k} +\sum_k \lamBar_k \hat{Y}_{-\,k})
\end{align}
where $\lamBar_k= g_{k,n}/(\omega_q + n_k \omega_r).$ Then, to second order in all perturbation parameters, the non-RWA multiphoton multimode dispersive Hamiltonian is
\begin{align}
\hat{H}_{\text{Disp}}^{\text{MMO}}&=\hat{U}_{\text{Disp}}^\dagger \hat{H}_{\text{MMR}}\hat{U}_{\text{Disp}}   \nonumber\\ &\simeq  \sum_{k}\hat{H}^k_{\text{Disp}}\nonumber\\&\,\,\,\,\,\,\, +\sum_{l> k}\frac{(\widetilde{\chi}_{k,l}+\widetilde{\xi}_{k,l})}{2}\sigZ(\aop_{k}^{\dagger n_k} + \aop_{k}^{n_k})(\aop_{l}^{\dagger n_l} + \aop_{l}^{n_l}),
\end{align}
where $\hat{H}_{\text{Disp}}^{k}$ is the $k$th oscillator non-RWA dispersive Hamiltonian of Eq.~\eqref{eq:nonRWA_DispHam} with $n_k$ replacing $n$. In this case, the qubit-conditional interaction arising comprises both excitation-perserving and non-preserving terms.

It is interesting to contrast the multiqubit and multimode cases. In the multiqubit case, the mediated qubit-qubit interaction is photon-number dependent and involves sums over the powers of the photon number operators, whereas in the multimode case, the mediated oscillator-oscillator interaction depends on $\sigZ$ and not higher powers of $\sigZ$ \footnote{If it were the case, summing over higher powers of $\sigZ$ would result in an additional unconditional interaction since $\sigZ^{k}=\hat{\mathbb{I}}$ for even $k$ and stronger conditional interactions since $\sigZ^k=\sigZ$ for odd $k.$}. The reason for the summation over photon-number powers in the multiqubit scenario is due to the higher-order commutators, e.g. $[\aop^n,\adagn]$, attached to the effective qubit-qubit interaction term in the perturbative expansion. The resulting qubit-conditional interactions can be utilized for creating interesting non-Gaussian states, qubit-controlled interferometry and multiphoton down-conversion \cite{CWilsonPRX}. Additionally, these interactions can be leveraged for faster multimode unitary synthesis over the total Hilbert space \cite{FastUnivControlDisp,ImplementationSC0}. Using similar parameters for the superconducting circuits example in Sec~\ref{sec:ExpSignatures}, we can find estimates for these mediated interactions to be on the order of a few MHz.

    \section{Summary and Conclusions}\label{Sec:Conc}
    Typical SW-type perturbation theory transformations are used to decouple low-energy and high-energy subspaces; this allows us to obtain an effective description of a complicated many-body system within its low-energy subspaces \cite{BRAVYI20112793}. The presented generalization provides an accurate description for the lower-energy states, and the accuracy becomes ever-vanishing for higher-energy states, as expected from a low-energy effective description of the system. {In the usual linear interaction qubit-oscillator RWA dispersive regime, the accuracy is commonly restricted by a critical photon number, $n_{\text{ph,c}}=\Delta^2/4g^2$, which signifies the growing deviation between the perturbative and full system Hamiltonians for higher photon occupation numbers. We derived a generalized heuristic critical photon number for the multiphoton RWA dispersive regime, $n_{\text{ph},c}^{(n)}\propto (\Delta_n / g_n)^{2/n},$ that defines a narrowing regime of validity for higher-energy states. In the RWA regime, we found that our analytic spectra were in excellent agreement with the numerical results. In the non-RWA regime ($|\Delta_n|\sim\Sigma_n$), the RWA analytic spectra become less accurate of an approximation, and we resorted to an extension of the SW transformation that takes the counter-rotating interaction terms, $\sigM\adagn +\sigP\aop^n$, into account. In this regime, the non-RWA spectra are an excellent match for the numerical spectra.}
    
    In the generalized case of multiple qubits nonlinearly coupled to a shared oscillator, we found similar trends to the single qubit-oscillator case with some differences due to the cooperative effects.  {Generally, the perturbative spectra presented provide an accurate description for the multiqubit case when the qubits' collective cooperative effects are weak. A key difference is that in the presence of stronger cooperative effects, i.e. stronger individual qubit coupling strengths, the analytic RWA and non-RWA spectra deviate at smaller coupling strengths in their respective regimes of validty.} The second key difference is the emergence of a photon-number-dependent qubit-qubit interaction. Incidentally, this allows for an effective tuning of the two-qubit system parameters . The state of the oscillator can then be used to tune the effective multiqubit system parameters.

    Furthermore, we highlighted important experimental signatures that distinguish the different dispersive interaction orders. In the single qubit-oscillator case, the effective dressed qubit gets shifted frequency when the oscillator is populated with a coherent state. We found the shift in dressed frequency to be a polynomial function of the coherent state amplitude with order $n$. In the case of multiple qubits, the tunable photon-number dependent qubit-qubit interaction, which also happens to be a polynomial but of degree $n-1$, serves as a signature of the multiphoton dispersive regime since its strength can significantly alter the qubits' transitions, e.g. $\ket{ge}\mapsto\ket{eg}$.

    In Sec.~\ref{sec:Multimode}, we further extended the multiphoton dispersive regime to the converse multipartite scenario where $N$ oscillators are nonlinearly coupled to a single qubit. Here, we find qubit-conditional oscillator-oscillator nonlinear interactions. This can be of great interest for bosonically-encoded quantum computation research. The interest stems from the nonlinear terms being useful for control and readout since in this scenario, one relies on a qubit to control and readout the logical oscillator modes. 

    In summary, we developed a generalized SW perturbation theory for the dispersive regime of a qubit-oscillator system interacting through an $n$-photon interaction. We considered a generalization of the one-photon RWA and non-RWA dispersive-regime SW transformations, and we examined the accuracy of their predictions (to second order in the perturbation parameters) compared to numerical (and semi-analytical) results. This framework was then generalized to two multipartite scenarios, multiple qubits coupled to a shared oscillator, and multiple oscillators coupled to a shared qubit. The multiphoton SW transformation presented here and its multipartite generalizations are important in their relation to the rising number of applications of collective cooperative phenomena for sensing \cite{SuperradiantSensing}, unitary synthesis and universal control \cite{ImplementationSC0,FastUnivControlDisp}, and nonlinear interferometry and multiphoton spontaneous parametric down-conversion \cite{CWilsonPRX}. We believe the work presented here can serve as a guide for designing experimental implementations exploiting the effects of dispersive multiphoton qubit-oscillator interactions.

\begin{acknowledgements}
{We thank the anonymous referees for their useful comments that significantly helped improve the manuscript.}
M.A. was supported by the Institute for Quantum Computing (IQC) through funding provided by Transformative Quantum Technologies (TQT). S.A. was supported by Japan's Ministry of Education, Culture, Sports, Science and Technology's Quantum Leap Flagship Program Grant No. JPMXS0120319794.
\end{acknowledgements}

\appendix

\section{Combinatorial aspects of multiphoton perturbation theory}\label{app:ComComb}

The perturbative RWA and non-RWA expansions used in the main text rely on the commutators arising from the first and second order BCH expansion terms. These terms rely on the commutators $[\xpn,\xmn],\,[\ypn,\xmn],\,[\xpn,\ymn],$ and $[\ypn,\ymn].$ In this appendix, we explicitly express these commutators in the diagonal form shown in the main text.

We begin by showing all the steps involved in evaluating $[\xpn,\xmn]$ from which $[\ypn,\ymn]$ immediately follows; $[\xpn, \ymn]$ and $[\ypn,\xmn]$ can be immediately evaluated and both are equal to $ \sigZ(\aop^{\dagger 2n}-\aop^{2n})$. The commutator reads as
\begin{align}
    [\xpn,\xmn]=&[\sigM\adagn +\sigP\aop^n,\sigM\adagn - \sigP\aop^n]\nonumber\\=&\sigZ\{\aop^n,\adagn\} + [\aop^n,\adagn],
\end{align}
where $\{,\}$ is the anti-commutator.
We now apply the normal-ordering procedure, $\mathcal{N}[...]$, that allows us to rewrite $\aop^n\adagn$ as \cite{ReorderingIntro}
\begin{align}
    \aop^n\adagn=\mathcal{N}[\aop^n\adagn]= \sum_{k=0}^n \begin{pmatrix}n\\ k\end{pmatrix}n^{\underline{k}}\aop^{\dagger n-k}\aop^{n-k},   
\end{align}
where $\begin{pmatrix}n\\ k\end{pmatrix}=n!/(k!(n-k)!)$ and $n^{\underline{k}}=n(n-1)...(n-k+1)$ is the falling factorial with the convention $n^{\underline{0}}=1.$
Next, we use the identity \cite{KATRIEL2000159},
\begin{align}
    \adagn \aop^n= \sum_{k=0}^n s_1(n,k)(\adag\aop)^k,
\end{align}
to obtain a power series in the photon-number operator. Here, we recall that $s_1(n,k)$ are Stirling numbers of the first kind. Thus, we readily have the desired expression for $\adagn\aop^{n}$, but we must further simplify the term $\aop^n\adagn$. It can be shown using some combinatorial tricks that \cite{Reordering2}
\begin{align}
     \aop^n\adagn=&\sum_{k=0}^n \begin{pmatrix}n\\ k\end{pmatrix}n^{\underline{k}}\aop^{\dagger n-k}\aop^{n-k}\nonumber\\ =&\sum_{k=0}^n (-1)^{(n+k)} s_1(n+1,k+1) (\adag\aop)^k.
\end{align}
Thus, we may now express the commutator $[\xpn,\xmn]$ in diagonal form as
\begin{align}
    [\xpn,\xmn]=&\sigZ\{\aop^n,\adagn\} + [\aop^n,\adagn]\nonumber\\ 
    =&\sigZ\sum_{k=0}^n C_{n,k}^+ (\adag\aop)^k + \sum_{k=0}^{n-1} C_{n,k}^- (\adag\aop)^k
\end{align}
where
\begin{align}
    C_{n,k}^{\pm}=(-1)^{n+k} s_1(n+1,k+1)\pm s_1(n,k).
\end{align}
The qubit-state-independent sum terminates at $k=n-1$ because $C_{n,n}^-=0$. 
We can now simply reuse these results for the other commutator, $[\ypn,\ymn]$:
\begin{align}
    [\ypn,\ymn]=&[\sigM\aop^n +\sigP\adagn,\sigM\aop^n - \sigP\adagn]\nonumber\\=&\sigZ\{\aop^n,\adagn\} - [\aop^n,\adagn]\nonumber\\ =&\sigZ\sum_{k=0}^n C_{n,k}^+ (\adag\aop)^k - \sum_{k=0}^{n-1} C_{n,k}^- (\adag\aop)^k.
\end{align}
This last equation explains why the non-RWA qubit-state-independent oscillator Kerr terms come with a minus sign $(-\xi_n).$

We may now find the second-order RWA and non-RWA dispersive Hamiltonians presented in the main text. The BCH expansion truncated at second order for the RWA Hamiltonian reads
\begin{align}
    \hat{H}_{\text{Disp,RWA}}^{(n),2}&= \hat{H}_{n\text{JC}} + \lambda_n[\hat{H}_{n\text{JC}},\xmn] \nonumber\\ &\,\,\,\,\,\,\,+\frac{\lambda_n^2}{2}[[\hat{H}_{n\text{JC}},\xmn],\xmn]\nonumber\\& = \hat{H}_{n\text{JC}} -\Delta_n\lambda_n\xmn\nonumber\\&\,\,\,\,\,\,\,  +\left( g_n\lambda_n-\frac{\Delta_n\lambda_n^2}{2}\right)[\xpn,\xmn]\nonumber\\&\,\,\,\,\,\,\, + \frac{g_n\lambda_n^2}{2}[[\xpn,\xmn],\xmn]   \nonumber\\ &=\omega_o\adag\aop + \frac{\omega_q}{2}\sigZ +\frac{\chi_n}{2}\sum_{k=0}^n C_{n,k}^{+}\sigZ(\adag\aop)^k \nonumber\\ &\,\,\,\,\,\,\, +\frac{\chi_n}{2}\sum_{k=1}^{n-1} C_{n,k}^{-}(\adag\aop)^k \nonumber\\&\,\,\,\,\,\,\, +\frac{g_n^3}{2\Delta_n}[[\xpn,\xmn],\xmn],    
\end{align}
where we used $[\hat{H}_{n\text{JC}},\xmn]=-\Delta_n\xmn + g_n[\xpn,\xmn].$ The last term is on the order of $\lambda_n^3$ and, thus, neglected from the Hamiltonian of Eq.~\eqref{eq:RWA_DispHam} presented in the main text. A similar calculation for the non-RWA Hamiltonian yields
\begin{widetext}
\begin{align*}
    \hat{H}_{\text{Disp}}^{(n),2}&= \hat{H}_{n\text{R}}+\lambda_n[\hat{H}_{\text{n-R}},\xmn]+\lamBar_n[\hat{H}_{\text{n-R}},\ymn]+\frac{\lambda_n^2}{2}[[\hat{H}_{n\text{R}},\xmn],\xmn]+\frac{\lamBar_n^2}{2}[[\hat{H}_{n\text{R}},\ymn],\ymn]\nonumber\\ &\,\,\,\,\,\,\,+\frac{\lambda_n\lamBar_n}{2}[[\hat{H}_{n\text{R}},\xmn],\ymn]+\frac{\lamBar_n\lambda_n}{2}[[\hat{H}_{n\text{R}},\ymn],\xmn]\nonumber\\&=\hat{H}_{n\text{R}}-\Delta_n\lambda_n\xmn +\left(g_n\lambda_n -\frac{\Sigma_n\lambda_n\lamBar_n}{2}\right)[\ypn,\xmn]+\left(g_n\lambda_n-\frac{-\Delta_n\lambda_n^2}{2}\right)[\xpn,\xmn] -\Sigma_n\lamBar_n\ymn\nonumber\\ &\,\,\,\,\,\,\,+\left(g_n\lamBar_n-\frac{\Sigma_n\lamBar_n^2}{2}\right)[\ypn,\ymn]+\left(g_n\lamBar_n -\frac{\Delta_n \lamBar_n\lambda_n}{2}\right)[\xpn,\ymn]\nonumber\\&\,\,\,\,\,\,\,+\frac{g_n}{2}\bigg(\lambda_n^2 ([[\xpn,\xmn],\xmn]+[[\ypn,\xmn],\xmn])+\lamBar_n^2 ([[\xpn,\ymn],\ymn]+[[\ypn,\ymn],\ymn])\nonumber\\&\,\,\,\,\,\,\,+\lambda_n\lamBar_n([[\xpn,\ymn],\xmn]+[[\ypn,\xmn],\ymn]+[[\xpn,\xmn],\ymn]+[[\ypn,\ymn],\xmn])\bigg) 
    \end{align*}
    \begin{align}
    &=\omega_o \adag\aop + \frac{\omega_q}{2}\sigZ + \frac{(\chi_n+\xi_n)}{2}\sum_{k=0}^n C_{n,k}^{+}\sigZ(\adag\aop)^k +\frac{(\chi_n-\xi_n)}{2}\sum_{k=1}^{n-1} C_{n,k}^{-}(\adag\aop)^k+\frac{(\chi_n+\xi_n)}{2}\sigZ(\hat{a}^{\dagger 2n} +\aop^{2n})\nonumber\\&\,\,\,\,\,\,\,+\frac{g_n}{2}\bigg(\lambda_n^2 ([[\xpn,\xmn],\xmn]+[[\ypn,\xmn],\xmn])+\lamBar_n^2 ([[\xpn,\ymn],\ymn]+[[\ypn,\ymn],\ymn])\nonumber\\&\,\,\,\,\,\,\,+\lambda_n\lamBar_n([[\xpn,\ymn],\xmn]+[[\ypn,\xmn],\ymn]+[[\xpn,\xmn],\ymn]+[[\ypn,\ymn],\xmn])\bigg),
\end{align}
\end{widetext}
where we used $[\hat{H}_{\text{n-R}},\xmn]=-\Delta_n \xpn + g_n[\ypn,\xmn]+g_n[\xpn,\xmn]$ and $[\hat{H}_{\text{n-R}},\ymn]=-\Sigma_n \ypn + g_n[\ypn,\ymn]+g_n[\xpn,\ymn].$ The last eight terms comprising nested commutators are of third order in $\lambda_n$ and $\lamBar_n$, and, thus, are not included in the non-RWA Hamiltonian of Eq.~\eqref{eq:nonRWA_DispHam}.

\section{{Stabilization of multiphoton qubit-oscillator interaction models}}\label{app:Stab}

\begin{figure}[t]
        \includegraphics[scale=1.3,trim={.75cm 0cm 0cm 0cm}]{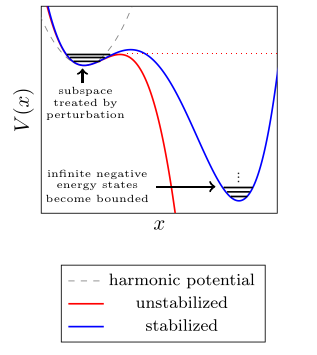}
        \caption{{Schematic diagram that illustrates the stabilization of the Hamiltonian. The gray dashed line represents the unperturbed harmonic oscillator potential. The red line shows the unstabilized potential with harmonic, $x^2$, and cubic, $x^3$, contributions. In the unstabilized case, we obtain a continuous spectrum corresponding to unbounded states, with energies extending to $-\infty$. The blue line is the stabilized potential where a small positive quartic term, $x^4$, is added to the unstabilized potential to ensure that it is bounded from below. The infinite negative energy states in the unstabilized case are replaced by bounded well-behaved states in the stabilized potential. The horizontal red dotted line represents the energy barrier that helps isolate the metastable states in the shallow well and make them largely insensitive to the details of the potential in the deep well.}}
        \label{fig:StabilizationSchematic}
\end{figure}
\begin{figure}[t]
        \includegraphics[scale=1,trim={0.25cm 0cm 0cm 0cm}]{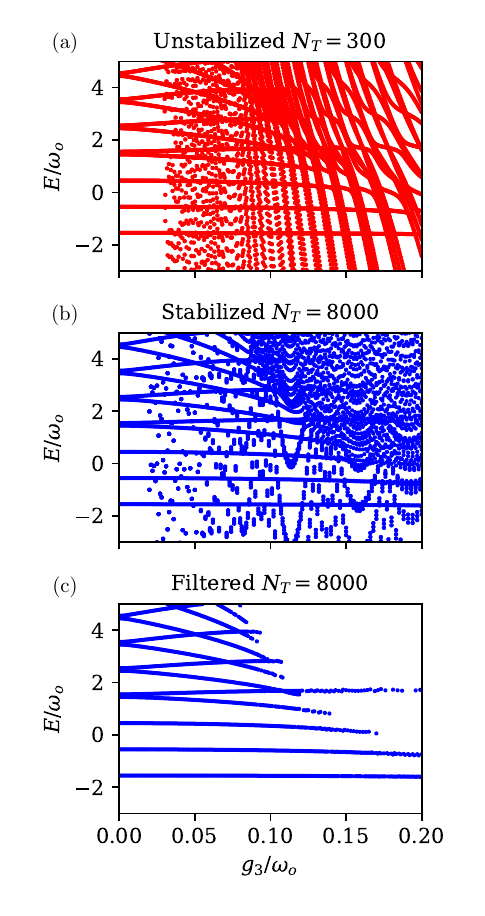}
        \caption{ {Spectrum of the three-photon Rabi model. (a) Unstabilized spectrum of the three-photon Rabi Hamiltonian [Eq.~\eqref{eq:Mod_nRHam} with $\eta=0$] with an oscillator Hilbert space truncation of $N_T=300$. (b) Stabilized spectrum with $\eta=0.02$ and truncation size  $N_T=8000.$ (c) Filtered spectrum in which we take the spectrum in (b) and keep only states with an average photon number less than 20.}}
        \label{fig:NumericalStabilization}
\end{figure}

{In the main text, we discussed the serious stability issues with the $n$-photon Rabi model. For $n\geq 3$, the case $g=0$ is an isolated point where the spectrum is discrete and energetically stable. The spectrum becomes continuous and unbounded from below for any $g\neq 0$. In this appendix, we discuss how higher-order stabilizing terms arise in physical implementations of the n-photon Rabi model, using a realistic superconducting circuit implementation as an example. Then, we present some numerical results that demonstrate how the stabilization works and why our perturbative approach is justified. It is worth noting here that a recent study used a similar idea, in which the two-photon quantum Rabi model in the unstable parameter regime was stabilized by adding an $\left( \aop + \adag \right)^4$ interaction term in the Hamiltonian \cite{Ying_CritMetroStab2025}.}

{\subsection{Stabilizing terms in physical implementations}}

{We consider a superconducting circuit implementation of an effective $n$-photon qubit-oscillator Hamiltonian. One specific example is the circuit proposed in Ref.~\cite{ImplementationSC0}, where a transmon qubit is coupled to an LC oscillator by means of a direct-current superconducting quantum interference device (dc SQUD). The transmon and oscillator are represented by their conjugate phase and charge variables, $\hat{\varphi}_k$ and $\hat{n}_k$, that satisfy the usual commutation relation $[\hat{\varphi}_k,\hat{n}_k]=i$ with $k=t$ for the transmon and $k=o$ for the oscillator, and $[\hat{\varphi}_t,\hat{n}_o]=[\hat{\varphi}_o,\hat{n}_t]=0$. These conjugate variables can be written in terms of creation and annihilation operators as
\begin{align}
    \hat{\varphi}_t=\varphi_t(\bdag +\bop),
\end{align}
\begin{align}
    \hat{n}_t=in_t(\bdag -\bop),
\end{align}
\begin{align}
    \hat{\varphi}_o=\varphi_o(\adag +\aop),
\end{align}
and
\begin{align}
    \hat{n}_o=in_o(\adag -\aop),
\end{align}
where $\varphi_k$ and $n_k$ are the zero-point fluctuation amplitudes in the phase and charge for the respective systems.
In theory, an asymmetric dc SQUID mediates all possible combinations of nonlinear interaction terms between the transmon and oscillator, i.e.~terms that take the form $\hat{\varphi}_t^m\hat{\varphi}_o^n$ for nonnegative integers $m$ and $n$, all occurring together with varying effects depending on the coupling strengths and resonance conditions. For our purposes, we typically seek to implement and utilize only one interaction term of the form $\hat{\varphi}_t\hat{\varphi}_o^n$. Under the two-level approximation, we have $\hat{\varphi}_t\propto (\bdag+\bop)\simeq \sigP + \sigM$, which effectively leads to an interaction Hamiltonian $\propto(\sigP+\sigM)(\adag+\aop)^n$. The interaction term $(\sigP\adagn+\sigM\aop^n)$ is an approximation obtained under additional weak coupling and specific resonance conditions. To arrive at such an interaction Hamiltonian, we assume that all other interactions (e.g.~those of the form $\hat{\varphi}_t^m\hat{\varphi}_o^{l}$ with $l\neq n$) have negligible effects on the spectrum and dynamics. A standard approximation in this context is that of low photon numbers and short evolution times. When one considers states with higher photon numbers and/or dynamics at long timescales, the effects of other interactions are generally no longer negligible. Hence, one must always keep in mind that a realistic system, as described here, has higher-order terms that invalidate the model when the photon number becomes large. The exact way in which the higher-order terms alter the behavior of the system depends on the specifics of the physical system. In this appendix, we consider a stabilizing term of the form $\hat{\varphi}_o^m$ with an even power $m>n$.}

{\subsection{Intuitive picture of spectrum stabilization}}

{To give an intuitive picture of how the stabilization of the spectrum works, it is convenient to consider the $n$-photon full Rabi Hamiltonian stabilized by an even $m$th-order term
\begin{align}\label{eq:Mod_nfRHam}
    \hat{H}_{n\text{fR}}=&\omega_o\adag\aop + \frac{\omega_q}{2}\sigZ + g_n \sigX(\aop+\adag)^n \nonumber\\&+ \eta g_n (\aop+\adag)^m,
\end{align}
with $m>n$ and $g_n,\eta>0$. In the regime $\omega_q\ll\omega_o$, a good approximation for the spectrum is obtained by focusing on one of the two cases $\sigX=\pm1$ \cite{Ashhab_QO_USC_PRA2010}}. {If we take $n=3$ and $m=4$, and we set $\sigX=-1$, we obtain the Hamiltonian
\begin{align}\label{eq:Mod_nfRHamM1}
    \hat{H}_{n\text{fR}}\Big|_{\sigma_x=-1}=&-\frac{\omega_o}{4} (\adag-\aop)^2 + \frac{\omega_o}{4} (\aop+\adag)^2  \nonumber\\& - g_3 (\aop+\adag)^3 + \eta g_3 (\aop+\adag)^4,
\end{align}
up to a constant. The Hamiltonian in Eq.~(\ref{eq:Mod_nfRHamM1}) describes a single-particle problem in which $i(\adag-\aop)$ plays the role of the momentum variable $p$ while $(\adag+\aop)$ plays the role of the position variable $x$. The first term represents the kinetic energy, while the last three terms represent the potential energy. The potential energy of this Hamiltonian is illustrated in Fig.~\ref{fig:StabilizationSchematic}. If we ignore the last term, we obtain a function that is unbounded from below. The potential energy can be made infinite and negative by going to large values of $x$. Furthermore, the spectrum of such an unbounded Hamiltonian is continuous; there are propagating wave states that correspond to all energies from $-\infty$ to $+\infty$. The stabilizing term ensures that this situation does not occur. For sufficiently large $x$, the $x^4$ term will always be the dominant term in the potential energy. As a result, the Hamiltonian is bounded from below and the spectrum is discrete. There will be negative energy solutions. The smaller the value of the stabilizing term coefficient $\eta$, the larger (in absolute value) these energies. However, the most serious consequences of having a Hamiltonian that is unbounded from below are eliminated.}

{Perhaps more importantly, the potential energy has a local minimum at $x=0$. There is an energy barrier that separates the local minimum (shallow well) from the global minimum (deep well). As a result, one can expect that the energy levels in the shallow well that are well below the barrier height will still exist and only slightly be affected by the existence of a deep well relatively far away in $x$. Furthermore, for small values of $\eta$, the stabilizing term is negligible at small values of $x$. As a result, one can expect a perturbative treatment that ignores the stabilizing term to produce accurate results at small values of $x$, even if the spectrum as a whole exhibits singular behavior in this case.}

{A final point that is worth mentioning here is that the quantum states in the shallow well can be considered metastable states. As long as the barrier is high, the energy levels are close to those of the unperturbed Hamiltonian. As the coupling strength $g$ increases and the energy barrier decreases, the states in the shallow well become increasingly prone to tunneling into the deep well, and the corresponding energy levels gradually disappear from the spectrum. A simple analysis of the potential energy function in Eq.~(\ref{eq:Mod_nfRHamM1}) shows that the disappearance of the lowest energy levels in the shallow well occurs when $g_3\sim\eta\omega_o$, which is an expression specific to our definition of the coefficient in the stabilizing term. This behavior will be observed in the spectra shown in the next subsection of this appendix (up to a minor quantitative difference, considering that the next subsection uses a different Hamiltonian). Even in the absence of the stabilizing term ($\eta=0$), one could say that these metastable states of the Hamiltonian exist and that they have a finite lifetime to tunnel into the continuum, after which the system relaxes into a qualitatively different state that is not described by the model Hamiltonian $\hat{H}_{n\text{fR}}$.}

{\subsection{Numerical results on stabilization}}

{For purposes of analyzing the spectrum of $\hat{H}_{n \rm JC}$ in the low-photon-number regime, it is convenient to consider a stabilizing term of the form $a^{\dagger m/2} a^{m/2}$ which is obtained from an oscillator phase term $\hat{\varphi}_o^m$. It turns out that the stabilizing term $a^{\dagger m/2} a^{m/2}$ is more effective in stabilizing the Hamiltonian than the term $(\aop+\adag)^m$. The reason is that there are states with large photon numbers but small $(\aop+\adag)$, which makes it possible for some states to have large values of $(\aop^n+\adagn)$ but small values of $(\aop+\adag)^m$.}

{In Fig.~\ref{fig:NumericalStabilization}, we compare energy levels obtained using the unstabilized Hamiltonian with those obtained using the stabilized Hamiltonian. The vast majority of the energy levels behave differently in the two cases. Furthermore, the energy levels around zero energy (and negative energies) never converge if we increase the number of photons (i.e.~the Hilbert space size) in the simulation. In contrast, the small-energy part of the spectrum of the stabilized Hamiltonian is independent of simulation size, provided a sufficient Hilbert space size is used. Focusing on the small-slope energy levels in the weak-coupling regime, we find very good agreement between the spectra of the stabilized and unstabilized Hamiltonians. These energy levels correspond to the metastable states at low photon numbers described in the previous subsection.}

{As a further illustration of the behavior of these metastable states, in Fig.~\ref{fig:NumericalStabilization}(c) we plot a spectrum in which we keep only the energy levels that correspond to eigenstates with $\left\langle \adag \aop \right\rangle < 20$, i.e.~states with fewer than 20 photons on average. This way we keep only states that are localized near $x=0$. As expected, all the steep lines in the spectrum are eliminated, because they correspond to states that are either localized at large $x$ values or traverse a large range of $x$ values. Only the states that are confined in the shallow well remain in this spectrum. these energy levels are the ones for which we can expect the perturbative treatment to be valid. Furthermore, as the coupling strength increases and starts to approach $\eta\omega_o$, most metastable states disappear in rapid succession, as expected.}

{We finally note that we did similar simulations for $\omega_q=3.5\omega_o$ and found that, while the details of the weak-coupling spectrum change in accordance with the dispersive regime formulas in the main text, the point at which the states become unstable and disappear from the spectrum is almost the same as in the case $\omega_q=3.1\omega_o$, i.e.~as shown in Fig.~\ref{fig:NumericalStabilization}(c). Indeed, the dispersive regime energy shifts and the stability of states in the shallow well of the effective trapping potential are two separate issues.}

\bibliography{paper}

\begin{thebibliography}{71}%
\makeatletter
\providecommand \@ifxundefined [1]{%
 \@ifx{#1\undefined}
}%
\providecommand \@ifnum [1]{%
 \ifnum #1\expandafter \@firstoftwo
 \else \expandafter \@secondoftwo
 \fi
}%
\providecommand \@ifx [1]{%
 \ifx #1\expandafter \@firstoftwo
 \else \expandafter \@secondoftwo
 \fi
}%
\providecommand \natexlab [1]{#1}%
\providecommand \enquote  [1]{``#1''}%
\providecommand \bibnamefont  [1]{#1}%
\providecommand \bibfnamefont [1]{#1}%
\providecommand \citenamefont [1]{#1}%
\providecommand \href@noop [0]{\@secondoftwo}%
\providecommand \href [0]{\begingroup \@sanitize@url \@href}%
\providecommand \@href[1]{\@@startlink{#1}\@@href}%
\providecommand \@@href[1]{\endgroup#1\@@endlink}%
\providecommand \@sanitize@url [0]{\catcode `\\12\catcode `\$12\catcode `\&12\catcode `\#12\catcode `\^12\catcode `\_12\catcode `\%12\relax}%
\providecommand \@@startlink[1]{}%
\providecommand \@@endlink[0]{}%
\providecommand \url  [0]{\begingroup\@sanitize@url \@url }%
\providecommand \@url [1]{\endgroup\@href {#1}{\urlprefix }}%
\providecommand \urlprefix  [0]{URL }%
\providecommand \Eprint [0]{\href }%
\providecommand \doibase [0]{https://doi.org/}%
\providecommand \selectlanguage [0]{\@gobble}%
\providecommand \bibinfo  [0]{\@secondoftwo}%
\providecommand \bibfield  [0]{\@secondoftwo}%
\providecommand \translation [1]{[#1]}%
\providecommand \BibitemOpen [0]{}%
\providecommand \bibitemStop [0]{}%
\providecommand \bibitemNoStop [0]{.\EOS\space}%
\providecommand \EOS [0]{\spacefactor3000\relax}%
\providecommand \BibitemShut  [1]{\csname bibitem#1\endcsname}%
\let\auto@bib@innerbib\@empty
\bibitem [{\citenamefont {Jaynes}\ and\ \citenamefont {Cummings}(1963)}]{JaynesCummings}%
  \BibitemOpen
  \bibfield  {author} {\bibinfo {author} {\bibfnamefont {E.}~\bibnamefont {Jaynes}}\ and\ \bibinfo {author} {\bibfnamefont {F.}~\bibnamefont {Cummings}},\ }\href {https://doi.org/10.1109/PROC.1963.1664} {\bibfield  {journal} {\bibinfo  {journal} {Proceedings of the IEEE}\ }\textbf {\bibinfo {volume} {51}},\ \bibinfo {pages} {89} (\bibinfo {year} {1963})}\BibitemShut {NoStop}%
\bibitem [{\citenamefont {Rabi}(1936)}]{Rabi_PhysRev.49.324}%
  \BibitemOpen
  \bibfield  {author} {\bibinfo {author} {\bibfnamefont {I.~I.}\ \bibnamefont {Rabi}},\ }\href {https://doi.org/10.1103/PhysRev.49.324} {\bibfield  {journal} {\bibinfo  {journal} {Phys. Rev.}\ }\textbf {\bibinfo {volume} {49}},\ \bibinfo {pages} {324} (\bibinfo {year} {1936})}\BibitemShut {NoStop}%
\bibitem [{\citenamefont {Rabi}(1937)}]{Rabi2_PhysRev.51.652}%
  \BibitemOpen
  \bibfield  {author} {\bibinfo {author} {\bibfnamefont {I.~I.}\ \bibnamefont {Rabi}},\ }\href {https://doi.org/10.1103/PhysRev.51.652} {\bibfield  {journal} {\bibinfo  {journal} {Phys. Rev.}\ }\textbf {\bibinfo {volume} {51}},\ \bibinfo {pages} {652} (\bibinfo {year} {1937})}\BibitemShut {NoStop}%
\bibitem [{\citenamefont {Haroche}\ and\ \citenamefont {Raimond}(2006)}]{Haroche_Raimond}%
  \BibitemOpen
  \bibfield  {author} {\bibinfo {author} {\bibfnamefont {S.}~\bibnamefont {Haroche}}\ and\ \bibinfo {author} {\bibfnamefont {J.-M.}\ \bibnamefont {Raimond}},\ }\href@noop {} {\emph {\bibinfo {title} {Exploring the Quantum: Atoms, Cavities, and Photons}}}\ (\bibinfo  {publisher} {Oxford University Press},\ \bibinfo {year} {2006})\BibitemShut {NoStop}%
\bibitem [{\citenamefont {Blais}\ \emph {et~al.}(2021)\citenamefont {Blais}, \citenamefont {Grimsmo}, \citenamefont {Girvin},\ and\ \citenamefont {Wallraff}}]{CircuitQEDReview}%
  \BibitemOpen
  \bibfield  {author} {\bibinfo {author} {\bibfnamefont {A.}~\bibnamefont {Blais}}, \bibinfo {author} {\bibfnamefont {A.~L.}\ \bibnamefont {Grimsmo}}, \bibinfo {author} {\bibfnamefont {S.~M.}\ \bibnamefont {Girvin}},\ and\ \bibinfo {author} {\bibfnamefont {A.}~\bibnamefont {Wallraff}},\ }\href {https://doi.org/10.1103/RevModPhys.93.025005} {\bibfield  {journal} {\bibinfo  {journal} {Rev. Mod. Phys.}\ }\textbf {\bibinfo {volume} {93}},\ \bibinfo {pages} {025005} (\bibinfo {year} {2021})}\BibitemShut {NoStop}%
\bibitem [{\citenamefont {Leibfried}\ \emph {et~al.}(2003)\citenamefont {Leibfried}, \citenamefont {Blatt}, \citenamefont {Monroe},\ and\ \citenamefont {Wineland}}]{TrappedIonsReview}%
  \BibitemOpen
  \bibfield  {author} {\bibinfo {author} {\bibfnamefont {D.}~\bibnamefont {Leibfried}}, \bibinfo {author} {\bibfnamefont {R.}~\bibnamefont {Blatt}}, \bibinfo {author} {\bibfnamefont {C.}~\bibnamefont {Monroe}},\ and\ \bibinfo {author} {\bibfnamefont {D.}~\bibnamefont {Wineland}},\ }\href {https://doi.org/10.1103/RevModPhys.75.281} {\bibfield  {journal} {\bibinfo  {journal} {Rev. Mod. Phys.}\ }\textbf {\bibinfo {volume} {75}},\ \bibinfo {pages} {281} (\bibinfo {year} {2003})}\BibitemShut {NoStop}%
\bibitem [{\citenamefont {Buck}\ and\ \citenamefont {Sukumar}(1981)}]{IntensityDep_BUCK1981132}%
  \BibitemOpen
  \bibfield  {author} {\bibinfo {author} {\bibfnamefont {B.}~\bibnamefont {Buck}}\ and\ \bibinfo {author} {\bibfnamefont {C.}~\bibnamefont {Sukumar}},\ }\href {https://doi.org/https://doi.org/10.1016/0375-9601(81)90042-6} {\bibfield  {journal} {\bibinfo  {journal} {Physics Letters A}\ }\textbf {\bibinfo {volume} {81}},\ \bibinfo {pages} {132} (\bibinfo {year} {1981})}\BibitemShut {NoStop}%
\bibitem [{\citenamefont {Singh}(1982)}]{IntensityDep_Multiph_Singh_PhysRevA.25.3206}%
  \BibitemOpen
  \bibfield  {author} {\bibinfo {author} {\bibfnamefont {S.}~\bibnamefont {Singh}},\ }\href {https://doi.org/10.1103/PhysRevA.25.3206} {\bibfield  {journal} {\bibinfo  {journal} {Phys. Rev. A}\ }\textbf {\bibinfo {volume} {25}},\ \bibinfo {pages} {3206} (\bibinfo {year} {1982})}\BibitemShut {NoStop}%
\bibitem [{\citenamefont {Bu\ifmmode~\check{z}\else \v{z}\fi{}ek}(1989)}]{IntensityDep_Buzek_PhysRevA.39.3196}%
  \BibitemOpen
  \bibfield  {author} {\bibinfo {author} {\bibfnamefont {V.}~\bibnamefont {Bu\ifmmode~\check{z}\else \v{z}\fi{}ek}},\ }\href {https://doi.org/10.1103/PhysRevA.39.3196} {\bibfield  {journal} {\bibinfo  {journal} {Phys. Rev. A}\ }\textbf {\bibinfo {volume} {39}},\ \bibinfo {pages} {3196} (\bibinfo {year} {1989})}\BibitemShut {NoStop}%
\bibitem [{\citenamefont {Abdalla}\ \emph {et~al.}(1986)\citenamefont {Abdalla}, \citenamefont {Hassan},\ and\ \citenamefont {Obada}}]{Multiphoton_Seb}%
  \BibitemOpen
  \bibfield  {author} {\bibinfo {author} {\bibfnamefont {M.~S.}\ \bibnamefont {Abdalla}}, \bibinfo {author} {\bibfnamefont {S.~S.}\ \bibnamefont {Hassan}},\ and\ \bibinfo {author} {\bibfnamefont {A.~S.~F.}\ \bibnamefont {Obada}},\ }\href {https://doi.org/10.1103/PhysRevA.34.4869} {\bibfield  {journal} {\bibinfo  {journal} {Phys. Rev. A}\ }\textbf {\bibinfo {volume} {34}},\ \bibinfo {pages} {4869} (\bibinfo {year} {1986})}\BibitemShut {NoStop}%
\bibitem [{\citenamefont {Aliskenderov}\ \emph {et~al.}(1987)\citenamefont {Aliskenderov}, \citenamefont {Rustamov}, \citenamefont {Shumovsky},\ and\ \citenamefont {Quang}}]{Multiphoton_Aliskenderov_1987}%
  \BibitemOpen
  \bibfield  {author} {\bibinfo {author} {\bibfnamefont {E.~I.}\ \bibnamefont {Aliskenderov}}, \bibinfo {author} {\bibfnamefont {K.~A.}\ \bibnamefont {Rustamov}}, \bibinfo {author} {\bibfnamefont {A.~S.}\ \bibnamefont {Shumovsky}},\ and\ \bibinfo {author} {\bibfnamefont {T.}~\bibnamefont {Quang}},\ }\href {https://doi.org/10.1088/0305-4470/20/18/026} {\bibfield  {journal} {\bibinfo  {journal} {Journal of Physics A: Mathematical and General}\ }\textbf {\bibinfo {volume} {20}},\ \bibinfo {pages} {6265} (\bibinfo {year} {1987})}\BibitemShut {NoStop}%
\bibitem [{\citenamefont {Ashraf}\ \emph {et~al.}(1990)\citenamefont {Ashraf}, \citenamefont {Gea-Banacloche},\ and\ \citenamefont {Zubairy}}]{Multiphoton_Ashraf_PhysRevA.42.6704}%
  \BibitemOpen
  \bibfield  {author} {\bibinfo {author} {\bibfnamefont {I.}~\bibnamefont {Ashraf}}, \bibinfo {author} {\bibfnamefont {J.}~\bibnamefont {Gea-Banacloche}},\ and\ \bibinfo {author} {\bibfnamefont {M.~S.}\ \bibnamefont {Zubairy}},\ }\href {https://doi.org/10.1103/PhysRevA.42.6704} {\bibfield  {journal} {\bibinfo  {journal} {Phys. Rev. A}\ }\textbf {\bibinfo {volume} {42}},\ \bibinfo {pages} {6704} (\bibinfo {year} {1990})}\BibitemShut {NoStop}%
\bibitem [{\citenamefont {Larson}\ and\ \citenamefont {Mavrogordatos}(2021)}]{JCM_Descendants}%
  \BibitemOpen
  \bibfield  {author} {\bibinfo {author} {\bibfnamefont {J.}~\bibnamefont {Larson}}\ and\ \bibinfo {author} {\bibfnamefont {T.}~\bibnamefont {Mavrogordatos}},\ }\href {https://doi.org/10.1088/978-0-7503-3447-1} {\emph {\bibinfo {title} {The Jaynes–Cummings Model and Its Descendants}}},\ 2053-2563\ (\bibinfo  {publisher} {IOP Publishing},\ \bibinfo {year} {2021})\BibitemShut {NoStop}%
\bibitem [{\citenamefont {Ng}\ \emph {et~al.}(1999)\citenamefont {Ng}, \citenamefont {Lo},\ and\ \citenamefont {Liu}}]{SpecCollapseNg1999}%
  \BibitemOpen
  \bibfield  {author} {\bibinfo {author} {\bibfnamefont {K.~M.}\ \bibnamefont {Ng}}, \bibinfo {author} {\bibfnamefont {C.~F.}\ \bibnamefont {Lo}},\ and\ \bibinfo {author} {\bibfnamefont {K.~L.}\ \bibnamefont {Liu}},\ }\href {https://doi.org/10.1007/PL00021611} {\bibfield  {journal} {\bibinfo  {journal} {The European Physical Journal D - Atomic, Molecular, Optical and Plasma Physics}\ }\textbf {\bibinfo {volume} {6}},\ \bibinfo {pages} {119} (\bibinfo {year} {1999})}\BibitemShut {NoStop}%
\bibitem [{\citenamefont {Emary}\ and\ \citenamefont {Bishop}(2002)}]{SpecCollapse_CEmary_2002}%
  \BibitemOpen
  \bibfield  {author} {\bibinfo {author} {\bibfnamefont {C.}~\bibnamefont {Emary}}\ and\ \bibinfo {author} {\bibfnamefont {R.~F.}\ \bibnamefont {Bishop}},\ }\href {https://doi.org/10.1088/0305-4470/35/39/307} {\bibfield  {journal} {\bibinfo  {journal} {Journal of Physics A: Mathematical and General}\ }\textbf {\bibinfo {volume} {35}},\ \bibinfo {pages} {8231} (\bibinfo {year} {2002})}\BibitemShut {NoStop}%
\bibitem [{\citenamefont {Trav\ifmmode~\check{e}\else \v{e}\fi{}nec}(2012)}]{TwoPh_Travenec_PhysRevA.85.043805}%
  \BibitemOpen
  \bibfield  {author} {\bibinfo {author} {\bibfnamefont {I.}~\bibnamefont {Trav\ifmmode~\check{e}\else \v{e}\fi{}nec}},\ }\href {https://doi.org/10.1103/PhysRevA.85.043805} {\bibfield  {journal} {\bibinfo  {journal} {Phys. Rev. A}\ }\textbf {\bibinfo {volume} {85}},\ \bibinfo {pages} {043805} (\bibinfo {year} {2012})}\BibitemShut {NoStop}%
\bibitem [{\citenamefont {Duan}\ \emph {et~al.}(2016)\citenamefont {Duan}, \citenamefont {Xie}, \citenamefont {Braak},\ and\ \citenamefont {Chen}}]{SpecCollapse_Duan_2016}%
  \BibitemOpen
  \bibfield  {author} {\bibinfo {author} {\bibfnamefont {L.}~\bibnamefont {Duan}}, \bibinfo {author} {\bibfnamefont {Y.-F.}\ \bibnamefont {Xie}}, \bibinfo {author} {\bibfnamefont {D.}~\bibnamefont {Braak}},\ and\ \bibinfo {author} {\bibfnamefont {Q.-H.}\ \bibnamefont {Chen}},\ }\href {https://doi.org/10.1088/1751-8113/49/46/464002} {\bibfield  {journal} {\bibinfo  {journal} {Journal of Physics A: Mathematical and Theoretical}\ }\textbf {\bibinfo {volume} {49}},\ \bibinfo {pages} {464002} (\bibinfo {year} {2016})}\BibitemShut {NoStop}%
\bibitem [{\citenamefont {Braak}(2011)}]{Braak2011_PhysRevLett.107.100401}%
  \BibitemOpen
  \bibfield  {author} {\bibinfo {author} {\bibfnamefont {D.}~\bibnamefont {Braak}},\ }\href {https://doi.org/10.1103/PhysRevLett.107.100401} {\bibfield  {journal} {\bibinfo  {journal} {Phys. Rev. Lett.}\ }\textbf {\bibinfo {volume} {107}},\ \bibinfo {pages} {100401} (\bibinfo {year} {2011})}\BibitemShut {NoStop}%
\bibitem [{\citenamefont {Chen}\ \emph {et~al.}(2012)\citenamefont {Chen}, \citenamefont {Wang}, \citenamefont {He}, \citenamefont {Liu},\ and\ \citenamefont {Wang}}]{TwoPh_Chen_PhysRevA.86.023822}%
  \BibitemOpen
  \bibfield  {author} {\bibinfo {author} {\bibfnamefont {Q.-H.}\ \bibnamefont {Chen}}, \bibinfo {author} {\bibfnamefont {C.}~\bibnamefont {Wang}}, \bibinfo {author} {\bibfnamefont {S.}~\bibnamefont {He}}, \bibinfo {author} {\bibfnamefont {T.}~\bibnamefont {Liu}},\ and\ \bibinfo {author} {\bibfnamefont {K.-L.}\ \bibnamefont {Wang}},\ }\href {https://doi.org/10.1103/PhysRevA.86.023822} {\bibfield  {journal} {\bibinfo  {journal} {Phys. Rev. A}\ }\textbf {\bibinfo {volume} {86}},\ \bibinfo {pages} {023822} (\bibinfo {year} {2012})}\BibitemShut {NoStop}%
\bibitem [{\citenamefont {Cui}\ \emph {et~al.}(2017)\citenamefont {Cui}, \citenamefont {Cao}, \citenamefont {Fan},\ and\ \citenamefont {Amico}}]{TwoPh_Cui_2017}%
  \BibitemOpen
  \bibfield  {author} {\bibinfo {author} {\bibfnamefont {S.}~\bibnamefont {Cui}}, \bibinfo {author} {\bibfnamefont {J.-P.}\ \bibnamefont {Cao}}, \bibinfo {author} {\bibfnamefont {H.}~\bibnamefont {Fan}},\ and\ \bibinfo {author} {\bibfnamefont {L.}~\bibnamefont {Amico}},\ }\href {https://doi.org/10.1088/1751-8121/aa6a6f} {\bibfield  {journal} {\bibinfo  {journal} {Journal of Physics A: Mathematical and Theoretical}\ }\textbf {\bibinfo {volume} {50}},\ \bibinfo {pages} {204001} (\bibinfo {year} {2017})}\BibitemShut {NoStop}%
\bibitem [{\citenamefont {Drechsler}\ \emph {et~al.}(2020)\citenamefont {Drechsler}, \citenamefont {Bel\'en~Far\'{\i}as}, \citenamefont {Freitas}, \citenamefont {Schmiegelow},\ and\ \citenamefont {Paz}}]{StateDepSqz_PhysRevA.101.052331}%
  \BibitemOpen
  \bibfield  {author} {\bibinfo {author} {\bibfnamefont {M.}~\bibnamefont {Drechsler}}, \bibinfo {author} {\bibfnamefont {M.}~\bibnamefont {Bel\'en~Far\'{\i}as}}, \bibinfo {author} {\bibfnamefont {N.}~\bibnamefont {Freitas}}, \bibinfo {author} {\bibfnamefont {C.~T.}\ \bibnamefont {Schmiegelow}},\ and\ \bibinfo {author} {\bibfnamefont {J.~P.}\ \bibnamefont {Paz}},\ }\href {https://doi.org/10.1103/PhysRevA.101.052331} {\bibfield  {journal} {\bibinfo  {journal} {Phys. Rev. A}\ }\textbf {\bibinfo {volume} {101}},\ \bibinfo {pages} {052331} (\bibinfo {year} {2020})}\BibitemShut {NoStop}%
\bibitem [{\citenamefont {Ayyash}\ \emph {et~al.}(2024)\citenamefont {Ayyash}, \citenamefont {Xu}, \citenamefont {Ashhab},\ and\ \citenamefont {Mariantoni}}]{ImplementationSC0}%
  \BibitemOpen
  \bibfield  {author} {\bibinfo {author} {\bibfnamefont {M.}~\bibnamefont {Ayyash}}, \bibinfo {author} {\bibfnamefont {X.}~\bibnamefont {Xu}}, \bibinfo {author} {\bibfnamefont {S.}~\bibnamefont {Ashhab}},\ and\ \bibinfo {author} {\bibfnamefont {M.}~\bibnamefont {Mariantoni}},\ }\href@noop {} {\bibfield  {journal} {\bibinfo  {journal} {Phys. Rev. A}\ }\textbf {\bibinfo {volume} {104}},\ \bibinfo {pages} {003700} (\bibinfo {year} {2024})}\BibitemShut {NoStop}%
\bibitem [{\citenamefont {Saner}\ \emph {et~al.}(2024)\citenamefont {Saner}, \citenamefont {Băzăvan}, \citenamefont {Webb}, \citenamefont {Araneda}, \citenamefont {Lucas}, \citenamefont {Ballance},\ and\ \citenamefont {Srinivas}}]{SanerNonclassicalHOStates}%
  \BibitemOpen
  \bibfield  {author} {\bibinfo {author} {\bibfnamefont {S.}~\bibnamefont {Saner}}, \bibinfo {author} {\bibfnamefont {O.}~\bibnamefont {Băzăvan}}, \bibinfo {author} {\bibfnamefont {D.~J.}\ \bibnamefont {Webb}}, \bibinfo {author} {\bibfnamefont {G.}~\bibnamefont {Araneda}}, \bibinfo {author} {\bibfnamefont {D.~M.}\ \bibnamefont {Lucas}}, \bibinfo {author} {\bibfnamefont {C.~J.}\ \bibnamefont {Ballance}},\ and\ \bibinfo {author} {\bibfnamefont {R.}~\bibnamefont {Srinivas}},\ }\href {https://arxiv.org/abs/2409.03482} {\bibinfo {title} {Generating arbitrary superpositions of nonclassical quantum harmonic oscillator states}} (\bibinfo {year} {2024}),\ \Eprint {https://arxiv.org/abs/2409.03482} {arXiv:2409.03482 [quant-ph]} \BibitemShut {NoStop}%
\bibitem [{\citenamefont {Felicetti}\ \emph {et~al.}(2018{\natexlab{a}})\citenamefont {Felicetti}, \citenamefont {Rossatto}, \citenamefont {Rico}, \citenamefont {Solano},\ and\ \citenamefont {Forn-D\'{\i}az}}]{ImplementationSC1}%
  \BibitemOpen
  \bibfield  {author} {\bibinfo {author} {\bibfnamefont {S.}~\bibnamefont {Felicetti}}, \bibinfo {author} {\bibfnamefont {D.~Z.}\ \bibnamefont {Rossatto}}, \bibinfo {author} {\bibfnamefont {E.}~\bibnamefont {Rico}}, \bibinfo {author} {\bibfnamefont {E.}~\bibnamefont {Solano}},\ and\ \bibinfo {author} {\bibfnamefont {P.}~\bibnamefont {Forn-D\'{\i}az}},\ }\href {https://doi.org/10.1103/PhysRevA.97.013851} {\bibfield  {journal} {\bibinfo  {journal} {Phys. Rev. A}\ }\textbf {\bibinfo {volume} {97}},\ \bibinfo {pages} {013851} (\bibinfo {year} {2018}{\natexlab{a}})}\BibitemShut {NoStop}%
\bibitem [{\citenamefont {Felicetti}\ \emph {et~al.}(2018{\natexlab{b}})\citenamefont {Felicetti}, \citenamefont {Hwang},\ and\ \citenamefont {Le~Boit\'e}}]{ImplementationSC2}%
  \BibitemOpen
  \bibfield  {author} {\bibinfo {author} {\bibfnamefont {S.}~\bibnamefont {Felicetti}}, \bibinfo {author} {\bibfnamefont {M.-J.}\ \bibnamefont {Hwang}},\ and\ \bibinfo {author} {\bibfnamefont {A.}~\bibnamefont {Le~Boit\'e}},\ }\href {https://doi.org/10.1103/PhysRevA.98.053859} {\bibfield  {journal} {\bibinfo  {journal} {Phys. Rev. A}\ }\textbf {\bibinfo {volume} {98}},\ \bibinfo {pages} {053859} (\bibinfo {year} {2018}{\natexlab{b}})}\BibitemShut {NoStop}%
\bibitem [{\citenamefont {Zou}\ \emph {et~al.}(2020)\citenamefont {Zou}, \citenamefont {Zhang}, \citenamefont {Xu}, \citenamefont {Huang},\ and\ \citenamefont {Liao}}]{ImplementationSC3}%
  \BibitemOpen
  \bibfield  {author} {\bibinfo {author} {\bibfnamefont {F.}~\bibnamefont {Zou}}, \bibinfo {author} {\bibfnamefont {X.-Y.}\ \bibnamefont {Zhang}}, \bibinfo {author} {\bibfnamefont {X.-W.}\ \bibnamefont {Xu}}, \bibinfo {author} {\bibfnamefont {J.-F.}\ \bibnamefont {Huang}},\ and\ \bibinfo {author} {\bibfnamefont {J.-Q.}\ \bibnamefont {Liao}},\ }\href {https://doi.org/10.1103/PhysRevA.102.053710} {\bibfield  {journal} {\bibinfo  {journal} {Phys. Rev. A}\ }\textbf {\bibinfo {volume} {102}},\ \bibinfo {pages} {053710} (\bibinfo {year} {2020})}\BibitemShut {NoStop}%
\bibitem [{\citenamefont {Puebla}\ \emph {et~al.}(2017)\citenamefont {Puebla}, \citenamefont {Hwang}, \citenamefont {Casanova},\ and\ \citenamefont {Plenio}}]{ImplementationTI2}%
  \BibitemOpen
  \bibfield  {author} {\bibinfo {author} {\bibfnamefont {R.}~\bibnamefont {Puebla}}, \bibinfo {author} {\bibfnamefont {M.-J.}\ \bibnamefont {Hwang}}, \bibinfo {author} {\bibfnamefont {J.}~\bibnamefont {Casanova}},\ and\ \bibinfo {author} {\bibfnamefont {M.~B.}\ \bibnamefont {Plenio}},\ }\href {https://doi.org/10.1103/PhysRevA.95.063844} {\bibfield  {journal} {\bibinfo  {journal} {Phys. Rev. A}\ }\textbf {\bibinfo {volume} {95}},\ \bibinfo {pages} {063844} (\bibinfo {year} {2017})}\BibitemShut {NoStop}%
\bibitem [{\citenamefont {von L{\"u}pke}\ \emph {et~al.}(2024)\citenamefont {von L{\"u}pke}, \citenamefont {Rodrigues}, \citenamefont {Yang}, \citenamefont {Fadel},\ and\ \citenamefont {Chu}}]{vonLüpke2024_multimode}%
  \BibitemOpen
  \bibfield  {author} {\bibinfo {author} {\bibfnamefont {U.}~\bibnamefont {von L{\"u}pke}}, \bibinfo {author} {\bibfnamefont {I.~C.}\ \bibnamefont {Rodrigues}}, \bibinfo {author} {\bibfnamefont {Y.}~\bibnamefont {Yang}}, \bibinfo {author} {\bibfnamefont {M.}~\bibnamefont {Fadel}},\ and\ \bibinfo {author} {\bibfnamefont {Y.}~\bibnamefont {Chu}},\ }\href {https://doi.org/10.1038/s41567-023-02377-w} {\bibfield  {journal} {\bibinfo  {journal} {Nature Physics}\ }\textbf {\bibinfo {volume} {20}},\ \bibinfo {pages} {564} (\bibinfo {year} {2024})}\BibitemShut {NoStop}%
\bibitem [{\citenamefont {Marti}\ \emph {et~al.}(2024)\citenamefont {Marti}, \citenamefont {von L{\"u}pke}, \citenamefont {Joshi}, \citenamefont {Yang}, \citenamefont {Bild}, \citenamefont {Omahen}, \citenamefont {Chu},\ and\ \citenamefont {Fadel}}]{Marti2024_mechSqz}%
  \BibitemOpen
  \bibfield  {author} {\bibinfo {author} {\bibfnamefont {S.}~\bibnamefont {Marti}}, \bibinfo {author} {\bibfnamefont {U.}~\bibnamefont {von L{\"u}pke}}, \bibinfo {author} {\bibfnamefont {O.}~\bibnamefont {Joshi}}, \bibinfo {author} {\bibfnamefont {Y.}~\bibnamefont {Yang}}, \bibinfo {author} {\bibfnamefont {M.}~\bibnamefont {Bild}}, \bibinfo {author} {\bibfnamefont {A.}~\bibnamefont {Omahen}}, \bibinfo {author} {\bibfnamefont {Y.}~\bibnamefont {Chu}},\ and\ \bibinfo {author} {\bibfnamefont {M.}~\bibnamefont {Fadel}},\ }\href {https://doi.org/10.1038/s41567-024-02545-6} {\bibfield  {journal} {\bibinfo  {journal} {Nature Physics}\ }\textbf {\bibinfo {volume} {20}},\ \bibinfo {pages} {1448} (\bibinfo {year} {2024})}\BibitemShut {NoStop}%
\bibitem [{\citenamefont {Blais}\ \emph {et~al.}(2007)\citenamefont {Blais}, \citenamefont {Gambetta}, \citenamefont {Wallraff}, \citenamefont {Schuster}, \citenamefont {Girvin}, \citenamefont {Devoret},\ and\ \citenamefont {Schoelkopf}}]{Blais_QIP_cQED}%
  \BibitemOpen
  \bibfield  {author} {\bibinfo {author} {\bibfnamefont {A.}~\bibnamefont {Blais}}, \bibinfo {author} {\bibfnamefont {J.}~\bibnamefont {Gambetta}}, \bibinfo {author} {\bibfnamefont {A.}~\bibnamefont {Wallraff}}, \bibinfo {author} {\bibfnamefont {D.~I.}\ \bibnamefont {Schuster}}, \bibinfo {author} {\bibfnamefont {S.~M.}\ \bibnamefont {Girvin}}, \bibinfo {author} {\bibfnamefont {M.~H.}\ \bibnamefont {Devoret}},\ and\ \bibinfo {author} {\bibfnamefont {R.~J.}\ \bibnamefont {Schoelkopf}},\ }\href {https://doi.org/10.1103/PhysRevA.75.032329} {\bibfield  {journal} {\bibinfo  {journal} {Phys. Rev. A}\ }\textbf {\bibinfo {volume} {75}},\ \bibinfo {pages} {032329} (\bibinfo {year} {2007})}\BibitemShut {NoStop}%
\bibitem [{\citenamefont {Zueco}\ \emph {et~al.}(2009)\citenamefont {Zueco}, \citenamefont {Reuther}, \citenamefont {Kohler},\ and\ \citenamefont {H\"anggi}}]{Zueco_DispersiveRegime}%
  \BibitemOpen
  \bibfield  {author} {\bibinfo {author} {\bibfnamefont {D.}~\bibnamefont {Zueco}}, \bibinfo {author} {\bibfnamefont {G.~M.}\ \bibnamefont {Reuther}}, \bibinfo {author} {\bibfnamefont {S.}~\bibnamefont {Kohler}},\ and\ \bibinfo {author} {\bibfnamefont {P.}~\bibnamefont {H\"anggi}},\ }\href {https://doi.org/10.1103/PhysRevA.80.033846} {\bibfield  {journal} {\bibinfo  {journal} {Phys. Rev. A}\ }\textbf {\bibinfo {volume} {80}},\ \bibinfo {pages} {033846} (\bibinfo {year} {2009})}\BibitemShut {NoStop}%
\bibitem [{\citenamefont {Beaudoin}\ \emph {et~al.}(2011)\citenamefont {Beaudoin}, \citenamefont {Gambetta},\ and\ \citenamefont {Blais}}]{DissipAndUSC}%
  \BibitemOpen
  \bibfield  {author} {\bibinfo {author} {\bibfnamefont {F.}~\bibnamefont {Beaudoin}}, \bibinfo {author} {\bibfnamefont {J.~M.}\ \bibnamefont {Gambetta}},\ and\ \bibinfo {author} {\bibfnamefont {A.}~\bibnamefont {Blais}},\ }\href {https://doi.org/10.1103/PhysRevA.84.043832} {\bibfield  {journal} {\bibinfo  {journal} {Phys. Rev. A}\ }\textbf {\bibinfo {volume} {84}},\ \bibinfo {pages} {043832} (\bibinfo {year} {2011})}\BibitemShut {NoStop}%
\bibitem [{\citenamefont {Schrieffer}\ and\ \citenamefont {Wolff}(1966)}]{OriginalSWPaper}%
  \BibitemOpen
  \bibfield  {author} {\bibinfo {author} {\bibfnamefont {J.~R.}\ \bibnamefont {Schrieffer}}\ and\ \bibinfo {author} {\bibfnamefont {P.~A.}\ \bibnamefont {Wolff}},\ }\href {https://doi.org/10.1103/PhysRev.149.491} {\bibfield  {journal} {\bibinfo  {journal} {Phys. Rev.}\ }\textbf {\bibinfo {volume} {149}},\ \bibinfo {pages} {491} (\bibinfo {year} {1966})}\BibitemShut {NoStop}%
\bibitem [{\citenamefont {Boissonneault}\ \emph {et~al.}(2010{\natexlab{a}})\citenamefont {Boissonneault}, \citenamefont {Gambetta},\ and\ \citenamefont {Blais}}]{ReadoutQubitInducedNonlinearity}%
  \BibitemOpen
  \bibfield  {author} {\bibinfo {author} {\bibfnamefont {M.}~\bibnamefont {Boissonneault}}, \bibinfo {author} {\bibfnamefont {J.~M.}\ \bibnamefont {Gambetta}},\ and\ \bibinfo {author} {\bibfnamefont {A.}~\bibnamefont {Blais}},\ }\href {https://doi.org/10.1103/PhysRevLett.105.100504} {\bibfield  {journal} {\bibinfo  {journal} {Phys. Rev. Lett.}\ }\textbf {\bibinfo {volume} {105}},\ \bibinfo {pages} {100504} (\bibinfo {year} {2010}{\natexlab{a}})}\BibitemShut {NoStop}%
\bibitem [{\citenamefont {Forn-D\'{\i}az}\ \emph {et~al.}(2010)\citenamefont {Forn-D\'{\i}az}, \citenamefont {Lisenfeld}, \citenamefont {Marcos}, \citenamefont {Garc\'{\i}a-Ripoll}, \citenamefont {Solano}, \citenamefont {Harmans},\ and\ \citenamefont {Mooij}}]{USC_FornDiaz2010}%
  \BibitemOpen
  \bibfield  {author} {\bibinfo {author} {\bibfnamefont {P.}~\bibnamefont {Forn-D\'{\i}az}}, \bibinfo {author} {\bibfnamefont {J.}~\bibnamefont {Lisenfeld}}, \bibinfo {author} {\bibfnamefont {D.}~\bibnamefont {Marcos}}, \bibinfo {author} {\bibfnamefont {J.~J.}\ \bibnamefont {Garc\'{\i}a-Ripoll}}, \bibinfo {author} {\bibfnamefont {E.}~\bibnamefont {Solano}}, \bibinfo {author} {\bibfnamefont {C.~J. P.~M.}\ \bibnamefont {Harmans}},\ and\ \bibinfo {author} {\bibfnamefont {J.~E.}\ \bibnamefont {Mooij}},\ }\href {https://doi.org/10.1103/PhysRevLett.105.237001} {\bibfield  {journal} {\bibinfo  {journal} {Phys. Rev. Lett.}\ }\textbf {\bibinfo {volume} {105}},\ \bibinfo {pages} {237001} (\bibinfo {year} {2010})}\BibitemShut {NoStop}%
\bibitem [{\citenamefont {Niemczyk}\ \emph {et~al.}(2010{\natexlab{a}})\citenamefont {Niemczyk}, \citenamefont {Deppe}, \citenamefont {Huebl}, \citenamefont {Menzel}, \citenamefont {Hocke}, \citenamefont {Schwarz}, \citenamefont {Garcia-Ripoll}, \citenamefont {Zueco}, \citenamefont {H{\"u}mmer}, \citenamefont {Solano}, \citenamefont {Marx},\ and\ \citenamefont {Gross}}]{USC_Niemczyk2010}%
  \BibitemOpen
  \bibfield  {author} {\bibinfo {author} {\bibfnamefont {T.}~\bibnamefont {Niemczyk}}, \bibinfo {author} {\bibfnamefont {F.}~\bibnamefont {Deppe}}, \bibinfo {author} {\bibfnamefont {H.}~\bibnamefont {Huebl}}, \bibinfo {author} {\bibfnamefont {E.~P.}\ \bibnamefont {Menzel}}, \bibinfo {author} {\bibfnamefont {F.}~\bibnamefont {Hocke}}, \bibinfo {author} {\bibfnamefont {M.~J.}\ \bibnamefont {Schwarz}}, \bibinfo {author} {\bibfnamefont {J.~J.}\ \bibnamefont {Garcia-Ripoll}}, \bibinfo {author} {\bibfnamefont {D.}~\bibnamefont {Zueco}}, \bibinfo {author} {\bibfnamefont {T.}~\bibnamefont {H{\"u}mmer}}, \bibinfo {author} {\bibfnamefont {E.}~\bibnamefont {Solano}}, \bibinfo {author} {\bibfnamefont {A.}~\bibnamefont {Marx}},\ and\ \bibinfo {author} {\bibfnamefont {R.}~\bibnamefont {Gross}},\ }\href {https://doi.org/10.1038/nphys1730} {\bibfield  {journal} {\bibinfo  {journal} {Nature Physics}\ }\textbf {\bibinfo {volume} {6}},\ \bibinfo {pages} {772} (\bibinfo {year} {2010}{\natexlab{a}})}\BibitemShut {NoStop}%
\bibitem [{\citenamefont {Yoshihara}\ \emph {et~al.}(2017)\citenamefont {Yoshihara}, \citenamefont {Fuse}, \citenamefont {Ashhab}, \citenamefont {Kakuyanagi}, \citenamefont {Saito},\ and\ \citenamefont {Semba}}]{BeyondUSC_Yoshihara2017}%
  \BibitemOpen
  \bibfield  {author} {\bibinfo {author} {\bibfnamefont {F.}~\bibnamefont {Yoshihara}}, \bibinfo {author} {\bibfnamefont {T.}~\bibnamefont {Fuse}}, \bibinfo {author} {\bibfnamefont {S.}~\bibnamefont {Ashhab}}, \bibinfo {author} {\bibfnamefont {K.}~\bibnamefont {Kakuyanagi}}, \bibinfo {author} {\bibfnamefont {S.}~\bibnamefont {Saito}},\ and\ \bibinfo {author} {\bibfnamefont {K.}~\bibnamefont {Semba}},\ }\href {https://doi.org/10.1038/nphys3906} {\bibfield  {journal} {\bibinfo  {journal} {Nature Physics}\ }\textbf {\bibinfo {volume} {13}},\ \bibinfo {pages} {44} (\bibinfo {year} {2017})}\BibitemShut {NoStop}%
\bibitem [{\citenamefont {Joshi}\ \emph {et~al.}(2017)\citenamefont {Joshi}, \citenamefont {Irish},\ and\ \citenamefont {Spiller}}]{DispSqueezing}%
  \BibitemOpen
  \bibfield  {author} {\bibinfo {author} {\bibfnamefont {C.}~\bibnamefont {Joshi}}, \bibinfo {author} {\bibfnamefont {E.~K.}\ \bibnamefont {Irish}},\ and\ \bibinfo {author} {\bibfnamefont {T.~P.}\ \bibnamefont {Spiller}},\ }\href {https://doi.org/10.1038/srep45587} {\bibfield  {journal} {\bibinfo  {journal} {Scientific Reports}\ }\textbf {\bibinfo {volume} {7}},\ \bibinfo {pages} {45587} (\bibinfo {year} {2017})}\BibitemShut {NoStop}%
\bibitem [{\citenamefont {Dicke}(1954)}]{DickePaper}%
  \BibitemOpen
  \bibfield  {author} {\bibinfo {author} {\bibfnamefont {R.~H.}\ \bibnamefont {Dicke}},\ }\href {https://doi.org/10.1103/PhysRev.93.99} {\bibfield  {journal} {\bibinfo  {journal} {Phys. Rev.}\ }\textbf {\bibinfo {volume} {93}},\ \bibinfo {pages} {99} (\bibinfo {year} {1954})}\BibitemShut {NoStop}%
\bibitem [{\citenamefont {Hepp}\ and\ \citenamefont {Lieb}(1973)}]{Hepp_Lieb_DickeModel}%
  \BibitemOpen
  \bibfield  {author} {\bibinfo {author} {\bibfnamefont {K.}~\bibnamefont {Hepp}}\ and\ \bibinfo {author} {\bibfnamefont {E.~H.}\ \bibnamefont {Lieb}},\ }\href {https://doi.org/https://doi.org/10.1016/0003-4916(73)90039-0} {\bibfield  {journal} {\bibinfo  {journal} {Annals of Physics}\ }\textbf {\bibinfo {volume} {76}},\ \bibinfo {pages} {360} (\bibinfo {year} {1973})}\BibitemShut {NoStop}%
\bibitem [{\citenamefont {Wang}\ and\ \citenamefont {Hioe}(1973)}]{Wang_Hioe_DickeModel}%
  \BibitemOpen
  \bibfield  {author} {\bibinfo {author} {\bibfnamefont {Y.~K.}\ \bibnamefont {Wang}}\ and\ \bibinfo {author} {\bibfnamefont {F.~T.}\ \bibnamefont {Hioe}},\ }\href {https://doi.org/10.1103/PhysRevA.7.831} {\bibfield  {journal} {\bibinfo  {journal} {Phys. Rev. A}\ }\textbf {\bibinfo {volume} {7}},\ \bibinfo {pages} {831} (\bibinfo {year} {1973})}\BibitemShut {NoStop}%
\bibitem [{\citenamefont {Tavis}\ and\ \citenamefont {Cummings}(1968)}]{TavisCummingsModel}%
  \BibitemOpen
  \bibfield  {author} {\bibinfo {author} {\bibfnamefont {M.}~\bibnamefont {Tavis}}\ and\ \bibinfo {author} {\bibfnamefont {F.~W.}\ \bibnamefont {Cummings}},\ }\href {https://doi.org/10.1103/PhysRev.170.379} {\bibfield  {journal} {\bibinfo  {journal} {Phys. Rev.}\ }\textbf {\bibinfo {volume} {170}},\ \bibinfo {pages} {379} (\bibinfo {year} {1968})}\BibitemShut {NoStop}%
\bibitem [{\citenamefont {Kakuyanagi}\ \emph {et~al.}(2016)\citenamefont {Kakuyanagi}, \citenamefont {Matsuzaki}, \citenamefont {D\'eprez}, \citenamefont {Toida}, \citenamefont {Semba}, \citenamefont {Yamaguchi}, \citenamefont {Munro},\ and\ \citenamefont {Saito}}]{TC_Dicke_ExpKakuyanagi}%
  \BibitemOpen
  \bibfield  {author} {\bibinfo {author} {\bibfnamefont {K.}~\bibnamefont {Kakuyanagi}}, \bibinfo {author} {\bibfnamefont {Y.}~\bibnamefont {Matsuzaki}}, \bibinfo {author} {\bibfnamefont {C.}~\bibnamefont {D\'eprez}}, \bibinfo {author} {\bibfnamefont {H.}~\bibnamefont {Toida}}, \bibinfo {author} {\bibfnamefont {K.}~\bibnamefont {Semba}}, \bibinfo {author} {\bibfnamefont {H.}~\bibnamefont {Yamaguchi}}, \bibinfo {author} {\bibfnamefont {W.~J.}\ \bibnamefont {Munro}},\ and\ \bibinfo {author} {\bibfnamefont {S.}~\bibnamefont {Saito}},\ }\href {https://doi.org/10.1103/PhysRevLett.117.210503} {\bibfield  {journal} {\bibinfo  {journal} {Phys. Rev. Lett.}\ }\textbf {\bibinfo {volume} {117}},\ \bibinfo {pages} {210503} (\bibinfo {year} {2016})}\BibitemShut {NoStop}%
\bibitem [{\citenamefont {Jaako}\ \emph {et~al.}(2016)\citenamefont {Jaako}, \citenamefont {Xiang}, \citenamefont {Garcia-Ripoll},\ and\ \citenamefont {Rabl}}]{Jaako_BeyondUSCDicke2016}%
  \BibitemOpen
  \bibfield  {author} {\bibinfo {author} {\bibfnamefont {T.}~\bibnamefont {Jaako}}, \bibinfo {author} {\bibfnamefont {Z.-L.}\ \bibnamefont {Xiang}}, \bibinfo {author} {\bibfnamefont {J.~J.}\ \bibnamefont {Garcia-Ripoll}},\ and\ \bibinfo {author} {\bibfnamefont {P.}~\bibnamefont {Rabl}},\ }\href {https://doi.org/10.1103/PhysRevA.94.033850} {\bibfield  {journal} {\bibinfo  {journal} {Phys. Rev. A}\ }\textbf {\bibinfo {volume} {94}},\ \bibinfo {pages} {033850} (\bibinfo {year} {2016})}\BibitemShut {NoStop}%
\bibitem [{\citenamefont {Ma}\ \emph {et~al.}(2021)\citenamefont {Ma}, \citenamefont {Puri}, \citenamefont {Schoelkopf}, \citenamefont {Devoret}, \citenamefont {Girvin},\ and\ \citenamefont {Jiang}}]{BosonicControlReview}%
  \BibitemOpen
  \bibfield  {author} {\bibinfo {author} {\bibfnamefont {W.-L.}\ \bibnamefont {Ma}}, \bibinfo {author} {\bibfnamefont {S.}~\bibnamefont {Puri}}, \bibinfo {author} {\bibfnamefont {R.~J.}\ \bibnamefont {Schoelkopf}}, \bibinfo {author} {\bibfnamefont {M.~H.}\ \bibnamefont {Devoret}}, \bibinfo {author} {\bibfnamefont {S.}~\bibnamefont {Girvin}},\ and\ \bibinfo {author} {\bibfnamefont {L.}~\bibnamefont {Jiang}},\ }\href {https://doi.org/https://doi.org/10.1016/j.scib.2021.05.024} {\bibfield  {journal} {\bibinfo  {journal} {Science Bulletin}\ }\textbf {\bibinfo {volume} {66}},\ \bibinfo {pages} {1789} (\bibinfo {year} {2021})}\BibitemShut {NoStop}%
\bibitem [{\citenamefont {Liu}\ \emph {et~al.}(2024)\citenamefont {Liu}, \citenamefont {Singh}, \citenamefont {Smith}, \citenamefont {Crane}, \citenamefont {Martyn}, \citenamefont {Eickbusch}, \citenamefont {Schuckert}, \citenamefont {Li}, \citenamefont {Sinanan-Singh}, \citenamefont {Soley}, \citenamefont {Tsunoda}, \citenamefont {Chuang}, \citenamefont {Wiebe},\ and\ \citenamefont {Girvin}}]{HybridQubitOscillatorReview}%
  \BibitemOpen
  \bibfield  {author} {\bibinfo {author} {\bibfnamefont {Y.}~\bibnamefont {Liu}}, \bibinfo {author} {\bibfnamefont {S.}~\bibnamefont {Singh}}, \bibinfo {author} {\bibfnamefont {K.~C.}\ \bibnamefont {Smith}}, \bibinfo {author} {\bibfnamefont {E.}~\bibnamefont {Crane}}, \bibinfo {author} {\bibfnamefont {J.~M.}\ \bibnamefont {Martyn}}, \bibinfo {author} {\bibfnamefont {A.}~\bibnamefont {Eickbusch}}, \bibinfo {author} {\bibfnamefont {A.}~\bibnamefont {Schuckert}}, \bibinfo {author} {\bibfnamefont {R.~D.}\ \bibnamefont {Li}}, \bibinfo {author} {\bibfnamefont {J.}~\bibnamefont {Sinanan-Singh}}, \bibinfo {author} {\bibfnamefont {M.~B.}\ \bibnamefont {Soley}}, \bibinfo {author} {\bibfnamefont {T.}~\bibnamefont {Tsunoda}}, \bibinfo {author} {\bibfnamefont {I.~L.}\ \bibnamefont {Chuang}}, \bibinfo {author} {\bibfnamefont {N.}~\bibnamefont {Wiebe}},\ and\ \bibinfo {author} {\bibfnamefont {S.~M.}\ \bibnamefont {Girvin}},\ }\href {https://arxiv.org/abs/2407.10381} {\bibinfo {title} {Hybrid oscillator-qubit quantum
  processors: Instruction set architectures, abstract machine models, and applications}} (\bibinfo {year} {2024}),\ \Eprint {https://arxiv.org/abs/2407.10381} {arXiv:2407.10381 [quant-ph]} \BibitemShut {NoStop}%
\bibitem [{\citenamefont {Niemczyk}\ \emph {et~al.}(2010{\natexlab{b}})\citenamefont {Niemczyk}, \citenamefont {Deppe}, \citenamefont {Huebl}, \citenamefont {Menzel}, \citenamefont {Hocke}, \citenamefont {Schwarz}, \citenamefont {Garcia-Ripoll}, \citenamefont {Zueco}, \citenamefont {H{\"u}mmer}, \citenamefont {Solano}, \citenamefont {Marx},\ and\ \citenamefont {Gross}}]{Niemczyk2010}%
  \BibitemOpen
  \bibfield  {author} {\bibinfo {author} {\bibfnamefont {T.}~\bibnamefont {Niemczyk}}, \bibinfo {author} {\bibfnamefont {F.}~\bibnamefont {Deppe}}, \bibinfo {author} {\bibfnamefont {H.}~\bibnamefont {Huebl}}, \bibinfo {author} {\bibfnamefont {E.~P.}\ \bibnamefont {Menzel}}, \bibinfo {author} {\bibfnamefont {F.}~\bibnamefont {Hocke}}, \bibinfo {author} {\bibfnamefont {M.~J.}\ \bibnamefont {Schwarz}}, \bibinfo {author} {\bibfnamefont {J.~J.}\ \bibnamefont {Garcia-Ripoll}}, \bibinfo {author} {\bibfnamefont {D.}~\bibnamefont {Zueco}}, \bibinfo {author} {\bibfnamefont {T.}~\bibnamefont {H{\"u}mmer}}, \bibinfo {author} {\bibfnamefont {E.}~\bibnamefont {Solano}}, \bibinfo {author} {\bibfnamefont {A.}~\bibnamefont {Marx}},\ and\ \bibinfo {author} {\bibfnamefont {R.}~\bibnamefont {Gross}},\ }\href {https://doi.org/10.1038/nphys1730} {\bibfield  {journal} {\bibinfo  {journal} {Nature Physics}\ }\textbf {\bibinfo {volume} {6}},\ \bibinfo {pages} {772} (\bibinfo {year} {2010}{\natexlab{b}})}\BibitemShut {NoStop}%
\bibitem [{\citenamefont {Sundaresan}\ \emph {et~al.}(2015)\citenamefont {Sundaresan}, \citenamefont {Liu}, \citenamefont {Sadri}, \citenamefont {Sz\ifmmode~\mbox{\H{o}}\else \H{o}\fi{}cs}, \citenamefont {Underwood}, \citenamefont {Malekakhlagh}, \citenamefont {T\"ureci},\ and\ \citenamefont {Houck}}]{Sundaresan_Houck_nonRWAMR}%
  \BibitemOpen
  \bibfield  {author} {\bibinfo {author} {\bibfnamefont {N.~M.}\ \bibnamefont {Sundaresan}}, \bibinfo {author} {\bibfnamefont {Y.}~\bibnamefont {Liu}}, \bibinfo {author} {\bibfnamefont {D.}~\bibnamefont {Sadri}}, \bibinfo {author} {\bibfnamefont {L.~J.}\ \bibnamefont {Sz\ifmmode~\mbox{\H{o}}\else \H{o}\fi{}cs}}, \bibinfo {author} {\bibfnamefont {D.~L.}\ \bibnamefont {Underwood}}, \bibinfo {author} {\bibfnamefont {M.}~\bibnamefont {Malekakhlagh}}, \bibinfo {author} {\bibfnamefont {H.~E.}\ \bibnamefont {T\"ureci}},\ and\ \bibinfo {author} {\bibfnamefont {A.~A.}\ \bibnamefont {Houck}},\ }\href {https://doi.org/10.1103/PhysRevX.5.021035} {\bibfield  {journal} {\bibinfo  {journal} {Phys. Rev. X}\ }\textbf {\bibinfo {volume} {5}},\ \bibinfo {pages} {021035} (\bibinfo {year} {2015})}\BibitemShut {NoStop}%
\bibitem [{\citenamefont {Ao}\ \emph {et~al.}(2023)\citenamefont {Ao}, \citenamefont {Ashhab}, \citenamefont {Yoshihara}, \citenamefont {Fuse}, \citenamefont {Kakuyanagi}, \citenamefont {Saito}, \citenamefont {Aoki},\ and\ \citenamefont {Semba}}]{Ao2023}%
  \BibitemOpen
  \bibfield  {author} {\bibinfo {author} {\bibfnamefont {Z.}~\bibnamefont {Ao}}, \bibinfo {author} {\bibfnamefont {S.}~\bibnamefont {Ashhab}}, \bibinfo {author} {\bibfnamefont {F.}~\bibnamefont {Yoshihara}}, \bibinfo {author} {\bibfnamefont {T.}~\bibnamefont {Fuse}}, \bibinfo {author} {\bibfnamefont {K.}~\bibnamefont {Kakuyanagi}}, \bibinfo {author} {\bibfnamefont {S.}~\bibnamefont {Saito}}, \bibinfo {author} {\bibfnamefont {T.}~\bibnamefont {Aoki}},\ and\ \bibinfo {author} {\bibfnamefont {K.}~\bibnamefont {Semba}},\ }\href {https://doi.org/10.1038/s41598-023-36547-w} {\bibfield  {journal} {\bibinfo  {journal} {Scientific Reports}\ }\textbf {\bibinfo {volume} {13}},\ \bibinfo {pages} {11340} (\bibinfo {year} {2023})}\BibitemShut {NoStop}%
\bibitem [{\citenamefont {Mariantoni}\ \emph {et~al.}(2008)\citenamefont {Mariantoni}, \citenamefont {Deppe}, \citenamefont {Marx}, \citenamefont {Gross}, \citenamefont {Wilhelm},\ and\ \citenamefont {Solano}}]{MariantoniQSwitch}%
  \BibitemOpen
  \bibfield  {author} {\bibinfo {author} {\bibfnamefont {M.}~\bibnamefont {Mariantoni}}, \bibinfo {author} {\bibfnamefont {F.}~\bibnamefont {Deppe}}, \bibinfo {author} {\bibfnamefont {A.}~\bibnamefont {Marx}}, \bibinfo {author} {\bibfnamefont {R.}~\bibnamefont {Gross}}, \bibinfo {author} {\bibfnamefont {F.~K.}\ \bibnamefont {Wilhelm}},\ and\ \bibinfo {author} {\bibfnamefont {E.}~\bibnamefont {Solano}},\ }\href {https://doi.org/10.1103/PhysRevB.78.104508} {\bibfield  {journal} {\bibinfo  {journal} {Phys. Rev. B}\ }\textbf {\bibinfo {volume} {78}},\ \bibinfo {pages} {104508} (\bibinfo {year} {2008})}\BibitemShut {NoStop}%
\bibitem [{\citenamefont {Chapman}\ \emph {et~al.}(2023)\citenamefont {Chapman}, \citenamefont {de~Graaf}, \citenamefont {Xue}, \citenamefont {Zhang}, \citenamefont {Teoh}, \citenamefont {Curtis}, \citenamefont {Tsunoda}, \citenamefont {Eickbusch}, \citenamefont {Read}, \citenamefont {Koottandavida}, \citenamefont {Mundhada}, \citenamefont {Frunzio}, \citenamefont {Devoret}, \citenamefont {Girvin},\ and\ \citenamefont {Schoelkopf}}]{ConditionalBS}%
  \BibitemOpen
  \bibfield  {author} {\bibinfo {author} {\bibfnamefont {B.~J.}\ \bibnamefont {Chapman}}, \bibinfo {author} {\bibfnamefont {S.~J.}\ \bibnamefont {de~Graaf}}, \bibinfo {author} {\bibfnamefont {S.~H.}\ \bibnamefont {Xue}}, \bibinfo {author} {\bibfnamefont {Y.}~\bibnamefont {Zhang}}, \bibinfo {author} {\bibfnamefont {J.}~\bibnamefont {Teoh}}, \bibinfo {author} {\bibfnamefont {J.~C.}\ \bibnamefont {Curtis}}, \bibinfo {author} {\bibfnamefont {T.}~\bibnamefont {Tsunoda}}, \bibinfo {author} {\bibfnamefont {A.}~\bibnamefont {Eickbusch}}, \bibinfo {author} {\bibfnamefont {A.~P.}\ \bibnamefont {Read}}, \bibinfo {author} {\bibfnamefont {A.}~\bibnamefont {Koottandavida}}, \bibinfo {author} {\bibfnamefont {S.~O.}\ \bibnamefont {Mundhada}}, \bibinfo {author} {\bibfnamefont {L.}~\bibnamefont {Frunzio}}, \bibinfo {author} {\bibfnamefont {M.}~\bibnamefont {Devoret}}, \bibinfo {author} {\bibfnamefont {S.}~\bibnamefont {Girvin}},\ and\ \bibinfo {author} {\bibfnamefont {R.}~\bibnamefont {Schoelkopf}},\ }\href
  {https://doi.org/10.1103/PRXQuantum.4.020355} {\bibfield  {journal} {\bibinfo  {journal} {PRX Quantum}\ }\textbf {\bibinfo {volume} {4}},\ \bibinfo {pages} {020355} (\bibinfo {year} {2023})}\BibitemShut {NoStop}%
\bibitem [{\citenamefont {Menicucci}\ \emph {et~al.}(2006)\citenamefont {Menicucci}, \citenamefont {van Loock}, \citenamefont {Gu}, \citenamefont {Weedbrook}, \citenamefont {Ralph},\ and\ \citenamefont {Nielsen}}]{UniversalQCCV}%
  \BibitemOpen
  \bibfield  {author} {\bibinfo {author} {\bibfnamefont {N.~C.}\ \bibnamefont {Menicucci}}, \bibinfo {author} {\bibfnamefont {P.}~\bibnamefont {van Loock}}, \bibinfo {author} {\bibfnamefont {M.}~\bibnamefont {Gu}}, \bibinfo {author} {\bibfnamefont {C.}~\bibnamefont {Weedbrook}}, \bibinfo {author} {\bibfnamefont {T.~C.}\ \bibnamefont {Ralph}},\ and\ \bibinfo {author} {\bibfnamefont {M.~A.}\ \bibnamefont {Nielsen}},\ }\href {https://doi.org/10.1103/PhysRevLett.97.110501} {\bibfield  {journal} {\bibinfo  {journal} {Phys. Rev. Lett.}\ }\textbf {\bibinfo {volume} {97}},\ \bibinfo {pages} {110501} (\bibinfo {year} {2006})}\BibitemShut {NoStop}%
\bibitem [{\citenamefont {Wang}\ \emph {et~al.}(2016)\citenamefont {Wang}, \citenamefont {Gao}, \citenamefont {Reinhold}, \citenamefont {Heeres}, \citenamefont {Ofek}, \citenamefont {Chou}, \citenamefont {Axline}, \citenamefont {Reagor}, \citenamefont {Blumoff}, \citenamefont {Sliwa}, \citenamefont {Frunzio}, \citenamefont {Girvin}, \citenamefont {Jiang}, \citenamefont {Mirrahimi}, \citenamefont {Devoret},\ and\ \citenamefont {Schoelkopf}}]{Wang_CatTwoBoxes2016}%
  \BibitemOpen
  \bibfield  {author} {\bibinfo {author} {\bibfnamefont {C.}~\bibnamefont {Wang}}, \bibinfo {author} {\bibfnamefont {Y.~Y.}\ \bibnamefont {Gao}}, \bibinfo {author} {\bibfnamefont {P.}~\bibnamefont {Reinhold}}, \bibinfo {author} {\bibfnamefont {R.~W.}\ \bibnamefont {Heeres}}, \bibinfo {author} {\bibfnamefont {N.}~\bibnamefont {Ofek}}, \bibinfo {author} {\bibfnamefont {K.}~\bibnamefont {Chou}}, \bibinfo {author} {\bibfnamefont {C.}~\bibnamefont {Axline}}, \bibinfo {author} {\bibfnamefont {M.}~\bibnamefont {Reagor}}, \bibinfo {author} {\bibfnamefont {J.}~\bibnamefont {Blumoff}}, \bibinfo {author} {\bibfnamefont {K.~M.}\ \bibnamefont {Sliwa}}, \bibinfo {author} {\bibfnamefont {L.}~\bibnamefont {Frunzio}}, \bibinfo {author} {\bibfnamefont {S.~M.}\ \bibnamefont {Girvin}}, \bibinfo {author} {\bibfnamefont {L.}~\bibnamefont {Jiang}}, \bibinfo {author} {\bibfnamefont {M.}~\bibnamefont {Mirrahimi}}, \bibinfo {author} {\bibfnamefont {M.~H.}\ \bibnamefont {Devoret}},\ and\ \bibinfo {author} {\bibfnamefont {R.~J.}\
  \bibnamefont {Schoelkopf}},\ }\href {https://doi.org/10.1126/science.aaf2941} {\bibfield  {journal} {\bibinfo  {journal} {Science}\ }\textbf {\bibinfo {volume} {352}},\ \bibinfo {pages} {1087} (\bibinfo {year} {2016})},\ \Eprint {https://arxiv.org/abs/https://www.science.org/doi/pdf/10.1126/science.aaf2941} {https://www.science.org/doi/pdf/10.1126/science.aaf2941} \BibitemShut {NoStop}%
\bibitem [{\citenamefont {Rico}\ \emph {et~al.}(2020)\citenamefont {Rico}, \citenamefont {Maldonado-Villamizar},\ and\ \citenamefont {Rodriguez-Lara}}]{SpecCollapseTwoPh}%
  \BibitemOpen
  \bibfield  {author} {\bibinfo {author} {\bibfnamefont {R.~J.~A.}\ \bibnamefont {Rico}}, \bibinfo {author} {\bibfnamefont {F.~H.}\ \bibnamefont {Maldonado-Villamizar}},\ and\ \bibinfo {author} {\bibfnamefont {B.~M.}\ \bibnamefont {Rodriguez-Lara}},\ }\href {https://doi.org/10.1103/PhysRevA.101.063825} {\bibfield  {journal} {\bibinfo  {journal} {Phys. Rev. A}\ }\textbf {\bibinfo {volume} {101}},\ \bibinfo {pages} {063825} (\bibinfo {year} {2020})}\BibitemShut {NoStop}%
\bibitem [{\citenamefont {Braak}(2025)}]{Braak_kPhotonRabiModel}%
  \BibitemOpen
  \bibfield  {author} {\bibinfo {author} {\bibfnamefont {D.}~\bibnamefont {Braak}},\ }\href {https://arxiv.org/abs/2401.02370} {\bibinfo {title} {The $k$-photon quantum rabi model}} (\bibinfo {year} {2025}),\ \Eprint {https://arxiv.org/abs/2401.02370} {arXiv:2401.02370 [quant-ph]} \BibitemShut {NoStop}%
\bibitem [{\citenamefont {Chang}\ \emph {et~al.}(2020)\citenamefont {Chang}, \citenamefont {Sab\'{\i}n}, \citenamefont {Forn-D\'{\i}az}, \citenamefont {Quijandr\'{\i}a}, \citenamefont {Vadiraj}, \citenamefont {Nsanzineza}, \citenamefont {Johansson},\ and\ \citenamefont {Wilson}}]{CWilsonPRX}%
  \BibitemOpen
  \bibfield  {author} {\bibinfo {author} {\bibfnamefont {C.~W.~S.}\ \bibnamefont {Chang}}, \bibinfo {author} {\bibfnamefont {C.}~\bibnamefont {Sab\'{\i}n}}, \bibinfo {author} {\bibfnamefont {P.}~\bibnamefont {Forn-D\'{\i}az}}, \bibinfo {author} {\bibfnamefont {F.}~\bibnamefont {Quijandr\'{\i}a}}, \bibinfo {author} {\bibfnamefont {A.~M.}\ \bibnamefont {Vadiraj}}, \bibinfo {author} {\bibfnamefont {I.}~\bibnamefont {Nsanzineza}}, \bibinfo {author} {\bibfnamefont {G.}~\bibnamefont {Johansson}},\ and\ \bibinfo {author} {\bibfnamefont {C.~M.}\ \bibnamefont {Wilson}},\ }\href {https://doi.org/10.1103/PhysRevX.10.011011} {\bibfield  {journal} {\bibinfo  {journal} {Phys. Rev. X}\ }\textbf {\bibinfo {volume} {10}},\ \bibinfo {pages} {011011} (\bibinfo {year} {2020})}\BibitemShut {NoStop}%
\bibitem [{\citenamefont {Bravyi}\ \emph {et~al.}(2011)\citenamefont {Bravyi}, \citenamefont {DiVincenzo},\ and\ \citenamefont {Loss}}]{BRAVYI20112793}%
  \BibitemOpen
  \bibfield  {author} {\bibinfo {author} {\bibfnamefont {S.}~\bibnamefont {Bravyi}}, \bibinfo {author} {\bibfnamefont {D.~P.}\ \bibnamefont {DiVincenzo}},\ and\ \bibinfo {author} {\bibfnamefont {D.}~\bibnamefont {Loss}},\ }\href {https://doi.org/https://doi.org/10.1016/j.aop.2011.06.004} {\bibfield  {journal} {\bibinfo  {journal} {Annals of Physics}\ }\textbf {\bibinfo {volume} {326}},\ \bibinfo {pages} {2793} (\bibinfo {year} {2011})}\BibitemShut {NoStop}%
\bibitem [{\citenamefont {Boissonneault}\ \emph {et~al.}(2010{\natexlab{b}})\citenamefont {Boissonneault}, \citenamefont {Gambetta},\ and\ \citenamefont {Blais}}]{NonlinearReadout}%
  \BibitemOpen
  \bibfield  {author} {\bibinfo {author} {\bibfnamefont {M.}~\bibnamefont {Boissonneault}}, \bibinfo {author} {\bibfnamefont {J.~M.}\ \bibnamefont {Gambetta}},\ and\ \bibinfo {author} {\bibfnamefont {A.}~\bibnamefont {Blais}},\ }\href {https://doi.org/10.1103/PhysRevLett.105.100504} {\bibfield  {journal} {\bibinfo  {journal} {Phys. Rev. Lett.}\ }\textbf {\bibinfo {volume} {105}},\ \bibinfo {pages} {100504} (\bibinfo {year} {2010}{\natexlab{b}})}\BibitemShut {NoStop}%
\bibitem [{\citenamefont {Miranowicz}\ \emph {et~al.}(2013)\citenamefont {Miranowicz}, \citenamefont {Paprzycka}, \citenamefont {Liu}, \citenamefont {Bajer},\ and\ \citenamefont {Nori}}]{Miranowicz_PhotonBlockade}%
  \BibitemOpen
  \bibfield  {author} {\bibinfo {author} {\bibfnamefont {A.}~\bibnamefont {Miranowicz}}, \bibinfo {author} {\bibfnamefont {M.}~\bibnamefont {Paprzycka}}, \bibinfo {author} {\bibfnamefont {Y.-x.}\ \bibnamefont {Liu}}, \bibinfo {author} {\bibfnamefont {J.~c.~v.}\ \bibnamefont {Bajer}},\ and\ \bibinfo {author} {\bibfnamefont {F.}~\bibnamefont {Nori}},\ }\href {https://doi.org/10.1103/PhysRevA.87.023809} {\bibfield  {journal} {\bibinfo  {journal} {Phys. Rev. A}\ }\textbf {\bibinfo {volume} {87}},\ \bibinfo {pages} {023809} (\bibinfo {year} {2013})}\BibitemShut {NoStop}%
\bibitem [{\citenamefont {Yurke}\ and\ \citenamefont {Stoler}(1986)}]{Yurke_AmplitudeDispersion}%
  \BibitemOpen
  \bibfield  {author} {\bibinfo {author} {\bibfnamefont {B.}~\bibnamefont {Yurke}}\ and\ \bibinfo {author} {\bibfnamefont {D.}~\bibnamefont {Stoler}},\ }\href {https://doi.org/10.1103/PhysRevLett.57.13} {\bibfield  {journal} {\bibinfo  {journal} {Phys. Rev. Lett.}\ }\textbf {\bibinfo {volume} {57}},\ \bibinfo {pages} {13} (\bibinfo {year} {1986})}\BibitemShut {NoStop}%
\bibitem [{\citenamefont {Smith}\ \emph {et~al.}(2021)\citenamefont {Smith}, \citenamefont {Bhattacharya},\ and\ \citenamefont {Masiello}}]{KCSmith_ExactKBodyJC}%
  \BibitemOpen
  \bibfield  {author} {\bibinfo {author} {\bibfnamefont {K.~C.}\ \bibnamefont {Smith}}, \bibinfo {author} {\bibfnamefont {A.}~\bibnamefont {Bhattacharya}},\ and\ \bibinfo {author} {\bibfnamefont {D.~J.}\ \bibnamefont {Masiello}},\ }\href {https://doi.org/10.1103/PhysRevA.104.013707} {\bibfield  {journal} {\bibinfo  {journal} {Phys. Rev. A}\ }\textbf {\bibinfo {volume} {104}},\ \bibinfo {pages} {013707} (\bibinfo {year} {2021})}\BibitemShut {NoStop}%
\bibitem [{\citenamefont {Katriel}(2000)}]{KATRIEL2000159}%
  \BibitemOpen
  \bibfield  {author} {\bibinfo {author} {\bibfnamefont {J.}~\bibnamefont {Katriel}},\ }\href {https://doi.org/https://doi.org/10.1016/S0375-9601(00)00488-6} {\bibfield  {journal} {\bibinfo  {journal} {Physics Letters A}\ }\textbf {\bibinfo {volume} {273}},\ \bibinfo {pages} {159} (\bibinfo {year} {2000})}\BibitemShut {NoStop}%
\bibitem [{\citenamefont {Ashhab}\ and\ \citenamefont {Semba}(2017)}]{Ashhab_SuperradianceParameterFluctuation}%
  \BibitemOpen
  \bibfield  {author} {\bibinfo {author} {\bibfnamefont {S.}~\bibnamefont {Ashhab}}\ and\ \bibinfo {author} {\bibfnamefont {K.}~\bibnamefont {Semba}},\ }\href {https://doi.org/10.1103/PhysRevA.95.053833} {\bibfield  {journal} {\bibinfo  {journal} {Phys. Rev. A}\ }\textbf {\bibinfo {volume} {95}},\ \bibinfo {pages} {053833} (\bibinfo {year} {2017})}\BibitemShut {NoStop}%
\bibitem [{\citenamefont {Garbe}\ \emph {et~al.}(2017)\citenamefont {Garbe}, \citenamefont {Egusquiza}, \citenamefont {Solano}, \citenamefont {Ciuti}, \citenamefont {Coudreau}, \citenamefont {Milman},\ and\ \citenamefont {Felicetti}}]{Garbe_TwoPhDicke_SPT}%
  \BibitemOpen
  \bibfield  {author} {\bibinfo {author} {\bibfnamefont {L.}~\bibnamefont {Garbe}}, \bibinfo {author} {\bibfnamefont {I.~L.}\ \bibnamefont {Egusquiza}}, \bibinfo {author} {\bibfnamefont {E.}~\bibnamefont {Solano}}, \bibinfo {author} {\bibfnamefont {C.}~\bibnamefont {Ciuti}}, \bibinfo {author} {\bibfnamefont {T.}~\bibnamefont {Coudreau}}, \bibinfo {author} {\bibfnamefont {P.}~\bibnamefont {Milman}},\ and\ \bibinfo {author} {\bibfnamefont {S.}~\bibnamefont {Felicetti}},\ }\href {https://doi.org/10.1103/PhysRevA.95.053854} {\bibfield  {journal} {\bibinfo  {journal} {Phys. Rev. A}\ }\textbf {\bibinfo {volume} {95}},\ \bibinfo {pages} {053854} (\bibinfo {year} {2017})}\BibitemShut {NoStop}%
\bibitem [{\citenamefont {Chen}\ and\ \citenamefont {Zhang}(2018)}]{Chen_TwoPhDicke_SPT}%
  \BibitemOpen
  \bibfield  {author} {\bibinfo {author} {\bibfnamefont {X.-Y.}\ \bibnamefont {Chen}}\ and\ \bibinfo {author} {\bibfnamefont {Y.-Y.}\ \bibnamefont {Zhang}},\ }\href {https://doi.org/10.1103/PhysRevA.97.053821} {\bibfield  {journal} {\bibinfo  {journal} {Phys. Rev. A}\ }\textbf {\bibinfo {volume} {97}},\ \bibinfo {pages} {053821} (\bibinfo {year} {2018})}\BibitemShut {NoStop}%
\bibitem [{\citenamefont {Eickbusch}\ \emph {et~al.}(2022)\citenamefont {Eickbusch}, \citenamefont {Sivak}, \citenamefont {Ding}, \citenamefont {Elder}, \citenamefont {Jha}, \citenamefont {Venkatraman}, \citenamefont {Royer}, \citenamefont {Girvin}, \citenamefont {Schoelkopf},\ and\ \citenamefont {Devoret}}]{FastUnivControlDisp}%
  \BibitemOpen
  \bibfield  {author} {\bibinfo {author} {\bibfnamefont {A.}~\bibnamefont {Eickbusch}}, \bibinfo {author} {\bibfnamefont {V.}~\bibnamefont {Sivak}}, \bibinfo {author} {\bibfnamefont {A.~Z.}\ \bibnamefont {Ding}}, \bibinfo {author} {\bibfnamefont {S.~S.}\ \bibnamefont {Elder}}, \bibinfo {author} {\bibfnamefont {S.~R.}\ \bibnamefont {Jha}}, \bibinfo {author} {\bibfnamefont {J.}~\bibnamefont {Venkatraman}}, \bibinfo {author} {\bibfnamefont {B.}~\bibnamefont {Royer}}, \bibinfo {author} {\bibfnamefont {S.~M.}\ \bibnamefont {Girvin}}, \bibinfo {author} {\bibfnamefont {R.~J.}\ \bibnamefont {Schoelkopf}},\ and\ \bibinfo {author} {\bibfnamefont {M.~H.}\ \bibnamefont {Devoret}},\ }\href {https://doi.org/10.1038/s41567-022-01776-9} {\bibfield  {journal} {\bibinfo  {journal} {Nature Physics}\ }\textbf {\bibinfo {volume} {18}},\ \bibinfo {pages} {1464} (\bibinfo {year} {2022})}\BibitemShut {NoStop}%
\bibitem [{\citenamefont {Koppenh\"ofer}\ \emph {et~al.}(2022)\citenamefont {Koppenh\"ofer}, \citenamefont {Groszkowski}, \citenamefont {Lau},\ and\ \citenamefont {Clerk}}]{SuperradiantSensing}%
  \BibitemOpen
  \bibfield  {author} {\bibinfo {author} {\bibfnamefont {M.}~\bibnamefont {Koppenh\"ofer}}, \bibinfo {author} {\bibfnamefont {P.}~\bibnamefont {Groszkowski}}, \bibinfo {author} {\bibfnamefont {H.-K.}\ \bibnamefont {Lau}},\ and\ \bibinfo {author} {\bibfnamefont {A.}~\bibnamefont {Clerk}},\ }\href {https://doi.org/10.1103/PRXQuantum.3.030330} {\bibfield  {journal} {\bibinfo  {journal} {PRX Quantum}\ }\textbf {\bibinfo {volume} {3}},\ \bibinfo {pages} {030330} (\bibinfo {year} {2022})}\BibitemShut {NoStop}%
\bibitem [{\citenamefont {Blasiak}\ \emph {et~al.}(2007)\citenamefont {Blasiak}, \citenamefont {Horzela}, \citenamefont {Penson}, \citenamefont {Solomon},\ and\ \citenamefont {Duchamp}}]{ReorderingIntro}%
  \BibitemOpen
  \bibfield  {author} {\bibinfo {author} {\bibfnamefont {P.}~\bibnamefont {Blasiak}}, \bibinfo {author} {\bibfnamefont {A.}~\bibnamefont {Horzela}}, \bibinfo {author} {\bibfnamefont {K.~A.}\ \bibnamefont {Penson}}, \bibinfo {author} {\bibfnamefont {A.~I.}\ \bibnamefont {Solomon}},\ and\ \bibinfo {author} {\bibfnamefont {G.~H.~E.}\ \bibnamefont {Duchamp}},\ }\href {https://doi.org/10.1119/1.2723799} {\bibfield  {journal} {\bibinfo  {journal} {American Journal of Physics}\ }\textbf {\bibinfo {volume} {75}},\ \bibinfo {pages} {639} (\bibinfo {year} {2007})},\ \Eprint {https://arxiv.org/abs/https://pubs.aip.org/aapt/ajp/article-pdf/75/7/639/13083703/639\_1\_online.pdf} {https://pubs.aip.org/aapt/ajp/article-pdf/75/7/639/13083703/639\_1\_online.pdf} \BibitemShut {NoStop}%
\bibitem [{\citenamefont {Blasiak}\ \emph {et~al.}(2003)\citenamefont {Blasiak}, \citenamefont {Penson},\ and\ \citenamefont {Solomon}}]{Reordering2}%
  \BibitemOpen
  \bibfield  {author} {\bibinfo {author} {\bibfnamefont {P.}~\bibnamefont {Blasiak}}, \bibinfo {author} {\bibfnamefont {K.~A.}\ \bibnamefont {Penson}},\ and\ \bibinfo {author} {\bibfnamefont {A.~I.}\ \bibnamefont {Solomon}},\ }\href {https://doi.org/10.1007/s00026-003-0177-z} {\bibfield  {journal} {\bibinfo  {journal} {Annals of Combinatorics}\ }\textbf {\bibinfo {volume} {7}},\ \bibinfo {pages} {127} (\bibinfo {year} {2003})}\BibitemShut {NoStop}%
\bibitem [{\citenamefont {Ying}\ \emph {et~al.}(2025)\citenamefont {Ying}, \citenamefont {Han}, \citenamefont {Li}, \citenamefont {Felicetti},\ and\ \citenamefont {Braak}}]{Ying_CritMetroStab2025}%
  \BibitemOpen
  \bibfield  {author} {\bibinfo {author} {\bibfnamefont {Z.-J.}\ \bibnamefont {Ying}}, \bibinfo {author} {\bibfnamefont {H.-H.}\ \bibnamefont {Han}}, \bibinfo {author} {\bibfnamefont {B.-J.}\ \bibnamefont {Li}}, \bibinfo {author} {\bibfnamefont {S.}~\bibnamefont {Felicetti}},\ and\ \bibinfo {author} {\bibfnamefont {D.}~\bibnamefont {Braak}},\ }\href {https://arxiv.org/abs/2503.19198} {\bibinfo {title} {Critical quantum metrology in a stabilized two-photon rabi model}} (\bibinfo {year} {2025}),\ \Eprint {https://arxiv.org/abs/2503.19198} {arXiv:2503.19198 [quant-ph]} \BibitemShut {NoStop}%
\bibitem [{\citenamefont {Ashhab}\ and\ \citenamefont {Nori}(2010)}]{Ashhab_QO_USC_PRA2010}%
  \BibitemOpen
  \bibfield  {author} {\bibinfo {author} {\bibfnamefont {S.}~\bibnamefont {Ashhab}}\ and\ \bibinfo {author} {\bibfnamefont {F.}~\bibnamefont {Nori}},\ }\href {https://doi.org/10.1103/PhysRevA.81.042311} {\bibfield  {journal} {\bibinfo  {journal} {Phys. Rev. A}\ }\textbf {\bibinfo {volume} {81}},\ \bibinfo {pages} {042311} (\bibinfo {year} {2010})}\BibitemShut {NoStop}%
\end{thebibliography}%

\end{document}
%